\documentclass[showpacs,preprintnumbers,amssymb,twocolum,aps,reprint,superscriptaddress,showkeys,amsmath,floatfix]{revtex4-2}

\usepackage{graphicx}
\usepackage{dcolumn}
\usepackage[dvipsnames]{xcolor}
\usepackage{bm}
\usepackage{amsmath}
\usepackage{amssymb}
\usepackage{epsfig}
\usepackage{amsfonts}
\usepackage{lineno,hyperref}
\usepackage{array}
\usepackage{tikz}
\usepackage{float}
\usepackage{microtype}
\usepackage{multirow}
\usepackage{adjustbox}
\usepackage[english]{babel}
\usepackage{epstopdf}
\usepackage{blindtext}
\usepackage{booktabs}
\usepackage{subcaption}
\usepackage[utf8]{inputenc}
\usepackage{tikz}
\usepackage{soul}
\usepackage[a4paper, total={6.8in, 10in}]{geometry}
\usepackage{appendix}

\def \a{\alpha}
\def \b{\beta}
\def \l{\lambda}

\def \g{\gamma}

\def \be{\begin{equation}}
\def \ee{\end{equation}}
\def \ben{\begin{eqnarray}}
\def \een{\end{eqnarray}}

\def \n{\nonumber\\}
\def \G{\bar{G}}
\def \Ga{\bar{\Gamma}}

\def \T{\bar{T}}
\def \La{\mathcal{L}}

\def \f{f({\bf x}, \bar{{\bf p}}, t)}
\def \E{\widetilde{E}}
\def \p{\bar{\textbf{p}}}
\def \q{\bar{\textbf{q}}}
\def \x{\textbf{x}}
\def \pa{\bar{p}}
\def \qa{\bar{q}}

\def \ua{\textbf{u}}

\begin{document}

\title{Boltzmann Dynamics in K-essence Cosmology: Photon Propagation in an Emergent Spacetime}

\author{Krishnendu Roy}
\email{krishnenduroy20294@gmail.com}
\affiliation{Department of Physics, University of Calcutta, 92, A.P.C. Road, Kolkata-700009, India}

\author{Debashis Gangopadhyay}
\email{debashis.g@snuniv.ac.in}
\affiliation{Department of Physics, School of Natural Sciences, Sister Nivedita University, DG 1/2, Action Area 1, Newtown, Kolkata 700156, India.}

\author{Chiranjeeb Singha}
\email{chiranjeeb.singha@iucaa.in}
\affiliation{Inter-University Center for Astronomy and Astrophysics, Post Bag 4, Pune 411 007, India}

\author{Goutam Manna$^a$}
\email{goutammanna.pkc@gmail.com \\$^a$Corresponding author}
\affiliation{Department of Physics, Prabhat Kumar College, Contai, Purba Medinipur 721404, India} 
\affiliation{Institute of Astronomy, Space and Earth Science, Kolkata 700054, India}

\date{\today}

\begin{abstract}
Recent cosmological tensions, notably the Hubble and $S_{8}$ tensions, motivate extensions of the conventional FLRW framework in which additional dynamical fields can alter the effective spacetime experienced by matter and radiation. In K-essence cosmology, the scalar field induces an emergent FLRW geometry disformally related to the gravitational metric, leading to a \emph{tilted causal structure} in which the photon light cone differs from that of gravity. We develop a covariant Boltzmann formalism in a homogeneous K-essence background and derive the modified mass-shell condition, geodesic equations, and collision integrals for both massless and massive species. The photon distribution remains thermal in the emergent frame, while appearing geometrically rescaled in the gravitational frame. Microphysical Thomson/Compton scattering is unmodified; instead, the emergent geometry induces effective rest-frame energies and a geometrically rescaled scattering rate, conveniently encoded in a transport coefficient $\sigma_T^{(\mathrm{geom})}$. Focusing on the tightly coupled pre-recombination era, we obtain modified acoustic dynamics and, for a purely kinetic DBI-type model, find an effective interaction rate scaling $n_{e}\sigma_T^{(\mathrm{geom})}a\propto a^{-8}$ and a diffusion scale scaling $k_{D}^{-2}\propto a^{29/2}$, enhancing the sensitivity of small-scale anisotropies to the evolving geometry. Our results provide a self-consistent kinetic description of particle transport in the emergent K-essence spacetime, and they lay the groundwork for a full numerical Einstein-Boltzmann implementation to compute post-recombination free streaming and the CMB angular power spectrum. A detailed numerical analysis will be pursued in future work.
\end{abstract}



\maketitle

\section{Introduction}

The standard cosmological model, FLRW, has successfully explained a broad spectrum of observational evidence, covering cosmic microwave background (CMB) anisotropies, large-scale structure, baryon acoustic oscillations, and primordial nucleosynthesis. In recent years, several statistically significant disputes have arisen between independent observational probes \cite{Riess1, Verde, Abdalla, Valentino, Riess2, Tegmark, Spergel, Perlmutter}. The most significant issue is the Hubble tension, which refers to the discrepancy between the locally measured value of the Hubble constant and that inferred from early-universe observations, such as the CMB. Furthermore, the $S_8$ tension associated with the magnitude of matter density fluctuations reveals a discrepancy between weak lensing surveys and expectations derived from the CMB. These persistent anomalies suggest that either unknown systematic
effects are present in the data or that the minimal FLRW framework may require
extensions at the level of fundamental physics.

From a theoretical perspective, these tensions generated significant interest in modifications to general relativity and in models incorporating additional dynamical degrees of freedom, such as scalar fields, early dark energy, or interacting dark sectors \cite{Clifton, Joyce, Koyama, Valentino1}. These kinds of extensions can alter the history of the background expansion and the evolution of cosmic perturbations over time. This could leave marks on the CMB and large-scale structure that we might observe. To test these ideas against empirical evidence, though, one has to look at non-equilibrium processes, particle propagation, and scattering in an expanding and perturbed spacetime in a consistent and systematic way.

The Boltzmann equation, which describes how the phase-space distributions of photons, neutrinos, baryons, and dark matter evolve over time in cosmology, provides an appropriate theoretical foundation for addressing the above issues \cite{Dodelson, Ma}. The Boltzmann hierarchy is the basis of contemporary cosmological perturbation theory and enables high-precision predictions of CMB anisotropies, polarization, and the matter power spectrum. Compton scattering, recombination, decoupling, and free-streaming are only a few of the important physical processes that are naturally included in the kinetic description.

In the conventional treatment, the Boltzmann equations are formulated within a perturbed FLRW geometry derived from Einstein's gravity, with all particle species following geodesics of the gravitational metric $g_{\mu\nu}$. This method has been quite successful and is used in many popular numerical codes, such as \texttt{CAMB} and \texttt{CLASS}. However, if gravity is altered or if supplementary fields significantly influence the causal structure of spacetime, the conventional Boltzmann framework necessitates appropriate generalization. In these instances, the effective metric governing particle propagation and interactions may differ from the gravitational metric, leading to modified dispersion relations and kinetic dynamics.

Scalar-field theories incorporating non-canonical kinetic terms, exemplified by K-essence, have emerged as well-founded extensions of conventional cosmology, with significant implications in inflationary dynamics, dark energy phenomenology, and emergent gravity frameworks \cite{Garriga, Picon1, Picon2, Picon3, Padmanabhan, Scherrer, Babichev, Chimento, Visser, Vikman, Deffayet}. K-essence was first used to explain how non-standard scalar fields work \cite{Garriga}. It naturally leads to an emergent or effective spacetime geometry that is disformally connected to the underlying gravitational metric $g_{\mu\nu}$ \cite{Bekenstein_1993}. In this framework, perturbations and matter fields propagate on an effective metric $\bar{G}_{\mu\nu}$ instead of the usual gravitational metric $g_{\mu\nu}$ \cite{Babichev, Deffayet, Sawicki, Mukohyama, gm1, gm2, gm3, Vikman}. This emergent geometry typically displays an altered causal structure and a tilted light cone, indicating that physical trajectories, signal propagation, and causal relationships are regulated by $\bar{G}_{\mu\nu}$ rather than by the background spacetime $g_{\mu\nu}$.

Because the emergent geometry is deformed, physically meaningful quantities, such as particle energies, temperatures, and the effective rest-frame energy, acquire a time-dependent rescaling governed by the scalar-field background. While the background cosmological evolution of K-essence models has been studied extensively in the literature \cite{Garriga, Picon1, Picon2, Picon3, Padmanabhan, Scherrer, Babichev, Chimento, Visser, Vikman, Deffayet, Babichev, Deffayet, Sawicki, Mukohyama, gm1, gm2, gm3, gm4, gm5, Panda1, Panda2, Panda3, Ganguly}, a systematic kinetic-theory treatment, especially a Boltzmann formulation fully adapted to the emergent geometry, remains comparatively underdeveloped. In particular, the consequences of the modified causal structure for distribution functions, scattering, and cosmological transport have not yet been incorporated in a uniform, geometrically consistent way. The primary objective of this work is, therefore, to construct a self-consistent Boltzmann framework for photons and matter propagating in the emergent K-essence spacetime and to assess, within a controlled, covariant setup, how effective rest-frame energies, rescaled interaction rates, and tilted light cones feed into cosmological observables.

The primary goal of this work is to construct a consistent Boltzmann framework for cosmology in a homogeneous K-essence background, where all particle species propagate according to the \emph{emergent FLRW geometry}. Within this setting, we derive the corresponding mass-shell conditions, geodesic equations, and collision terms and show that standard kinetic processes, including Thomson and Compton scattering, retain their usual structure while acquiring geometrically induced effective rest-frame energies and rescaled interaction rates as a direct consequence of the tilted causal structure of the emergent spacetime.

This approach enables a consistent connection between scalar-field dynamics and cosmic transport phenomena, including temperature evolution, acoustic oscillations, and diffusion damping, within a unified covariant framework. The primary aim is to establish a theoretically rigorous framework that may allow for the precise confrontation of emergent gravity and K-essence cosmologies with cosmological observations, especially those related to the cosmic microwave background and large-scale structure, thereby offering a potential avenue to probe departures from the standard FLRW paradigm.\\

It is essential to clarify the conceptual framework of the current study concerning disformal transformations and K-essence theories. A disformal transformation with $(X,\phi)$-dependent coefficients is known to transfer a K-essence scalar field theory into another theory of the same class, without introducing fundamentally new dynamical degrees of freedom \cite{Bekenstein_1993, Zumalacarregui}. In this context, the fundamental action examined in this study operates within the conventional K-essence paradigm and is minimally connected to Einsteinian gravity. Still, the primary objective of the current study is not the formulation of a new gravitational theory, but the establishment of a coherent kinetic (Boltzmann) framework for characterizing matter and radiation inside the emergent spacetime generated by the scalar field. In our formulation, the disformally associated metric $\bar{G}_{\mu\nu}$ is regarded as the effective geometry that governs particle propagation, thereby endowing it with intrinsic physical significance beyond a mere reinterpretation of the field.

In this interpretation, the emergent metric alters the causal structure, dispersion relations, and geodesic motion of particles, thereby significantly generalizing the conventional cosmological Boltzmann framework. The modified mass-shell condition, the effective energy $\widetilde{E}$, and the interaction rates introduce corrections to the Liouville operator and collision terms that are absent in the standard FLRW framework. Although scalar-field theory is part of the K-essence class, the kinetic description of photons and matter fields differs in a physically meaningful way. This could lead to observable effects such as modified sound horizons, diffusion-damping scales, and interaction rates. This divergence aligns with the broader understanding that different metric frames connected by disformal transformations may provide non-equivalent physical interpretations when matter interacts with an effective geometry \cite{Bettoni, Domenech2015}. Consequently, the innovation of the current study resides in the formulation of a covariant and observationally pertinent Boltzmann framework within the emergent K-essence spacetime.\\

This paper is organized as follows. In Sec. \ref{K-essence} we briefly review the K-essence emergent geometry and its relation to the background gravitational spacetime. In Sec. \ref{S3} we derive the modified Boltzmann equations for both massless and massive particles in a spatially flat FLRW background influenced by the K-essence scalar field. In Sec. \ref{S4}, we develop the evolution of perturbations within the newly formed modified Boltzmann equation in the K-essence geometry. In Sec. \ref{S5} we specialize to photon dynamics and obtain the perturbed Boltzmann hierarchy in the K-essence FLRW spacetime. In Sec. \ref{S6} we analyze the solutions of these equations in the pre-recombination era, focusing on tight coupling, acoustic oscillations, and diffusion damping. Finally, Sec. \ref{S7} summarizes our results and conclusions.

For the solutions of the modified perturbed Boltzmann equation, we restrict our analysis to the pre-recombination epoch and concentrate on photon dynamics in this regime, leaving post-recombination evolution and observational applications for future investigation.

\section{Overview of the K-essence geometry}\label{K-essence}
In this section, we provide a brief overview of the K-essence theory. The K-essence geometry describes the interaction between a scalar field and gravity, which may be expressed by the following action when the scalar field is minimally coupled to gravity~\cite{Picon1, Picon2, Picon3, Scherrer, Chimento, Babichev, Vikman, Visser, Mukohyama, gm1, gm2, gm3}, which is expressed as,
\ben
S_{k}[\phi,g_{\mu\nu}]= \int d^{4}x {\sqrt -g} \La(X,\phi),
\label{1}
\een
where $X=\frac{1}{2}g^{\mu\nu}\nabla_{\mu}\phi\nabla_{\nu}\phi$ is the canonical kinetic term and $\La(X,\phi)$ is the non-canonical Lagrangian.

The corresponding energy-momentum tensor is given by,
\ben
&&T_{\mu\nu}\equiv \frac{-2}{\sqrt {-g}}\frac{\delta S_{k}}{\delta g^{\mu\nu}}=-2\frac{\partial \La}{\partial g^{\mu\nu}}+g_{\mu\nu}\La\nonumber\\&&=-\La_{X}\nabla_{\mu}\phi\nabla_{\nu}\phi
+g_{\mu\nu}\La. 
\label{2}
\een
The covariant derivative $\nabla_{\mu}$ is defined with respect to the standard gravitational metric $g_{\mu\nu}$. 

The associated equation of motion (EOM) for the scalar field is 
\ben
-\frac{1}{\sqrt {-g}}\frac{\delta S_{k}}{\delta \phi}= \widetilde{G}^{\mu\nu}\nabla_{\mu}\nabla_{\nu}\phi +2X\La_{X\phi}-\La_{\phi}=0,~~
\label{3}
\een
where the effective emergent metric is given by,
\ben
\widetilde{G}^{\mu\nu}\equiv \frac{c_{s}}{\La_{X}^{2}}[\La_{X} g^{\mu\nu} + \La_{XX} \nabla ^{\mu}\phi\nabla^{\nu}\phi],
\label{4}
\een
where $\La_{\mathrm X}= \frac{d\La}{dX},~\La_{\mathrm XX}= \frac{d^{2}\La}{dX^{2}}$, and $\La_{\mathrm\phi}=\frac{d\La}{d\phi}$ with $1+ \frac{2X\La_{XX}}{\La_{X}} > 0$ and $c_s^{2}(X,\phi)\equiv{(1+2X\frac{\La_{XX}} {\La_{X}})^{-1}}$.

After a transformation $\bar G_{\mu\nu}\equiv \frac{c_{s}}{\La_{X}}G_{\mu\nu}$  \cite{gm1,gm2,gm3}, we may express the inverse \cite{Miller} of the effective (emergent) metric (\ref{4}) as 
\ben
\bar{G}_{\mu\nu}=g_{\mu\nu}-\frac{\La_{XX}}{\La_{X}+2X\La_{XX}}\nabla_{\mu}\phi\nabla_{\nu}\phi.
\label{5}
\een

Here, we apply two successive conformal rescalings to bring the (covariant) emergent metric into the simple and compact form~\eqref{5}, which streamlines subsequent calculations; see Refs.~\cite{gm1,gm2} for details. Note that $\widetilde{G}^{\mu\nu}$ and $G_{\mu\nu}$ are different tensors: $\widetilde{G}^{\mu\nu}$ is the contravariant effective metric governing the principal part of the perturbation equation, while $G_{\mu\nu}$ is its covariant inverse, defined by $\widetilde{G}^{\mu\alpha}G_{\alpha\nu}=\delta^{\mu}{}_{\nu}$ (see Refs.~\cite{Vikman, Babichev}). Finally, $\bar{G}_{\mu\nu}$ is not identical to $G_{\mu\nu}$; rather, they are related by a conformal factor (e.g., $\bar{G}_{\mu\nu}\propto G_{\mu\nu}$), and hence encode the same causal (null-cone) structure \cite{Wald}. 

Note that the Eqs. (\ref{3})-(\ref{5}) hold physical significance if $\La_{X}\neq 0$ is positive definite. In K-essence geometry, Eq. (\ref{5}) asserts that our emergent metric, $\G_{\mu\nu}$, is conformally distinct from $g_{\mu\nu}$ (disformally linked) \cite{Bekenstein_1993, Bettoni} for non-trivial spacetime configurations of $\phi$. Unlike canonical scalar fields, $\phi$ exhibits distinct local causal structural characteristics. They are also different from those given by $g_{\mu\nu}$. An extensive analysis of disformal transformations was performed by Bekenstein \cite{Bekenstein_1993} and Zumalacarregui et al. \cite{Zumalacarregui}.
The Eq. (\ref{5}) can also be recast as,
\ben
\bar{G}_{\mu\nu}=g_{\mu\nu}-A\nabla_{\mu}\phi\nabla_{\nu}\phi,
\label{6}
\een
where the (scalar) \emph{disformal factor}
\begin{equation}
A\equiv A(\phi,X)\equiv \frac{\La_{XX}}{\La_{X}+2X\La_{XX}}
\label{6a}
\end{equation}
controls the measure of the disformal term in Eq.~(\ref{6}). In other words, $A$ quantifies the deviation of the emergent geometry from the background gravitational metric by measuring how strongly the field gradients $\nabla_{\mu}\phi$ deform the effective metric $\bar{G}_{\mu\nu}$ relative to $g_{\mu\nu}$. Since the characteristic surfaces of the perturbations are determined by $\bar{G}_{\mu\nu}$, changing $A$ changes the effective causal cone (and hence the propagation speed) of fluctuations, which then feeds into the effective energies and interaction rates computed in the emergent spacetime.

The following limits are useful to note:
\begin{itemize}
\item $A\to 0$: the disformal term switches off and $\bar{G}_{\mu\nu}\to g_{\mu\nu}$.
\item $A>0$ and large: the disformal contribution dominates, producing a strong deformation of the effective causal structure.
\end{itemize}
Moreover, the condition $A\dot\phi^{2}<1$ ensures that $\bar{G}_{\mu\nu}$ remains Lorentzian (i.e., its signature does not become ill-defined). For example, in the purely kinetic DBI-type K-essence model ($\mathcal{L}(X) = -\sqrt{1+2X}$) \eqref{137} with $X=-\frac{1}{2}\dot\phi^{2}$, one finds $A=1$, so the disformal term in Eq.~(\ref{6}) takes a particularly simple and transparent form.

At this stage, we do not specify the explicit form of the non-canonical Lagrangian $\mathcal{L}(X,\phi)$; however, it may be considered later. Our goal is to use Eq.~(\ref{6}) to construct the generic Boltzmann equation.

As emphasised in Refs.~\cite{gm1,gm2}, once the effective metric is taken to be the emergent metric $\bar{G}_{\mu\nu}$ of Eq.~(\ref{6}), the corresponding Christoffel symbols $\bar\Gamma^{\lambda}{}_{\mu\nu}$ (and hence the geodesic equation for trajectories defined with $\bar{G}_{\mu\nu}$) are, in general, different from the standard gravitational ones built from $g_{\mu\nu}$. This simply reflects that the relevant geometry for fluctuations is the emergent geometry governed by Eq.~(\ref{5}) or (\ref{6}).

Following Refs.~\cite{Babichev,Panda3}, we introduce the covariant derivative $D_{\mu}$ compatible with $\bar{G}_{\mu\nu}$,
\ben
D_{\mu}V_{\nu}=\partial_{\mu}V_{\nu}-\bar\Gamma^{\lambda}{}_{\mu\nu}V_{\lambda},
\label{7}
\een
with inverse metric $\bar G^{\mu\nu}$ defined by $\bar G_{\mu\lambda}\bar G^{\lambda\nu}=\delta^{\nu}_{\mu}$. Imposing metric compatibility $D_{\lambda}\bar{G}_{\mu\nu}=0$ and vanishing torsion $\bar\Gamma^{\alpha}{}_{\mu\nu}=\bar\Gamma^{\alpha}{}_{\nu\mu}$, the Levi-Civita connection of the emergent metric is
\ben
\bar{\Gamma}^{\alpha}{}_{\mu\nu}
=
\frac{1}{2}
\bar{G}^{\alpha\beta}
\left(
\partial_{\mu}\bar{G}_{\beta\nu}
+
\partial_{\nu}\bar{G}_{\beta\mu}
-
\partial_{\beta}\bar{G}_{\mu\nu}
\right).
\label{7a}
\een
For comparison, the usual gravitational connection is
\ben
\Gamma^{\alpha}{}_{\mu\nu}
=
\frac{1}{2}
 g^{\alpha\beta}
\left(
\partial_{\mu}g_{\beta\nu}
+
\partial_{\nu}g_{\beta\mu}
-
\partial_{\beta}g_{\mu\nu}
\right).
\label{7b}
\een
Substituting $\bar{G}_{\mu\nu}=g_{\mu\nu}-A\nabla_{\mu}\phi\nabla_{\nu}\phi$ into Eq.~(\ref{7a}) leads to \cite{Rendall, Wald, gm1, gm2, gm3}
\ben
\bar{\Gamma}^{\alpha}{}_{\mu\nu}
&&=
\Gamma^{\alpha}{}_{\mu\nu}
+\frac{1}{2}
\left(\bar{G}^{\alpha\beta}-g^{\alpha\beta}\right)
\left(
\partial_{\mu}g_{\beta\nu}
+\partial_{\nu}g_{\beta\mu}
-\partial_{\beta}g_{\mu\nu}
\right)
\nonumber\\
&&
-\frac{1}{2}
\bar{G}^{\alpha\beta}
\Big[
\partial_{\mu}\!
\left(A\nabla_{\beta}\phi\nabla_{\nu}\phi\right)
+\partial_{\nu}\!
\left(A\nabla_{\beta}\phi\nabla_{\mu}\phi\right)
\nonumber\\ &&-\partial_{\beta}\!
\left(A\nabla_{\mu}\phi\nabla_{\nu}\phi\right)
\Big],
\label{7c}
\een
which makes explicit that $\bar\Gamma^{\alpha}{}_{\mu\nu}$ differs from $\Gamma^{\alpha}{}_{\mu\nu}$ through the disformal contribution involving $A$ and the scalar-field gradients. For the homogeneous background considered here, $\phi=\phi(t)$, only components with time indices acquire non-trivial corrections. For a special case, such as a DBI-type K-essence Lagrangian \eqref{136} or \eqref{137}, we get $A=1$, using Eqs. \eqref{6} and \eqref{7a}, we found (Refs. \cite{Babichev, Panda3, gm1, gm2}),
\ben
\bar{\Gamma}^{\alpha}_{\mu\nu}
=
\Gamma^{\alpha}_{\mu\nu}
-
\frac{1}{2(1-2X)}
\left(
\delta^{\alpha}_{\mu}\,\partial_{\nu}
+
\delta^{\alpha}_{\nu}\,\partial_{\mu}
\right)X.
\label{7d}
\een

For clarity, we emphasize that in a standard (minimally coupled) K-essence theory the gravitational field equations follow from varying the action with respect to the \emph{gravitational} metric $g_{\mu\nu}$ (together with the scalar-field equation of motion), as reviewed for example in Refs.~\cite{Vikman,Garriga,Picon3}. The emergent metric $\bar{G}_{\mu\nu}$ introduced in Eq.~\eqref{6} is instead a \emph{derived, effective} geometry that governs the propagation of perturbations/matter in our kinetic-theory description.

To prevent any ambiguity, we do \emph{not} treat the emergent metric $\bar{G}_{\mu\nu}$ as an independent gravitational field, and we do \emph{not} postulate field equations obtained by varying the action with respect to $\bar{G}_{\mu\nu}$. Whenever we introduce an ``emergent Einstein tensor (EET)($\bar{\mathcal{G}}_{\mu\nu}$),'' it is meant purely as a geometric quantity \emph{defined} from $\bar{G}_{\mu\nu}$:
\ben
\bar{\mathcal{G}}_{\mu\nu}\equiv \bar{\mathcal{R}}_{\mu\nu}-\frac{1}{2}\bar{G}_{\mu\nu}\,\bar{\mathcal{R}},
\label{8}
\een
where $\bar{\mathcal{R}}_{\mu\nu}$ and $\bar{\mathcal{R}}$ are the Ricci tensor and Ricci scalar constructed from $\bar{G}_{\mu\nu}$. In this paper, $\bar{G}_{\mu\nu}$ is used only as an \emph{effective propagation metric} for particle geodesics and the (modified) Boltzmann equation, and not as a metric whose dynamics are determined by an Einstein equation.

Finally, it is also important to say that we do not use any ``emergent Einstein equation'' in the present analysis: our results rely solely on the kinetic-theory framework and the associated modified Boltzmann hierarchy for photons in the emergent spacetime. Deriving (or motivating) a closed set of gravitational field equations for $\bar{G}_{\mu\nu}$ may require specifying an action principle for the emergent geometry, which is beyond the scope of this work and is left for future study.\\

\noindent

\textit{Matter coupling, gauge choice, and physical framework:}
A central ingredient of our Boltzmann construction is the assumption that the particle species entering the kinetic theory (photons, electrons, baryons, neutrinos, and cold dark matter) propagate according to the emergent (disformal) metric $\bar G_{\mu\nu}=g_{\mu\nu}-A(\phi,X)\nabla_\mu\phi\,\nabla_\nu\phi$. In ordinary K-essence, the effective metric is often introduced to describe the propagation of \emph{scalar} fluctuations, and this does not automatically imply that Standard Model fields follow $\bar G_{\mu\nu}$ when matter is minimally coupled to $g_{\mu\nu}$. For clarity, we state the effective matter coupling assumed throughout the present kinetic-theory construction:
\ben
S \;=\; S_{\rm EH}[g_{\mu\nu}] \;+\; S_{k}[\phi,g_{\mu\nu}] \;+\; S_{m}[\bar G_{\mu\nu}(\phi),\psi_m],
\label{Smatter_emergent}
\een
where $\psi_m$ collectively denotes the matter and radiation fields. Under this assumption, \eqref{Smatter_emergent}, the causal structure relevant for particle propagation and the phase-space kinematics are those of $\bar G_{\mu\nu}$; accordingly, the mass-shell relation, geodesic equation, phase-space measure, and collision operator entering the Boltzmann equation are constructed from $\bar G_{\mu\nu}$ throughout the paper.

In the perturbative treatment presented below we work in the unitary-gauge (effective) description $\delta\phi=0$, so that $\bar G_{\mu\nu}$ is evaluated on the (homogeneous) background scalar configuration and its metric perturbations, rather than introducing an independently propagating scalar fluctuation. This is precisely the sense in which \eqref{Smatter_emergent} is used in our kinetic theory: the matter sector is assumed to couple to the emergent geometry generated by the background K-essence field configuration that defines the tilted causal cone. This assumption is reflected consistently in Appendix~\ref{A}, where we illustrate the same coupling choice by writing the action of a probe massive field in the emergent metric (Eq.~\eqref{A1}) and discussing the massless (photon) dispersion relation when minimally coupled to $\bar G_{\mu\nu}$ (Eq.~\eqref{A12}). For completeness, we note that the alternative possibility $S_m[g_{\mu\nu},\psi_m]$ would lead to a different kinetic theory in which photons and electrons follow $g_{\mu\nu}$ rather than $\bar G_{\mu\nu}$; that case is not considered here. 

To avoid confusion, we use Eq.~\eqref{Smatter_emergent} as an \emph{effective} framework for kinetic theory: it specifies the metric that defines particle propagation and phase-space variables, but we do not attempt to derive or solve the full set of coupled field equations obtained by varying $S_{\rm EH}+S_k+S_m$. Instead, we take the homogeneous K-essence background $\phi(t)$ (and the associated emergent FLRW metric $\bar G_{\mu\nu}$) as given and develop the corresponding Boltzmann hierarchy for matter and radiation moving on this emergent geometry. The choice $\delta\phi=0$ is simply a gauge choice (unitary gauge) that removes an independent scalar perturbation; the K-essence background still affects the kinetic theory through the time-dependent lapse and connection of $\bar G_{\mu\nu}$. Within this setup, the mass-shell condition, geodesic equation, phase-space measure, and collision term are all defined with respect to $\bar G_{\mu\nu}$, while the intrinsic particle parameters entering microscopic amplitudes (e.g., $m_e$ and QED couplings) are unchanged. A fully dynamical treatment of the backreaction of $S_m[\bar G(\phi),\psi_m]$ on the scalar field and gravitational equations is left for future work. The consistency of this leading-order background approximation, namely, evolving matter and radiation on the homogeneous K-essence background while neglecting their \emph{background} backreaction on the scalar equation, is explained in Appendix~\ref{app:background_consistency}.

\subsection{Notations and Conventions}
In this subsection, we collect the notation and conventions (Table~\ref{tab-1}) used throughout the paper in order to keep the sign conventions unambiguous. We adopt the Lorentzian signature $(-,+,+,+)$ for the metrics used in this work, in particular for the emergent metric $\bar G_{\mu\nu}$ (and for the gravitational metric $g_{\mu\nu}$ when it appears), and we use natural units ($c=\hbar=k_{B}=1$) unless stated otherwise. When we display a metric in matrix form, the negative sign in the time-time component is understood according to this convention (e.g. $\bar G_{00}<0$ in a homogeneous FLRW background).

The K-essence kinetic term is defined by
$X \equiv \frac{1}{2}\,g^{\mu\nu}\nabla_{\mu}\phi\,\nabla_{\nu}\phi,$
which, for the homogeneous background $\phi=\phi(t)$ considered here, reduces to $X=\frac{1}{2}g^{00}\dot{\phi}^{\,2}=-\frac{1}{2}\dot{\phi}^{\,2}.$ With this convention, the commonly appearing combination becomes $1+2AX = 1-A\dot{\phi}^{\,2}$, consistent with the lapse factor in the emergent metric \eqref{12}. Unless explicitly stated otherwise, quantities marked with an `overbar' or a `tilde' refer to the emergent K-essence geometry (or to variables naturally defined in that emergent frame).

\begin{table}[t]
\centering
\caption{Notation and conventions adopted throughout this work.}
\label{tab-1}
\small\begin{tabular}{@{}l@{\quad}l@{}}
\hline\hline
Symbol & \parbox[t]{0.68\columnwidth} {Description}\\
\hline
$g_{\mu\nu}$ & Background gravitational metric.\\
$\bar{G}_{\mu\nu}$ & \parbox[t]{0.68\columnwidth}{Emergent K-essence metric governing particle propagation.}\\
$\Gamma^{\alpha}_{\mu\nu}$ & \parbox[t]{0.68\columnwidth}{Christoffel symbols constructed from $g_{\mu\nu}$.}\\\\
$\bar{\Gamma}^{\alpha}_{\mu\nu}$ & \parbox[t]{0.68\columnwidth}{Christoffel symbols constructed from $\bar{G}_{\mu\nu}$.}\\\\
$X\equiv\frac12 g^{\mu\nu}\nabla_{\mu}\phi\,\nabla_{\nu}\phi$ & \parbox[t]{0.68\columnwidth}{K-essence kinetic term (for $\phi=\phi(t)$ and mostly-plus signature, $X=-\dot\phi^{2}/2$).}\\\\
$\phi(t)$ & \parbox[t]{0.68\columnwidth}{Homogeneous K-essence scalar field.}\\\\
$A(\phi,X)$ & \parbox[t]{0.68\columnwidth}{Disformal/emergent coupling function entering $\bar{G}_{\mu\nu}$.}\\\\
$P^{\mu}$ & \parbox[t]{0.68\columnwidth}{Particle four-momentum (indices raised/lowered with the emergent metric in emergent-frame relations).}\\\\
$\widetilde{E}$ & \parbox[t]{0.68\columnwidth}{Particle energy measured in the emergent frame.}\\\\
$\bar{p}$ & \parbox[t]{0.68\columnwidth}{Emergent-frame physical momentum magnitude.}\\\\
$\widetilde{\eta}$ & \parbox[t]{0.68\columnwidth}{Emergent conformal time.}\\\\
$\widetilde{T}$ & \parbox[t]{0.68\columnwidth}{Emergent-frame photon temperature.}\\\\
$f(x^{\mu},p_{i})$ & \parbox[t]{0.68\columnwidth}{Phase-space distribution function in terms of spacetime position and covariant spatial momentum components.}\\\\
$M$ & \parbox[t]{0.68\columnwidth}{Invariant (rest) mass appearing in the matter action.}\\\\
$M_{\rm eff}$ & \parbox[t]{0.68\columnwidth}{Effective rest-frame energy scale induced by the emergent geometry (a kinematic rescaling, not a new fundamental mass).}\\\\
$m_e$ & \parbox[t]{0.68\columnwidth}{Invariant (physical) electron mass parameter in QED (not geometrically rescaled by the emergent metric).}\\\\
$\sigma_{T}$ & \parbox[t]{0.68\columnwidth}{Thomson scattering cross section from local QED.}\\\\
$\sigma_{T}^{(\mathrm{geom})}$ & \parbox[t]{0.68\columnwidth}{Effective (geometry-rescaled) transport coefficient used in the emergent-frame kinetic description.}\\\\
$g$ & \parbox[t]{0.68\columnwidth}{Internal degeneracy factor (number of discrete internal states such as spin/helicity/polarization, and color when relevant) of a particle species.}\\
\hline\hline
\end{tabular}
\end{table}

\section{Formulation of the Boltzmann equation in K-essence geometry within FLRW backgrounds}\label{S3}

In this section, we describe the formulation of the Boltzmann equation in an expanding universe under the K-essence framework, assuming the background gravitational spacetime ($g_{\mu\nu}$) is a flat FLRW model. For the development of the Boltzmann equation in the K-essence geometry, we follow the book ``Modern Cosmology" by Dodelson and Schmidt \cite{Dodelson}.

When particle-particle interactions are taken into account, or equivalently when a collision term is included, the Boltzmann equation assumes its most general form,
\ben
\frac{df}{dt}=C[f],
\label{9}
\een
where $f \equiv \f$ denotes the distribution function and $C[f]$ represents the collision term. The associated relativistic energy–momentum tensor for the field $\f$ in the K-essence geometry is given by \cite{Dodelson},
\ben
\T^{\mu}_{~\nu}({\bf x},t)&&=\frac{g}{\sqrt{-det(\G_{\a\b})}}\n &&\times \int \frac{dP_{1}dP_{2}dP_{3}}{(2\pi)^3}\frac{P^{\mu}P_{\nu}}{P^{0}}\f,
\label{10}
\een
where $g$ is the internal degeneracy factor of the particle species, i.e., the number of discrete internal states (spin/helicity/polarization and, when relevant, color) associated with that species \cite{Dodelson}. In the present work, we do not model additional quantum-statistical effects beyond this degeneracy accounting; $g$ appears only as a fixed integer multiplicity for each species.

The phase-space distribution function $\f\equiv f(x^{\mu},p_{i})$ is defined as the occupation number \emph{per} internal state, so that an explicit sum over those internal indices produces the overall multiplicative factor $g$ in Eq.~(\ref{10}). Accordingly, $g$ does not count distinct phase-space configurations or particle trajectories, which are already encoded by the dependence of $f$ on $(x^{\mu},p_{i})$ and by the momentum-space integral.

All kinematic quantities in Eq.~(\ref{10}) (momenta, measure, and index operations) are defined with respect to the emergent metric $\G_{\mu\nu}$ rather than the background metric $g_{\mu\nu}$; this geometric replacement does not modify the meaning of $g$, which is an intrinsic property of the particle species. The comoving momentum is defined as $P^{\mu}=\frac{dx^{\mu}}{d\l}$ and $P_{\mu}=\G_{\mu\nu}P^{\nu}$. For example, $g=2$ for photons, corresponding to the two transverse polarization states, while for electrons $g=2$ corresponds to the two spin states.\\

For our study, we consider the background gravitational spacetime to be flat FLRW, i.e., the line-element of the flat FLRW metric is,
\ben
ds^{2}=-dt^{2}+a^{2}(t)\sum_{i=1}^{3}(dx^{i})^{2}.
\label{11}
\een
Therefore, the K-essence emergent gravity (\ref{6}) line-element can be written as
\ben
dS^{2}=-(1-A\dot\phi^{2})dt^{2}+a^{2}(t)\sum_{i=1}^{3}(dx^{i})^{2},
\label{12}
\een
where $`dot'$ signifies the derivative with respect to coordinate time. We take a homogeneous background $\phi({\bf x},t)\equiv\phi(t)$, as commonly adopted for K-essence cosmologies \cite{Babichev, Vikman, gm1, gm2, Ganguly}. Then $X=-\dot{\phi}^{2}/2$ and the disformal coefficient reduces along the background to $A=A\big(\phi(t),X(t)\big)\equiv A(t)$ (or to a non-zero constant in specific models), so the emergent geometry remains FLRW and differs only through a time-dependent modulation of the light cone (see Appendix~\ref{A}). We require $A\dot\phi^{2}<1$ to keep the metric \eqref{12} Lorentzian.

Hence, we can write the components of the emergent gravity line-element (\ref{12}) as
\ben
\G_{00}=-(1-A\dot\phi^{2})~;~\G_{i0}=0~;~\G_{ij}=\delta_{ij}a^{2}(t)\equiv g_{ij}.\n~~~~
\label{13}
\een\\

\noindent
The distribution function of a system of particles is characterized within the entirety of six-dimensional phase space. Time is regarded as an external parameter that characterizes the evolution of this distribution. Given that the four-momentum of each particle adheres to the relativistic mass-shell condition, it follows that only three of the four momentum components can be considered independent. Consequently, the distribution function evolves within a six-dimensional framework, with its temporal variation illustrating the transformation of the particle ensemble in accordance with the universe's evolution.

The mass that appears in the microscopic matter action \eqref{A1} is set by microphysics and is therefore unaffected by the presence of a K-essence background. We denote this invariant rest-mass parameter by $M$ (for example, $M=m_e$ for electrons). If one instead describes particle kinematics using the K-essence-induced \emph{emergent} geometry, the only new feature is the non-trivial lapse in the time component of the emergent metric. This lapse rescales the particle's \emph{rest-frame coordinate-time energy} ($M_{\rm eff}$) \eqref{A7} defined in the emergent variables, while leaving the invariant mass parameter $M$ unchanged \cite{Babichev, Vikman, Winitzki} (see Appendix~\ref{A} for details). 

For the homogeneous background, the emergent line element is given by Eq.~\eqref{12}, so the lapse is
\begin{equation}
N(t)\equiv\sqrt{1-A\dot{\phi}^{2}}.
\end{equation}
Accordingly, for a particle at rest in the emergent frame, it is convenient to introduce the lapse-rescaled rest-energy scale
\begin{equation}
M_{\rm eff}\equiv MN=M\sqrt{1-A\dot{\phi}^{2}}.
\label{14}
\end{equation}
The quantity $M_{\rm eff}$ is not a new fundamental mass; rather, it is the lapse-rescaled rest-frame \emph{coordinate}-energy scale that appears when the particle is described in the emergent geometry. In the limit $A\dot\phi^{2}\to 0$ the emergent geometry reduces to the standard FLRW form and $M_{\rm eff}\to M$.

In the effective emergent-frame parametrization (cf.~Appendix~\ref{A}, Eqs.~\eqref{A12}--\eqref{A13}), the corresponding mass-shell relation is written as
\begin{equation}
\bar{G}_{\mu\nu}P^\mu P^\nu=-M_{\rm eff}^{2},
\label{15}
\end{equation}
where $P^{\mu}$ is the four-momentum used in the emergent-frame kinetic description, with index operations defined by $\bar G_{\mu\nu}$. Eq.~(\ref{15}) is thus simply a kinematic reformulation in terms of the emergent variables and does not alter the underlying microscopic mass parameter $M$.

The spatial momentum magnitude in the emergent frame is defined by
\begin{equation}
\bar{p}^{\,2}\equiv\bar{G}_{ij}P^{i}P^{j}.
\label{16}
\end{equation}
For the homogeneous K-essence background considered here, $\bar{G}_{ij}=g_{ij}=a^{2}(t)\delta_{ij}$, so the spatial metric is unchanged and $\bar p$ coincides with the usual FLRW physical momentum magnitude.

Using Eq.~(\ref{15}) together with the homogeneous-metric components $\bar G_{00}=-(1-A\dot\phi^{2})$ and $\bar G_{ij}=a^{2}(t)\delta_{ij}$, and the definition (\ref{16}), we define the emergent-frame energy variable by
\begin{equation}
\widetilde{E}^{\,2}\equiv\left(1-A\dot{\phi}^{2}\right)(P^{0})^{2},
\label{16a}
\end{equation}
so that Eqs.~(\ref{15}) and (\ref{16}) yield the effective dispersion relation
\begin{equation}
\boxed{\widetilde{E}^{\,2}=\bar{p}^{\,2}+M_{\rm eff}^{2}}.
\label{16b}
\end{equation}
Hence $\widetilde{E}=\sqrt{\bar{p}^{\,2}+M_{\rm eff}^{2}}$ for massive particles, while $\widetilde{E}=\bar p$ for massless ones. Here $E\equiv P^{0}$ denotes the usual FLRW (GR) \emph{coordinate-time} energy, whereas $\widetilde{E}$ is the corresponding energy variable adapted to the K-essence emergent geometry. Eq.~(\ref{16a}) then implies
\begin{equation}
P^{0}=\frac{\widetilde{E}}{\sqrt{1-A\dot{\phi}^{2}}}.
\label{16c}
\end{equation}

Eqs.~(\ref{14})--(\ref{16c}) summarize the two distinct scales used throughout:
\ben
&&M=\text{invariant microscopic mass},\nonumber\\ &&
M_{\rm eff}=M\sqrt{1-A\dot{\phi}^{2}}\nonumber\\ &&=\text{emergent-frame rest-energy scale.}
\label{16d}
\een
The former is fixed by microphysics, whereas the latter is the geometrical scale that enters the phase-space kinematics written in the tilted K-essence variables.


In the following, we work with the comoving four-momentum (defined with respect to the emergent metric) as
\ben
P^{\mu}\equiv \frac{dx^{\mu}}{d\lambda}=(P^{0},P^{i}),\quad P^{0}\equiv \frac{dt}{d\lambda},\quad P^{i}\equiv \frac{dx^{i}}{d\lambda},
\label{17a}
\een
where $\lambda$ is an affine parameter along the particle worldline in the emergent geometry. Eq.~\eqref{16c} gives the explicit form of the time component $P^{0}$ in the homogeneous K-essence FLRW background.

It is advantageous to decompose the dependence on the emergent-frame spatial momentum $\p$ into its magnitude $\pa\equiv\sqrt{\pa^{2}}$ and a corresponding unit vector $\hat p^{i}$, defined such that $\hat p_{i}\equiv\delta_{ij}\hat p^{j}$ and $\delta_{ij}\hat p^{i}\hat p^{j}=1$. We then write the comoving momentum as
\ben
P^{i}\equiv C\,\hat p^{i},
\label{18}
\een
with a proportionality factor $C$. Using Eqs.~(\ref{16}) and (\ref{13}) we obtain
\ben
\pa^{2}=a^{2}C^{2},
\label{19}
\een
so that $C=\pa/a$. Therefore,
\ben
P^{i}=\frac{\pa}{a}\hat p^{i}.
\label{20}
\een

It is worth noting that K-essence geometry differs from standard FLRW gravity primarily through its causal structure: null rays are defined by $\bar G_{\mu\nu}k^{\mu}k^{\nu}=0$ rather than by $g_{\mu\nu}k^{\mu}k^{\nu}=0$. In a homogeneous background this implies a modified coordinate-time slope for null curves, $dx/dt=\pm\sqrt{1-A\dot\phi^{2}}$ in a 2D $(t,x)$ diagram (equivalently, $v_{\rm ph}=\sqrt{1-A\dot\phi^{2}}$), so the emergent null cone is deformed relative to the gravitational one; see Appendix~\ref{A}. This ``tilt'' is a geometric way to quantify the mismatch between the two null cones and should not be interpreted as a literal rotation of spacetime \cite{Babichev, Vikman, Sawicki}. For the homogeneous FLRW backgrounds considered here, the deformation remains isotropic (it changes the opening of the cone but not its spatial direction dependence). A detailed discussion can be found in Appendix~\ref{A}.

\subsection{Boltzmann equation for K-essence induced expanding universe \label{S3.1}}
\noindent
According to Eq. \cite{Dodelson}, the Boltzmann equation can be written as
\ben
\frac{df}{dt}
=\frac{\partial f}{\partial t}
+\frac{\partial f}{\partial x^{i}}\frac{dx^{i}}{dt}
+\frac{\partial f}{\partial \pa}\frac{d\pa}{dt}
+\frac{\partial f}{\partial \hat{\pa}^{i}}\frac{d\hat{\pa}^{i}}{dt}.
\label{21}
\een
In this section, we derive the Boltzmann equation in a smooth, homogeneous, and isotropically expanding universe described by the K-essence geometry. In such a background, the direction of a particle's momentum remains unchanged during cosmic expansion, although its magnitude evolves with time. Consequently, the term proportional to $d\hat{\pa}^{i}/dt$, which accounts for variations in the momentum direction, vanishes. We therefore drop the last term in the general Boltzmann Eq. (\ref{21}). Using Eqs. (\ref{16c}) and (\ref{20}), we obtain
\ben
\frac{dx^i}{dt}
=\frac{\pa\,\hat{\pa}^i}{a}
\frac{\sqrt{1-A\dot\phi^{2}}}{\widetilde{E}}.
\label{22}
\een

Next, considering the time component of the geodesic equation,
\ben
\frac{dP^0}{d\lambda}
=-\bar{\Gamma}^{0}_{\a\b}P^{\a}P^{\b},
\label{23}
\een
and making use of Eqs. (\ref{16}) and (\ref{16c}), we find
\ben
\frac{d\pa}{dt}
=-
\frac{\pa\,\partial_{0}(A\dot\phi^{2})}{2(1-A\dot\phi^{2})}
-H\pa,
\label{24}
\een
where $H\equiv \dot a/a$ is the Hubble parameter.
Applying the above expressions, Eqs. (\ref{22}) and (\ref{24}) in Eq. (\ref{21}), we have {\it the Boltzmann equation in the K-essence geometry (BEKG) for flat FLRW backgrounds} is
\ben
&&\frac{\partial f}{\partial t}+\frac{\pa\hat{\pa}^i}{a}\frac{\sqrt{(1-A\dot\phi^{2})}}{\widetilde{E}}\frac{\partial f}{\partial x^{i}} - H\pa\frac{\partial f}{\partial \pa} \nonumber\\&&-\frac{\pa\,\partial_{0}(A\dot\phi^{2})}{2(1-A\dot\phi^{2})}\frac{\partial f}{\partial \pa}=C[f].
\label{25}
\een
This BEKG (Eq. (\ref{25})) is valid for all particles under the K-essence framework. 

It should be noted that in realistic scenarios, we often observe two separate limiting cases of the BEKG. The initial scenario relates to the relativistic limit, characterized by the condition where the momentum of the particle significantly exceeds its rest mass, $( \pa \gg M_{\rm eff} )$ i.e., $\widetilde{E} \propto {\pa}$. This simplification significantly impacts the dynamics and is especially pertinent for characterizing {\it massless or nearly massless} particles, such as {\it photons or neutrinos}, in the early universe. So in the {\it relativistic limit}, the BEKG (Eq. (\ref{25})) becomes
\ben
&&\frac{\partial f}{\partial t}+\frac{\hat{\pa}^i}{a}\sqrt{(1-A\dot\phi^{2})}~\frac{\partial f}{\partial x^{i}} - H\pa\frac{\partial f}{\partial \pa} \nonumber\\&&-\pa~\frac{\partial_{0}(A\dot\phi^{2})}{2(1-A\dot\phi^{2})}\frac{\partial f}{\partial \pa}=C[f].
\label{26}
\een

In the opposite regime, specifically the non-relativistic limit, the momentum of the particle is significantly less than its rest mass, $( \pa \ll M_{\rm eff})$, which implies that $(\widetilde{E} \simeq M_{\rm eff})$. As a result, Eq. (\ref{25}) simplifies accordingly, leading to a form that effectively characterizes {\it cold dark matter or massive particles} in motion at velocities significantly lower than the speed of light. Thus, in the {\it non-relativistic limit}, the BEKG (Eq. (\ref{25})) is
\ben
&&\frac{\partial f}{\partial t}+\frac{\pa}{M_{\rm eff}}\frac{\hat{\pa}^i}{a}\sqrt{(1-A\dot\phi^{2})}~\frac{\partial f}{\partial x^{i}} - H\pa\frac{\partial f}{\partial \pa} \nonumber\\&&-\frac{\pa\,\partial_{0}(A\dot\phi^{2})}{2(1-A\dot\phi^{2})}\frac{\partial f}{\partial \pa}=C[f].
\label{27}
\een\\

\noindent
We now focus on a significant application of the BEKG: analyzing how it governs the temporal evolution of {\it the number density} for a specific particle species. It is important to note that the number density $n(x,t)$ is derived by performing an integration of the distribution function $\f$ across the complete momentum space. To analyze its evolution, we start the process by integrating the BEKG (Eq. (\ref{25})) with respect to momentum. Also, we observe that in a homogeneous universe, spatial gradients are nullified, specifically, $\partial f/\partial x^i = 0$. This simplification assumes that the distribution function is homogeneous throughout space, leaving only its temporal and momentum dependencies as significant. The derived expression establishes a direct relationship between the microscopic collision dynamics described by the Boltzmann equation and the macroscopic variation in the number density. So the corresponding BEKG (Eq. (\ref{25})) becomes,

\ben
&&\int \frac{d^{3}\pa}{(2\pi)^{3}}\frac{\partial f}{\partial t}-H\int \frac{d^{3}\pa}{(2\pi)^{3}}\pa\frac{\partial f}{\partial \pa}\nonumber\\&&-\frac{\partial_{0}(A\dot\phi^{2})}{2(1-A\dot\phi^{2})}\int \frac{d^{3}\pa}{(2\pi)^{3}}\,\pa\,\frac{\partial f}{\partial \pa}\nonumber\\&=&\int \frac{d^{3}\pa}{(2\pi)^{3}}C[f].
\label{28}
\een
Performing the integration by parts in the 2nd and 3rd terms of the left-hand side of the above Eq. (\ref{28}), we obtain
\ben
&&\frac{dn(t)}{dt}+3Hn(t)+3n(t)~\frac{\partial_{0}(A\dot\phi^{2})}{2(1-A\dot\phi^{2})}\nonumber\\&&=\int \frac{d^{3}\pa}{(2\pi)^{3}}~C[f].\nonumber\\
\label{29}
\een
Eq. (\ref{29}) already has the standard ``number--density'' structure: after imposing homogeneity ($\partial f/\partial x^{i}=0$) and integrating over momentum, any explicit dependence on $M_{\rm eff}$ drops out. It is therefore convenient to rewrite it as
\ben
&&\frac{dn(t)}{dt}+3\Big[H+\frac{\partial_{0}(A\dot\phi^{2})}{2(1-A\dot\phi^{2})}\Big]n(t)=\int \frac{d^{3}\pa}{(2\pi)^{3}}~C[f].\nonumber\\
\label{30}
\een
In the GR limit ($A\dot\phi^{2}\to 0$), Eq.~(\ref{30}) reduces to the familiar result $\dot n+3Hn=\int d^{3}\pa\,C[f]/(2\pi)^{3}$, which is the same statement as in Dodelson \cite{Dodelson} that, in the absence of collisions, the comoving number $a^{3}n$ is conserved. In the K-essence emergent geometry, the additional term proportional to $\partial_{0}(A\dot\phi^{2})$ reflects the time-dependent lapse of the emergent metric and modifies the dilution law when the evolution is expressed in the gravitational-frame time coordinate.

For photons, one still has $M=0$ and hence $M_{\rm eff}=M\sqrt{1-A\dot\phi^{2}}=0$ identically; the condition $A\dot\phi^{2}<1$ is required only to keep the emergent metric Lorentzian. Thus, the K-essence background does not generate a mass gap for light; its effect is solely to distort the light cone (i.e., the propagation speed) while leaving the photon massless.

If we neglect the collision term ($C[f]=0$), then from the above Eq. (\ref{30}), we have
\ben
n(t)=a^{-3}~exp\Big[-\int \frac{3}{2}\Big(\frac{\partial_{0}(A\dot\phi^{2})}{(1-A\dot\phi^{2})}\Big)dt\Big].
\label{31}
\een
In the standard FLRW universe, once a species is collisionless, its comoving number is conserved and the physical number density decreases only because the physical volume grows, giving $n\propto a^{-3}$ \cite{Dodelson}. In our K-essence cosmology the same geometric dilution is present, but Eq.~(\ref{31}) shows an additional multiplicative factor controlled by the homogeneous scalar background. The reason is that particles propagate on the emergent metric, whose (time-dependent) lapse modifies how a fixed phase-space element is mapped into ``physical'' density when one uses the gravitational coordinate time in the Liouville operator. Equivalently, the evolution of $A\dot\phi^{2}$ changes the conversion between momentum/energy variables and the time parameter along trajectories. Depending on the sign of $\partial_{0}(A\dot\phi^{2})/(1-A\dot\phi^{2})$, this extra contribution makes the density fall faster than $a^{-3}$ or, conversely, partially compensates the usual dilution. Because interaction rates are compared to the expansion rate when computing decoupling, such a modification can shift freeze-out and thereby change relic abundances and aspects of the thermal history once a concrete K-essence model (Lagrangian) (i.e., a specific $A(\phi,X)$) is specified.

The ``inconsistency'' that one might infer from the extra term is only a matter of variables. The background collisionless Boltzmann equation is written using gravitational-frame coordinate time, so the emergent-metric lapse appears explicitly in the Liouville operator, while the perturbation hierarchy (Sec.~\ref{S4}) is formulated directly in emergent-frame variables. In that emergent description the same lapse factor is already absorbed into the definitions of conformal time \eqref{91} and of the kinetic variables $(\widetilde E,\widetilde T)$, so the equilibrium photon distribution remains exactly Planckian and the background temperature still scales as $\widetilde T\propto a^{-1}$ ({\it see} Subsec. \ref{Sub-V-1}). The extra Liouville contribution therefore encodes only the change of variables between the gravitational and emergent frames; it does not represent a distortion of the underlying blackbody spectrum.

\subsection{Derivation of the collision terms \label{S3.2}}
\noindent
The phenomenon of direct particle interactions is characterized, within the framework of Boltzmann statistics, as ``collisions". Interactions encompass scattering phenomena, as well as processes such as particle pair creation, annihilation, and particle decay. A typical category of process involves a reaction in which particles of type 1 and type 2 are involved in interactions, resulting in the formation of particles of type 3 and type 4 as,
\ben
(1)_{\p}+(2)_{\q}\leftrightarrow (3)_{\p'}+(4)_{\q'}.
\label{32}
\een
Indeed, all microscopic physical processes adhere to the principles of momentum and energy conservation which are given by, 
\ben
&&\p+\q=\p'+\q';\n &&
\E_{1}(\p)+\E_{2}(\q)=\E_{3}(\p')+\E_{4}(\q'),
\label{33}
\een
where $\E_s^{2} (\pa)= {\pa}^2 +M_{\rm eff(s)}^{2}$ represents the energy-momentum relationship for particles as indicated in Eq. (\ref{16}). Each type of particle possesses its corresponding distribution function $f_{s}(\x,\p,t),~ s = 1,2,3,4$. The collision term is expressed as,
\ben
C[f_{1}(\p)]&&=\frac{1}{2\widetilde{E}_{1}(\pa)}\int{\frac{d^3 \qa}{(2\pi)^3 2\widetilde{E}_{2}(\qa)
}}\int{\frac{d^3 {\pa}'}{(2\pi)^3 2\widetilde{E}_{3}({\pa}')
}}\times\nonumber\\ &&
\int{\frac{d^3 {\qa}'}{(2\pi)^3 2\widetilde{E}_{4}({\qa}')
}}|\mathcal{M}|^{2}\times\n && (2\pi)^{4}\delta_{D}^{(3)}\Big[\p+\q-\p'-\q'\Big]\times\nonumber\\ &&
\delta_{D}^{(1)}\Big[\widetilde{E}_{1}(\pa)+\widetilde{E}_{2}(\qa)-\widetilde{E}_{3}({\pa}')-\widetilde{E}_{4}({\qa}')\Big]\nonumber\\ && \times
\Big(f_{3}(\p')f_{4}(\q^{'})\big[1\pm f_{1}(\p)\big]\big[1\pm f_{2}(\q)\big]\nonumber\\&&-f_{1}(\p)f_{2}(\q)\big[1\pm f_{3}(\p')\big]\big[1\pm f_{4}(\q')\big]\Big).
\label{34}
\een
The expression of the collision term (Eq. (\ref{34})) is different from the usual one \cite{Dodelson}. The differences come from the energy part $\E$ (Eq. (\ref{16})). The spatial component of the momentum ($\p$) can remain unchanged, as in our modified spacetime (Eq. (\ref{13})), it ($P^{\mu}$) [Eqs. (\ref{16c}) and (\ref{20})] solely dependent on the homogeneous K-essence scalar field, rather than on the spatial component of the field. The overall microphysical characteristics of this interaction are encapsulated in the squared amplitude $|\mathcal{M}(\p,\q,\p',\q')|^2$, which typically varies with the momenta of the interacting particles and can be calculated using Feynman diagrams. 

One fundamental component is quantum phenomena, including stimulated emission (also known as Bose enhancement) and the Pauli exclusion principle (also known as Pauli blocking).  The reaction rate is influenced by these factors, which can either enhance or restrict the process based on the occupancy of the final state. Incorporating these elements corresponds to boosting the forward reaction with factors of $(1\pm f_3)(1 \pm f_4)$, while simultaneously modifying the reverse reaction with factors of $(1\pm f_1)(1\pm f_2)$.  In every instance, a positive sign is assigned when the associated particle is classified as a boson, whereas a negative sign is assigned when it is identified as a fermion.  Pauli blocking manifests distinctly: when the state of a fermion $1$ with momentum $\p$ is occupied, the term $(1-f_{1}(\p))$ becomes zero, preventing the occurrence of the inverse reaction with this final state, as required by the principles of quantum mechanics. Conversely, if the particle $1$ is a boson, the corresponding reaction rate is increased, as bosons tend to occupy the same quantum state. In particular, we {\it neglect the effects of Bose enhancement and Pauli blocking in our investigation}.

\subsubsection{Local nature of collision terms and validity of flat-space cross sections}
At first, it may seem conceptually contradictory to calculate the collision term of the modified Boltzmann equation with conventional quantum field theoretic scattering amplitudes while the particles traverse a curved cosmic spacetime. In the conventional FLRW framework \cite{Dodelson}, Thomson or Compton cross sections are computed precisely as in Minkowski spacetime, despite the expansion of the background geometry. The rationale arises from the locality of particle interactions when combined with the equivalence principle.

Particle collisions occur across microscopic spacetime intervals that are significantly smaller than the cosmological curvature scale. The standard interaction time and length scale of a Thomson scattering event are atomic, but the Hubble time and curvature radius are cosmic. Thus, in any particular interaction, the metric can be consistently approximated by a locally inertial frame in which spacetime simplifies to Minkowski form. Within this local framework, the scattering amplitude relies solely on Lorentz invariants and is hence equivalent to the flat-space outcome \cite{Weinberg, Kolb, Baumann, Ma1, deGroot, Ehlers, Peacock}. Consequently, in the standard cosmological Boltzmann equation, the curved geometry influences solely the Liouville (propagation) operator, whilst the collision term maintains the conventional special-relativistic cross section.

The same rationale applies under the current K-essence paradigm, albeit with significant variation. Particles propagate along the geodesics of the emergent metric $\bar G_{\mu\nu}$ instead of the gravitational metric $g_{\mu\nu}$. The local inertial frame pertinent to particle interactions is hence connected to the geometry that emerged. Consequently, the microphysics of scattering remains unaltered, and the fundamental interaction matrix elements are consistent with the conventional scenario, although the disformal factor modifies the relationship among energy, momentum, and physical time. Therefore, the cross sections maintain their flat-space configuration, whereas the effective interaction rate, energy transfer, and mass in the Boltzmann equation are rescaled.

As a result, our treatment aligns with the conventional FLRW framework and reverts to it when the disformal element is nullified. The distinction is not found in local particle physics but in the geometric correspondence between microscopic interactions and cosmological evolution: in FLRW, the geometry just affects particle propagation, while in the emergent K-essence spacetime, it also impacts the effective kinetic variables involved in the collision term. This is a physically justified generalization of the conventional Boltzmann formalism. For the detailed qualitative discussions of the aforementioned explanations, leave for \emph{Appendix \ref{E}}.\\

It is important to point out that the Liouville theorem is only true when the disformal transformation is used if the phase-space measure and Liouville operator in the emergent geometry are carefully specified. A detailed discussion is presented in \emph{Appendix \ref{F}}. In our case, the disformal transformation does not break the conservation of phase space in the context of K-essence geometry. Instead, it makes it easier to map between the phase-space variables defined in the gravitational frame and those defined in the emergent frame.

\section{The evolution of perturbations through the Boltzmann equation in K-essence cosmology}\label{S4}
\noindent
As discussed in Sec.~\ref{S3}, we take the homogeneous (isotropic) photon distribution in the emergent frame as the background solution. We now perturb about this equilibrium and work to linear order.

In practice, cosmology treats the universe as exactly homogeneous and isotropic only at zeroth order: this ``smooth'' FLRW background captures the mean expansion and average energy content. The observed universe, however, contains small inhomogeneities, seen most directly as temperature anisotropies in the CMB, which grow into the large-scale structure at late times. To follow this growth, we promote the background metric and distribution functions to perturbed quantities and linearize the Boltzmann equation. The resulting perturbed Boltzmann equation tracks how tiny fluctuations in the phase-space distribution evolve under gravity and, when relevant, collisions, providing the link between primordial perturbations and the anisotropies/inhomogeneities we observe today in a K-essence cosmology.\\

\subsection{Perturbed spacetime under the K-essence geometry with flat FLRW backgrounds \label{S4.1}}
\noindent
We consider perturbations around the K-essence-induced smooth universe (Eq. (\ref{13})), characterized by the scalar field ($\phi(t)$) and the scale factor $(a(t))$, with a functional relationship described by the EOM (Eq. (\ref{3})), which will be addressed subsequently. So that the perturbed universe requires two more functions, $\Psi$ and $\Phi$, both of which depend on space and time. Consequently, the perturbed metric may be expressed as
\ben
&&\G_{00}=-(1-A\dot\phi^{2})-2\Psi({\bf x},t)~;~\G_{i0}=0~;\nonumber\\&&\G_{ij}=a^{2}(t)\delta_{ij}[1+2\Phi({\bf x},t)].~~~~
\label{35}
\een
In the absence of the perturbations $\Psi$ and $\Phi$, Eq.~(\ref{35}) simplifies to the background K-essence-modified FLRW metric, characterizing a zeroth-order homogeneous and isotropic universe of Euclidean type. The fluctuations from this smooth background are represented within the perturbation fields. The perturbations to the metric $\Psi$ act as the K-essence-induced Newtonian potential, affecting the behavior of slow-moving, non-relativistic matter. The perturbation $\Phi$, conversely, modifies the spatial curvature and can be recognized as a localized variation of the scale factor, $a(t)\rightarrow a({\bf x},t)=a(t)\sqrt{1+2\Phi({\bf x},t)}$.
In general, both $\Psi$ and $\Phi$ exhibit a strong correlation with the dynamics of the K-essence scalar field ($\dot{\phi}$), a relationship that will be examined subsequently, after choosing the specific form of the non-canonical Lagrangian.

In our perturbative treatment, we take the K-essence background to be strictly homogeneous, $\phi=\phi(t)$ (so $A=A(t)~or~ \text{constant}$, along the background), and we do not introduce an independent scalar fluctuation $\delta\phi$ in the kinetic sector. This corresponds to an effective description in which the homogeneous field defines the preferred time slicing and its impact on matter and radiation enters through the (background and perturbed) emergent metric. In the context of the effective field theory of dark energy/inflation, this is the standard unitary slicing, $\delta\phi=0$, in which all scalar-sector perturbations are carried by the metric \cite{Cheung, Gleyzes, Zumalacarregui}. Accordingly, photons, neutrinos, baryons, cold dark matter, and electrons propagate on the perturbed emergent geometry \eqref{35}, and their Boltzmann hierarchies depend only on the metric potentials.

We also clarify the gauge choice implicit in Eq.~\eqref{35}. A general scalar perturbation of a flat FLRW metric involves four scalar functions $(\Psi,\Phi,B,E)$, while scalar coordinate transformations are generated by two functions (a time shift and a scalar spatial shift) \cite{MukhanovBook,Weinberg,MaBertschinger}. Fixing $\delta\phi=0$ uses the time-reparametrization freedom. We then use the remaining scalar spatial gauge freedom to set the scalar shear to zero ($E=0$). The shift perturbation $B$ is non-dynamical and is driven by the momentum constraint; in our setup it vanishes at linear order, so the metric reduces to the Newtonian/longitudinal form with $B=E=0$ and is fully characterized by the two Bardeen (gauge-invariant) potentials, which coincide with $\Psi$ and $\Phi$ in this gauge. In the absence of significant anisotropic stress from the matter sector, one further has $\Psi\simeq (-\Phi)$ as usual \cite{Dodelson}.

The non-vanishing components of the Christoffel symbols corresponding to the perturbed metric (Eq. (\ref{35})) are,
\ben
&&\bar{\Gamma}^{0}_{00}=\frac{-1}{[1-A\dot\phi^{2}+2\Psi]}\partial_{0}[-(1-A\dot\phi^{2})-2\Psi]\nonumber\\&&
=-\frac{1}{2}(1+A\dot\phi^{2})\partial_{0}(A\dot\phi^{2})+\dot\Psi (1+A\dot\phi^{2})\n &&+\Psi \partial_{0}(A\dot\phi^{2});\label{36}\\ &&
\bar{\Gamma}^{0}_{0i}=(1+A\dot\phi^{2})\partial_{i}\Psi;\label{37}\\ &&
\bar{\Gamma}^{0}_{ij}=a^{2}\delta_{ij}\Big[(H+\dot\Phi)(1+A\dot\phi^{2})\n &&+2H(\Phi(1+A\dot\phi^{2})-\Psi)\Big];~~~~~~\label{38}\\&&
\bar{\Gamma}^{i}_{00}=\frac{1}{a^{2}}\partial_{i}\Psi; \label{39}\\ &&
\bar{\Gamma}^{i}_{j0}=\delta_{ij}(H+\dot\Phi); \label{40}\\ &&
\bar{\Gamma}^{i}_{jk}=\delta_{ik}\partial_{j}\Phi+\delta_{ij}\partial_{k}\Phi-\delta_{jk}\partial_{i}\Phi. \label{41}
\een

\subsection{Geodesic equation for the perturbed spacetime ( Eq. (\ref{35})) \label{S4.2}}
\noindent
To derive the Boltzmann equation in a perturbed universe (Eq. (\ref{35})), it is essential to understand the dynamics of particle propagation within the context of the modified spacetime geometry. Their trajectory is controlled by the geodesic equation, which we shall now add to account for the influence of metric perturbations. Our objective is to calculate the derivatives $dx^{i}/dt$, $d\pa/dt$, and $d\hat{\pa}^{i}/dt$, as these are integral to the Boltzmann Eq. (\ref{21}) and to determine the evolution of the distribution function in relation to the underlying geometry. The mass-shell condition for a particle with effective rest-frame energy $M_{\rm eff}$ is expressed as,
\ben
\G_{\mu\nu}P^{\mu}P^{\nu}=-[(1-A\dot\phi^{2})+2\Psi](P^{0})^{2}+{\pa}^{2}=-M_{\rm eff}^{2},\nonumber\\
\label{42}
\een
so that
\ben
(P^{0})^{2}=\frac{{\pa}^2+M_{\rm eff}^2}{(1-A\dot\phi^{2})+2\Psi}~.
\label{43}
\een
where ${\pa}^2=\G_{ij}P^{i}P^{j}$, as defined in Eqs.~(\ref{15}) and (\ref{18}). 
For {\it massless particles}, one has $\widetilde{E}\simeq \pa$, whereas for {\it massive particles} the dispersion relation reads $\widetilde{E}^{2}=\pa^{2}+M_{\rm eff}^{2}$.  Using the relation given in Eq.~(\ref{16}), we obtain
 
\ben
P^{0}=\frac{\widetilde{E}}{\sqrt{(1-A\dot\phi^{2}+2\Psi)}}.
\label{44}
\een

Now using the definitions Eqs. (\ref{15}) and (\ref{18}) with the perturbed spacetime (Eq. (\ref{35})), we obtain
\ben
P^{i}\simeq\frac{{\pa}^{i}}{a}(1-\Phi)
\label{45}
\een
with $C\simeq \frac{\pa}{a}(1-\Phi)$ and the corresponding relations ${\pa}^{i}=\pa\hat{\pa}^{i}$, where $\hat{\pa}^{i}~(=\hat{\pa}_{i})$ satisfying the relation $\delta_{ij}\hat{\pa}^{i}\hat{\pa}^{j}=1$. The following expression (Eq. (\ref{46})) represents the four-momentum of a massive particle within the K-essence-induced perturbed FRLW spacetime, including the scenario of massless particles as well:
\ben
P^{\mu}=\Big[\frac{\widetilde{E}}{\sqrt{(1-A\dot\phi^{2}+2\Psi)}}, \frac{{\pa}^{i}}{a}(1-\Phi)\Big].
\label{46}
\een
In our setup, the K-essence field $\phi$ is \emph{minimally coupled} to the standard gravitational metric $g_{\mu\nu}$; hence the invariant particle rest mass entering the matter action is the constant $m$ (and we identify $M\equiv m$) (see Appendix \ref{A}, Eq. \eqref{A1}). When the same particle is described in the emergent K-essence geometry, the modified lapse induces a geometric rescaling of the mass-shell relation, which can be encoded as the effective rest-frame energy $M_{\rm eff}=m\sqrt{1-A\dot\phi^{2}}.$ The quantity $M_{\rm eff}$ is therefore not a new fundamental mass, but a kinematic energy scale associated with the emergent metric (equivalently, with the distortion of the emergent light cone). Accordingly, the Liouville operator and the kinetic equations are formulated in terms of the emergent energy $\widetilde{E}$ and the corresponding mass-shell condition.

For a homogeneous background scalar field, $M_{\rm eff}$ depends only on time, $M_{\rm eff}=M_{\rm eff}(t)$, and not on space. In this case, geodesics, distribution functions, and Boltzmann equations are naturally written with the emergent metric $\bar G_{\mu\nu}$, with particles satisfying $\bar G_{\mu\nu}\bar P^{\mu}\bar P^{\nu}=-M_{\rm eff}^{2}(t),$
while the intrinsic particle mass remains $m$.\\

We now compute $dx^{i}/dt$, $d\pa/dt$, and $d\hat{\pa}^{i}/dt$ using the non-vanishing Christoffel symbols in Eqs.~(\ref{36})--(\ref{41}). These expressions are needed to write the Boltzmann equation (Eq.~(\ref{21})) in the perturbed spacetime (Eq.~(\ref{35})), keeping only terms linear in the perturbations. We use the comoving momentum components $P^i\equiv dx^i/d\l$ and $P^0\equiv dt/d\l$ (Eq.~\eqref{44}), introduced below:
\ben
\frac{dx^{i}}{dt}=\frac{dx^i}{d\l}\frac{d\l}{dt}=\frac{\hat{\pa}^{i}}{a} \frac{\pa}{\widetilde{E}}\Big[1-\frac{1}{2}A\dot\phi^{2}+\Psi-\widetilde{\Phi}\Big].
\label{47}
\een
Using the Eq. (\ref{45}), $d/d\l=P^{\mu}\partial/\partial x^{\mu}$ and
\ben
&&\frac{dP^i}{d\l}=-\Ga^{i}_{\a\b}P^{\a}P^{\b}\nonumber\\&&
=-\Big[\Ga^{i}_{00}(P^0)^{2}+2\Ga^{i}_{0j}P^{0}P^{j}+\Ga^{i}_{jk}P^{j}P^{k}\Big],
\label{48}
\een
we obtain
\ben
&&\frac{d\pa^{i}}{d\lambda}=\frac{d}{d\l}[P^{i}(1+\Phi)a]\n &&=P^{i}\frac{d}{d\l}[(1+\Phi)a]+(1+\Phi)a\frac{dP^i}{d\l}\nonumber\\&&=\widetilde{E}\big(1+\frac{1}{2}A\dot\phi^{2}-\Psi\Big)\Big[{\pa}^{i}\Big(H+\dot{\Phi}\Big)\n &&+\frac{{\pa}^{i}}{a\widetilde{E}}\Big(1-\frac{1}{2}A\dot\phi^{2}\Big){\pa}^{k}\partial_{k}{\Phi}\Big]\nonumber\\&&-\widetilde{E}\Big[\widetilde{E}\frac{\partial_{i}\Psi}{a}\Big(1+A\dot{\phi}^{2}\Big)
+2{\pa}^{i}\Big(H+\dot{\Phi}\Big)\Big(1+\frac{1}{2}A\dot\phi^{2}-\Psi\Big)\nonumber\\&&+\frac{2{\pa}^{i}}{a\widetilde{E}}{\pa}^{k}\partial_{k}{\Phi}-\frac{{\pa}^{2}}{a\widetilde{E}}{\partial_{i}}{\Phi}\Big].\label{49}
\een
Now using the relation $d\pa^{i}/dt=(P^{0})^{-1}d\pa^{i}/d\l$, we have
\ben
\frac{d\pa^{i}}{dt}&&=-{\pa}^{i}\Big(H+\dot{\Phi}\Big)-\frac{\widetilde{E}}{a}\partial_{i}\widetilde{\Psi}-\frac{{\pa}^{i}}{a\widetilde{E}}{\pa}^{k}\partial_{k}\widetilde{\Phi}+\frac{{\pa}^{2}}{a\widetilde{E}}{\partial_{i}}\widetilde{\Phi}\n 
\label{50}
\een
with
\ben
&&\E=E\sqrt{1-A\dot\phi^{2}};\n && \widetilde{\Phi}=\Phi(1-\frac{1}{2}A\dot\phi^{2});\n &&
\widetilde{\Psi}=\Psi(1+\frac{1}{2}A\dot\phi^{2}-\frac{1}{2}A^{2}\dot\phi^{4}).
\label{51}
\een

The above Eq. (\ref{50}) is basically {\it the geodesic equation for the K-essence-induced FLRW universe (Eq. (\ref{35}))}. This is a compact relation that can be applied to many areas of diverse physics.

Now we calculate the rate of change of the magnitude of momentum as
\ben
\frac{d\pa}{dt}=\frac{d}{dt}\sqrt{\delta_{ij}\pa^{i}\pa^{j}}&&=-\pa\Big(H+\dot{\Phi}\Big)-\frac{\widetilde{E}}{a}\hat{\pa}^{i}\partial_{i}\widetilde{\Psi}. ~~~~~
\label{52}
\een
Finally, we calculate the rate of change of the direction of the momentum as
\ben
\frac{d\hat{\pa}^{i}}{dt}&&=\frac{\widetilde{E}}{a\pa}\big[\delta^{ik}-\hat{\pa}^{i}\hat{\pa}^{k}\Big]\partial_{k}\Big[\frac{\pa^{2}}{\widetilde{E}^{2}}\widetilde{\Phi}-\widetilde{\Psi}\Big].
\label{53}
\een
It is important to observe that from Eqs. (\ref{50}), (\ref{52}), and (\ref{53}), the terms $\frac{d\pa^{i}}{dt}$, $\frac{d\pa}{dt}$, and $\frac{d\hat{\pa}^{i}}{dt}$ are different from the conventional case \cite{Dodelson}; due to the presence of modified terms defined in Eq. (\ref{51}), which are influenced by the K-essence scalar field. It is evident that, for our analysis, we focus on the homogeneous K-essence scalar field ($\phi({\bf x},t)\equiv \phi(t)$). Consequently, the associated emergent metric (Eq. (\ref{12})) influenced by the K-essence scalar fields only in the `$00$' component and remains similar-type modifications in the perturbed spacetime (Eq. (\ref{35})). Thus, in our case, it indicates that $\partial_{i}\widetilde{\Psi}\equiv (1+\frac{1}{2}A\dot\phi^{2}-\frac{1}{2}A^{2}\dot\phi^{4})\partial_{i}\Psi$ and $\partial_{i}\widetilde{\Phi}\equiv (1-\frac{1}{2}A\dot\phi^{2})\partial_{i}\Phi$. Therefore, it can be concluded that in the context of our K-essence-induced perturbed universe as described in Eq. (\ref{35}), the metric perturbative functions $\Psi$, and $\Phi$ are modified to $\widetilde{\Psi}$ and $\widetilde{\Phi}$ as indicated in Eq. (\ref{51}), while the spatial derivatives of $\Phi$ and $\Psi$ are just rescaled by the K-essence scalar field. However, the fundamental equations from Eq. (\ref{47}) to (\ref{53}), which are associated with the description of particle motions, must be different from the conventional framework outlined in \cite{Dodelson}. This necessity arises from the influence of the modified perturbed spacetime represented in Eq. (\ref{35}) and the corresponding modifications articulated in Eq. (\ref{51}). 

The equation presented in Eq.~(\ref{52}) illustrates the evolution of the magnitude of a particle's comoving momentum $\pa$ within the context of a K-essence–induced perturbed FLRW universe. The initial term, $(-\pa^{i}H)$, denotes the cosmological redshift resulting from the expansion of the universe, whereas $(-\pa^{i}\dot{\Phi})$ accounts for the influence of the temporal variation of the gravitational potential, leading to energy shifts, which may be linked to the Integrated Sachs–Wolfe (ISW) effect \cite{Dodelson, Sachs, Fosalba}. The ISW effect manifests as photons travel through dynamic gravitational potentials from the surface of last scattering to the current epoch, particularly significant when the universe's energy density is primarily influenced by components other than matter. The final term, $(-(\widetilde{E}/a)\hat{\pa}^{i}\partial_{i}\widetilde{\Psi})$, arises from spatial inhomogeneities in the potential, governed by the perturbative field $\Psi$, and is associated with the local gravitational acceleration exerted on the particle. The terms collectively describe the mechanisms by which cosmic expansion and scalar perturbations in the K-essence framework influence the redshifting and energy transfer of particles moving through the perturbed spacetime.

\subsection{Boltzmann equation for the perturbed K-essence induced FLRW spacetime (Eq. (\ref{35})) \label{S4.3}}
\noindent
In this subsection, we formulate the perturbed Boltzmann equation for the K-essence induced FLRW metric (Eq. (\ref{35})), substituting all these expressions of $dx^{i}/dt$, $d\pa/dt$, and $d\hat{\pa}^{i}/dt$ in Eq. (\ref{21}) for the perturbed spacetime (Eq. (\ref{35})), we obtain the {\it perturbed BEKG} as follows,
\ben
&&\frac{df}{dt}=\frac{\partial f}{\partial t}+\frac{\partial f}{\partial{x}^{i}}\frac{\hat{\pa}^{i}}{a}\frac{\pa}{\widetilde{E}}\Big(1-\frac{1}{2}A\dot{\phi}^{2}+\Psi-\widetilde{\Phi}\Big)\nonumber\\&&
-\pa\frac{\partial f}{\partial \pa}\Big[H+\dot{\Phi}+\frac{\hat{\pa}^{i}}{a}\frac{\widetilde{E}}{\pa}\partial_{i}\widetilde{\Psi}\Big]\nonumber\\&&+\frac{\partial f}{\partial{\hat{\pa}^{i}}}\frac{\widetilde{E}}{a\pa}\Big[\partial_{i}\Big(\frac{\pa^2}{\widetilde{E}^2}\widetilde{\Phi}-\widetilde{\Psi}\Big)-\hat{\pa}^{i}\hat{\pa}^{k}\partial_{k}\Big(\frac{\pa^2}{\widetilde{E}^2}\widetilde{\Phi}-\widetilde{\Psi}\Big)\Big].~~~~~~
\label{54}
\een

First, we deduce {\it the collisionless perturbed Boltzmann equation for radiation, i.e., ultra-relativistic particles}, in the perturbed Universe (Eq. (\ref{35})), so that, in this case, $m=0 \Rightarrow M_{\rm eff}=0$, i.e., $\widetilde{E}\simeq \pa$. Therefore, the above Eq. (\ref{54}) can be written as 
\ben
\frac{df}{dt}=\frac{\partial f}{\partial t}+\frac{\partial f}{\partial{x}^{i}}\frac{\hat{\pa}^{i}}{a}\Big(1-\frac{1}{2}A\dot{\phi}^{2}\Big)\nonumber\\
-\pa\frac{\partial f}{\partial \pa}\Big[H+\dot{\Phi}+\frac{\hat{\pa}^{i}}{a}\partial_{i}\widetilde{\Psi}\Big].
\label{55}
\een
Finally, the linear-order {\it perturbed Boltzmann equation in K-essence-induced an FLRW universe (Eq. (\ref{35})), for massive particles} ($\widetilde{E}\simeq M_{\rm eff}$) can be written as
\ben
\frac{df}{dt}=\frac{\partial f}{\partial t}+\frac{\partial f}{\partial{x}^{i}}\frac{\hat{\pa}^{i}}{a}\frac{\pa}{\widetilde{E}}\Big(1-\frac{1}{2}A\dot{\phi}^{2}\Big)\nonumber\\
-\pa\frac{\partial f}{\partial \pa}\Big[H+\dot{\Phi}+\frac{\widetilde{E}}{a\pa}\hat{\pa}^{i}\partial_{i}\widetilde{\Psi}\Big].
\label{56}
\een
Here, for maintaining the linear order of the perturbed BEKG Eqs. (\ref{55}) and (\ref{56}), we neglect the term containing the product of $\partial f/\partial x^{i}$ and $\Psi$ or $\widetilde{\Phi}$, the second term in Eq. (\ref{54}), since it $\partial f/\partial x^{i}$ is first order \cite{Dodelson}, so product terms are basically second-order. Also, we drop the last term of Eq. (\ref{54}), since it $\partial f/\partial{\hat{\pa}^{i}}$ is a first-order \cite{Dodelson}, so the total last term of Eq. (\ref{54}) is a second-order term.

It is to important to note that these two scenarios, i.e, perturbed BEKG for radiations or photons Eq. (\ref{55}) and massive particles or cold dark matter Eq. (\ref{56}) are different from the conventional perturbed Boltzmann equation for the usual FLRW universe \cite{Dodelson}, due to the presence of $\widetilde{E}$ and $A\dot\phi^2$ terms. All other terms are the same due to our choice of homogeneous K-essence scalar field. However, the basic structure of the background geometry (K-essence) is different, where we deduce the Boltzmann equation, since in K-essence geometry, there exists a disformal relationship Eq. (\ref{5}) with the usual gravity \cite{Bekenstein_1993}.\\

To derive the second part of the Boltzmann Eq. (\ref{9}), the collision term $C[f]$ is on the right-hand side. This includes all interactions at the microscopic level, such as scattering, pair production, annihilation, and decay processes. Specifically, we consider a prescription of a two-particle scattering interaction as prescription in Eq. (\ref{32}) and correspondingly energy-momentum conservation in Eq. (\ref{33}), we derive the following collision term as
\ben
&& a\;C[f_{1}(\pa)]=\frac{a}{2\widetilde{E}_{1}(\pa)}\int{\frac{d^3 \qa}{(2\pi)^3 2\widetilde{E}_{2}(\qa)
}}\int{\frac{d^3 \pa'}{(2\pi)^3 2\widetilde{E}_{3}(\pa')
}}\times\nonumber\\ &&
\int{\frac{d^3 \qa'}{(2\pi)^3 2\widetilde{E}_{4}(\qa')
}}|\mathcal{M}|^{2}\times (2\pi)^{4}\delta_{D}^{(3)}\Big[\p+\q-\p'-\q'\Big]\times \nonumber\\ &&
\delta_{D}^{(1)}\Big[\widetilde{E}_{1}(\pa)+\widetilde{E}_{2}(\qa)-\widetilde{E}_{3}(\pa')-\widetilde{E}_{4}(\qa')\Big]\times\nonumber\\ && 
\Big(f_{3}(\p')f_{4}(\q^{'})\big[1\pm f_{1}(\p)\big]\big[1\pm f_{2}(\q)\big]\nonumber\\&&-f_{1}(\p)f_{2}(\q)\big[1\pm f_{3}(\p')\big]\big[1\pm f_{4}(\q')\big]\Big).
\label{57}
\een
This expression of the collision term in Eq.  (\ref{57}) is different from the usual one \cite{Dodelson}. The differences come as before, from the energy part $\E$. $|\mathcal{M}(\p,\q,\p',\q')|^2$ is the squared amplitude, which can be evaluated using Feynman diagrams. Later, we ignore the Bose enhancement and Pauli blocking terms. \\

The distribution function determines the energy-momentum tensor defined in Eq.~(\ref{10}). In a perturbed universe, the appropriate form of Eq.~(\ref{10}) is given in Eq.~(\ref{35}). Therefore, using Eq.~(\ref{35}), we can compute the energy-momentum tensor for the K-essence-induced perturbed universe as

\ben
\T_{\nu}^{\mu}(\x,t)&&=g(1+3\Phi)\big[1-\Psi-3\Phi(1-A\dot\phi^{2})+\frac{1}{2}A\dot\phi^{2}\big]\n &&\times\int\frac{dP_{1}dP_{2}dP_{3}}{(2\pi)^{3}}\frac{P^{\mu}P_{\nu}}{P^{0}} f(\x,\p,t)
\label{58}
\een
and components are
\ben
\T_{~~0}^{0}(\x,t)&&=-g\Big[\Big(1-\frac{3A^{2}\dot{\phi}^{4}}{4}\Big)+\Psi(3A\dot{\phi}^{2})+\Phi(\frac{9A\dot{\phi}^{2}}{2})\Big]\n && 
\times\int \frac{d^3 p}{(2\pi)^{3}}\widetilde{E}(\pa)\f;
\label{59}
\een

\ben
\T_{~~i}^{0}(\x,t)&&=g~a\Big[\Big(1+\frac{A\dot{\phi^{2}}}{2}\Big)-\Psi+\Phi\Big(1+5A\dot{\phi}^{2}\Big)\Big]\nonumber\\&&
\times\int\frac{d^{3}p}{(2\pi)^{3}}\pa_{i}\f;
\label{60}
\een

\ben
\T_{~~j}^{i}(\x,t)&&=g\Big[\Big(1-\frac{A^{2}\dot{\phi}^{4}}{4}\Big)+\Psi (A\dot{\phi}^{2})+\Big(\frac{9A\dot{\phi}^{2}}{2}\Big) \Phi\Big]\n &&
\times\int\frac{d^{3}p}{(2\pi)^3}\frac{\pa^{i}\pa_{j}}{\widetilde{E}}\f,
\label{61}
\een
where we eliminate all terms involving $\dot{\phi}^{6}$, $\Phi(A^{2}\dot{\phi}^{4})$, $\Psi(A^{2}\dot{\phi}^{4})$, and other higher-order contributions, as the emergent gravity metric in Eq. (\ref{12}) necessitates the condition $A\dot{\phi}^{2} < 1$ to maintain a well-defined metric signature. 

In the perturbative part of this analysis, we perform a controlled expansion in the dimensionless quantity $A\dot{\phi}^{2}$ and retain terms through $\mathcal{O}\!\big((A\dot{\phi}^{2})^{2}\big)\sim\mathcal{O}(\dot{\phi}^{4})$, discarding $\mathcal{O}(\dot{\phi}^{6})$ and higher. The inequality $A\dot{\phi}^{2}<1$ is required for the emergent metric to remain Lorentzian, but it does \emph{not} by itself guarantee that $A\dot{\phi}^{2}$ is a parametrically small quantity; it only bounds it. Our truncation is therefore understood as an approximation scheme: for $A\dot{\phi}^{2}\ll 1$ the series is rapidly convergent and one may further drop the quadratic (``second-order'') contribution in Eq.~\eqref{51}, while for $A\dot{\phi}^{2}$ approaching unity the quadratic term must be retained and higher orders may become important. We emphasize that this fixed-order expansion is used only where we explicitly work perturbatively in $A\dot\phi^2$; in the illustrative purely kinetic DBI example \eqref{137}, where the exact background evolution can approach the near-boundary regime $\dot\phi^{2}\to 1$ (for $A=1$) at very early times, we instead keep the full lapse factor $\sqrt{1-\dot\phi^{2}}$ without truncation.

\section{Radiation Dynamics in a Perturbed K-essence FLRW Universe}\label{S5}
\noindent
This section analyzes the behavior of radiation, specifically photons, in an inhomogeneous universe characterized by a perturbed FLRW metric within the K-essence framework. Metric perturbations, induced by the non-canonical scalar field, modify the geodesic motion of photons and alter their phase-space evolution. We analyze the influence of K-essence corrections on photon propagation, redshift, and the evolution of radiation anisotropies by applying the perturbed Boltzmann equation to this geometry. This formulation establishes a basis for addressing observational phenomena, including fluctuations in the cosmic microwave background, within the context of K-essence cosmology.

\subsection{Formulation of the Collisionless Boltzmann Equation for Photons}
\label{Sub-V-1}
\noindent
To derive the collisionless Boltzmann equation for photons, we initiate our analysis with the massless form of the perturbed Boltzmann equation (Eq. (\ref{55})) within the geometric framework of the K-essence. In this framework, the photon ($\E\simeq \pa$) distribution function $f$ is expressed as an expansion around its zeroth-order Bose–Einstein equilibrium state, incorporating small perturbations due to metric fluctuations and K-essence factors. Following Ref.~\cite{Dodelson}, the distribution function may be written as
\ben
f(\x,\pa,\hat{\p},t)
&=& \frac{1}{e^{\E/\widetilde{T}}-1}
\equiv \frac{1}{e^{E/T}-1} \nonumber \\
&=& \left[\exp\!\left(\frac{\pa}{\widetilde{T}(t)\,[1+\Theta(\x,\hat{\p},t)]}\right)-1\right]^{-1},\nonumber\\
\label{62}
\een
where $\Theta(\x,\hat{\p},t)$ denotes the temperature perturbation.
It is worth emphasizing that, in the K-essence–modified geometries
given by Eqs.~(\ref{12}) and (\ref{35}), the spacetime structure is explicitly
affected by the background scalar field $\phi(t)$. This leads to a modification of the particle energy, transforming it from $E$ to $\widetilde{E}~\big(=E\sqrt{1 - A\dot{\phi}^{2}}\big)$. According to the Tolman-Ehrenfest temperature law \cite{Lima, Tolman1, Tolman2}, the temperature associated with this modified spacetime ($\G_{\mu\nu}$) undergoes a transformation given by $\widetilde{T}\equiv T\sqrt{\G_{00}}= T\sqrt{1 - A\dot{\phi}^{2}}$, which preserves the ratio $\widetilde{E}/\widetilde{T} = E/T$. In K-essence geometry, energy and temperature can be expressed as $\widetilde{E}$ and $\widetilde{T}$, respectively, whereas in conventional geometry, they are denoted as $E$ and $T$, respectively. Consequently, the equilibrium photon distribution maintains its Bose-Einstein structure, $f_{\gamma}(\pa) = [e^{\widetilde{E}/\widetilde{T}} - 1]^{-1}$, leading to in the familiar Planck blackbody spectrum, $\rho_{\gamma}(\widetilde{E})d\widetilde{E} = \frac{8\pi \widetilde{E}^{3}}{(2\pi)^3}\frac{d\widetilde{E}}{e^{\widetilde{E}/\widetilde{T}} - 1}$, which indicates the thermal equilibrium condition of radiation within the context of a homogeneous K-essence background. The influence of the homogeneous scalar field on the expansion rate is solely through its non-canonical kinetic term, resulting in an unmodified spectral shape at zeroth order. In order to characterize an inhomogeneous universe, we consider small temperature perturbations, expressed as $\widetilde{T}(\mathbf{x}, \hat{\mathbf{p}}, t) = \widetilde{T}(t)[1 + \Theta(\mathbf{x}, \hat{\mathbf{p}}, t)]$, where $\Theta \equiv \delta\widetilde{T}/\widetilde{T} = \delta T/T$ represents the fractional temperature fluctuation, which is independent of the magnitude of momentum. The function $\Theta(\mathbf{x}, \hat{\mathbf{p}}, t)$ includes the spatial and directional variations in the photon field that emerge from metric perturbations $(\Phi, \Psi)$ and fluctuations induced by K-essence. These variations introduce slight distortions to the otherwise perfect Planck spectrum, resulting in observable imprints on the anisotropies of the CMB.

Given that the perturbation $\Theta$ is small and according to the assumption of small $\Phi$ and $\Psi$, we can perform a Taylor expansion of the distribution function $f$ in Eq. (\ref{62}) as follows:
\ben
f(\x,\p,t)&&\simeq f^{(0)}+ \frac{\partial f^{(0)} }{\partial \widetilde{T}}\widetilde{T} (t)\Theta(\x,\hat{\p},t)\n &&
=f^{(0)}(\pa,t)-\pa\frac{\partial f^{(0)}(\pa,t)}{\partial \pa}\Theta(\x,\hat{\p},t)~~~~
\label{63}
\een
with
\ben
f^{(0)} \equiv \frac{1}{exp\big(\frac{\pa}{\widetilde{T}}\big)-1}.
\label{64}
\een
We use the relation 
$\widetilde{T}\frac{\partial f^{(0)}}{\partial \widetilde{T}}=-\pa\frac{\partial f^{(0)}}{\partial \pa}$ in Eq. (\ref{63}). It is important to observe that Eq. (\ref{63}) is of a similar type with in the conventional scenario \cite{Dodelson}, as the ratio $\widetilde{E}/\widetilde{T}=E/T$ indicates that the K-essence rescaling does not alter the functional structure of the equilibrium distribution; it merely rescales both energy and temperature by an identical lapse factor. Consequently, at the zeroth order, the Bose–Einstein (Planck) distribution remains intact, and the linear perturbation in temperature results in the well-known correction given by ($-\pa\frac{\partial f^{(0)}(\pa,t)}{\partial \pa}\Theta(\x,\hat{\p},t)$). If $\Theta$ exhibits momentum dependence or when $\dot\phi$ varies spatially (here, it does not), additional terms develop, and the simple form may be generalized, which can be seen for the first order.\\

We can now separate the Boltzmann equation (Eq. (\ref{55})) into a zeroth-order equation for $f^{(0)}$, and a first-order equation for the perturbation $\Theta$ where
\ben
\frac{df}{dt}\Big|_{\text{zeroth order}}&&=\frac{\partial f^{(0)}}{\partial t}-H\pa\frac{\partial f^{(0)}}{\partial \pa}\nonumber\\&&
=-\pa\frac{\partial f^{(0)}}{\partial \pa}\Big[\frac{\frac{d\widetilde{T}}{dt}}{\widetilde{T}}+\frac{\frac{da}{dt}}{a}\Big]=0
\label{65}
\een
with
\ben
\frac{\partial f^{(0)}}{\partial t}=\frac{\partial f^{(0)}}{\partial \widetilde{T}}\frac{d\widetilde{T}}{dt}=-\frac{d\widetilde{T}/dt}{\widetilde{T}}\pa\frac{\partial f^{(0)}}{\partial \pa}.
\label{66}
\een
We have set $df/dt$ here equal to zero for collision-free particles. It should be noted that in the context of a K-essence-induced FLRW background, characterized by a homogeneous scalar field, the universe exhibits spatial uniformity and isotropy at the zeroth-order approximation. 
The modified effective metric (Eq. (\ref{12})), shaped by K-essence dynamics, impacts the background expansion rate solely while maintaining spatial homogeneity. As a result, the photon distribution function $f^{(0)}$ is determined exclusively by the magnitude of the momentum $\bar{p}$ and cosmic time, remaining unaffected by spatial coordinates or the direction of propagation. This results in $\partial f^{(0)}/\partial x^{i} = 0$, resulting in the term $\hat{\pa}^{i}\partial f/\partial x^{i}$ to vanish in the background Boltzmann equation. This indicates that, within a homogeneous K-essence universe, there is an absence of free-streaming or spatial transport of radiation, as all regions exhibit identical evolution. The relevance of the term emerges solely upon the inclusion of first-order perturbations within the K-essence geometric framework. From Eq. (\ref{65}), we obtain
\ben
&&\frac{d\widetilde{T}}{\widetilde{T}}=-\frac{da}{a}\n &&
\Rightarrow \widetilde{T}\propto \frac{1}{a}.
\label{67}
\een
The relationship $\widetilde{T} \propto a^{-1}$ indicates that inside the emergent K-essence geometry, the temperature $\widetilde{T}$ redshifts identically to that in conventional cosmology. This outcome derives from the definition of the zeroth-order photon distribution for the emergent metric $\bar{G}_{\mu\nu}$, whose altered lapse does not affect the conventional scaling in that framework.\\

When the temperature is expressed in the usual gravitational frame $g_{\mu\nu}$, the relationship is altered as
\ben
T \propto \frac{1}{a\sqrt{1 - A\dot{\phi}^{2}}}
\label{68}
\een 
indicating that the well-known $T\propto a^{-1}$ relationship is modified in the usual gravitational framework due to the inclination between the emergent and gravitational light cones. Therefore, {\it while the temperature redshifts conventionally in the emergent K-essence framework (where the dynamics are calculated), it seems altered when analyzed in the observational gravitational framework.}

This may be followed by explanations: In the K-essence-induced FLRW background, where the scalar field exhibits homogeneity, the dynamics of photons at zeroth order are consistent with those observed in conventional cosmological frameworks. The uniform K-essence field influences the background expansion rate solely via its contribution to the effective energy-momentum tensor, without directly impacting the local motion of photons at this order. As a result, the wavelengths of photons continue to expand in accordance with the scale factor ($\lambda \propto a$), leading to a redshift in their energies described by $\widetilde{E} \propto 1/a$. This results in the well-known adiabatic cooling law $\widetilde{T}\propto 1/a$, which demonstrates that the radiation temperature decreases inversely with expansion, while preserving a perfect blackbody spectrum. In the zeroth-order Boltzmann equation (Eq. (\ref{65})), this scaling guarantees the conservation of the photon distribution function $f^{(0)}(\pa,t)$ with respect to comoving momentum in the context of K-essence geometry. This indicates that the homogeneous K-essence background affects the overall expansion history solely, leaving the intrinsic behavior of photons unaffected up to the zeroth order.\\

We proceed to derive the equation that describes the deviation of the photon temperature ($\widetilde{T}$) from its zeroth-order value (Eq. (\ref{67})), specifically an equation for $\Theta$ based on the perturbed BEKG (Eq. (\ref{55})). To achieve this, we substitute the expansion of Eq. (\ref{63}) at every instance where $\f$ appears:
\ben
&&\frac{df}{dt}\Big|_{\text{first -order}}=-\pa\frac{\partial^2 f^{(0)}}{\partial \pa \partial t}\Theta-\pa\frac{\partial f^{(0)}}{\partial \pa}\dot{\Theta}\n &&
-\frac{\pa\hat{\pa}^i}{a}\frac{\partial f^{(0)}}{\partial \pa}\frac{\partial\Theta}{\partial x^{i}}\Big(1-\frac{1}{2}A\dot{\phi}^2\Big)
+H\pa\frac{\partial f^{(0)}}{\partial \pa}\Theta \n && -\pa\dot{\Phi}\frac{\partial f^{(0)}}{\partial \pa}-\pa\frac{\partial f^{(0)}}{\partial \pa}\frac{\hat{\pa}^{i}}{a}\partial_{i}\Psi+H\pa^2\frac{\partial^2 f^{(0)}}{\partial^2 \pa}\Theta
\n &&
=-\pa\frac{\partial f^{(0)}}{\partial \pa}\Big[\dot{\Theta}+\frac{\hat{\pa}}{a}\frac{\partial\widetilde{\Theta}}{\partial x^{i}}+\dot{\Phi}
+\frac{\hat{\pa}^{i}}{a}\frac{\partial \widetilde{\Psi}}{\partial x^{i}}\Big]
\label{69}
\een
with 
\ben
\widetilde{\Theta}=\Theta\Big(1-\frac{1}{2}A\dot{\phi}^2\Big)
\label{70}
\een
where we have used the following relations
\ben
&&\pa\Theta\frac{\partial^2 f^{(0)}}{\partial \pa \partial t}=\pa\Theta\frac{\partial^2 f^{(0)}}{\partial \pa \partial \widetilde{T}}\frac{\partial \widetilde{T}}{\partial t}\nonumber\\&&
=-\pa\Theta\frac{d\widetilde{T}/dt}{\widetilde{T}}\frac{\partial f^{(0)}}{\partial \pa}-\frac{\pa^2}{\widetilde{T}}\Theta \frac{d\widetilde{T}}{dt}\frac{\partial^2 f^{(0)}}{\partial^2 \pa},
\label{71}
\een
and from Eq. (\ref{67})
\ben
H=\frac{\dot{a}}{a}=-\frac{d\widetilde{T}/dt}{\widetilde{T}}.
\label{72}
\een

The first-order Boltzmann equation (Eq. (\ref{69})) within our K-essence framework exhibits fundamental differences from the conventional formulation \cite{Dodelson}. The first two terms denote derivatives taken along the trajectory of photons, or null geodesics, within the homogeneous background induced by K-essence. The terms indicate the evolution of the photon distribution function in a non-interactive context, representing a manifestation of {\it `free streaming induced by K-essence'}. 
The remaining terms represent the gravitational influence of perturbations induced by the K-essence scalar field, which modify photon propagation through metric fluctuations. However, the Eq. (\ref{69}) for $\Theta$ remains incomplete, as it is necessary to incorporate a nonvanishing collision term at first order in perturbations to account for the interactions between photons and matter adequately. In the subsequent section, we will analyze the terms associated with this collision with an example.

\subsection{Compton Scattering and the Collision Term in the Boltzmann Equation}
\noindent
This section examines the impact of Compton scattering on the photon distribution function within the perturbed K-essence cosmological framework (Eq. (\ref{35})). We consider processes that occur without assuming chemical or kinetic equilibrium, thereby allowing for full nonequilibrium behavior of photons. This expansion is driven by the observation that, as the universe grows, especially around the epoch of recombination, the scattering rate becomes inadequate to maintain kinetic equilibrium, resulting in deviations from the idealized equilibrium distribution. Incorporating these effects is crucial for precisely characterizing the evolution of the photon distribution in the perturbed K-essence-induced universe and for evaluating how the fundamental K-essence dynamics affect photon propagation, recombination processes, and the consequent characteristics of the observed CMB anisotropies.\\

We consider the following scattering process:

\ben
e^{-}(\q)+\gamma(\p)\leftrightarrow e^{-}(\q')+\gamma(\p'),
\label{73}
\een
where the momentum of each particle is represented within the first bracket, our focus is on the photon distribution assessed at momentum $\p$, characterized by its magnitude $\pa$ and direction $\hat{\p}$. Consequently, it is essential to perform integration across all other momenta $(\q,\q',\p')$ that influence $f(\p)$. Therefore, from Eq. (\ref{34}), we obtain,

\ben
C[f_{\g}(\pa)]&&=\frac{1}{2\widetilde{E}_{\g}(\pa)}\int{\frac{d^3 \qa}{(2\pi)^3 2\widetilde{E}_{e}(\qa)
}}\int{\frac{d^3 \qa'}{(2\pi)^3 2\widetilde{E}_{e}(\qa')
}}\n &&\times \int{\frac{d^3 \pa'}{(2\pi)^3 2\widetilde{E}_{\g}(\pa')
}}
\sum_{\text{3 spins}}|\mathcal{M}|^{2}\n &&\times (2\pi)^{4}\delta_{D}^{(3)}\Big[\p+\q-\p'-\q'\Big]\nonumber\\ &&
\times  \delta_{D}^{(1)}\Big[\widetilde{E}_{\g}(\pa)+\widetilde{E}_{e}(\qa)-\widetilde{E}_{\g}(\pa')-\widetilde{E}_{e}(\qa')\Big]\nonumber\\ && \times
\Big[f_{e}(\q')f_{\g}(\p')-f_{e}(\q)f_{\g}(\p)\Big],
\label{74}
\een
where we neglect the stimulated emission and Pauli blocking, which would result in factors of $1+f_{\g}$ (for the photons) and $1-f_{e}$ (for the electrons) at the corresponding momenta. Pauli blocking becomes insignificant after electron-positron annihilation due to the minimal occupation numbers $f_{e}$, and we will discuss below the reasons for the stimulated emission factors dropping out. As before, the expression of the above collision term (Eq. (\ref{74})) is different from the usual case of Compton scattering \cite{Dodelson}, due to the presence of modified energy $\E$, modified by the K-essence scalar field.

For the photon energies in the collision term (Eq.~(\ref{74})), we use the emergent-frame relation
$\widetilde{E}_{\gamma}(\pa)=\pa$ (and similarly $\widetilde{E}_{\gamma}(\pa')=\pa'$). Near equilibrium, typical photons satisfy $\pa\sim\widetilde T$.

For electrons, we work in the non-relativistic (NR) regime, appropriate around recombination because $\widetilde T\ll m_e$. In the emergent frame we write
\ben
&&\textbf{Photons:}~\widetilde{E}_{\gamma}(\pa)=\pa,\qquad (m_{\gamma}=0),\nonumber\\
&&\textbf{Electrons:}~\widetilde{E}_{e}(\qa)=\sqrt{\qa^{2}+M_{\rm (e)eff}^{2}},\nonumber\\ && M_{\rm (e)eff}=m_e\sqrt{1-A\dot\phi^{2}},\nonumber\\
&&\textbf{NR expansion:}~ \widetilde{E}_{e}(\qa)-M_{\rm (e)eff}\simeq\frac{\qa^{2}}{2M_{\rm (e)eff}}\sim\widetilde T, \nonumber\\ &&\Rightarrow\quad
\qa\sim\sqrt{2M_{\rm (e)eff}\,\widetilde T}.
\label{75}
\een
This scaling is dimensionally consistent in natural units (mass, momentum, and temperature all have dimensions of energy). Equivalently, using $\widetilde T=T\sqrt{1-A\dot\phi^{2}}$, one may write $\qa\sim\sqrt{2m_eT}\,\sqrt{1-A\dot\phi^{2}}$, where for the usual standard case $q=\sqrt{2m_{e}T}$ \cite{Dodelson}.

Here, we have utilized the principle that, in proximity to equilibrium, the characteristic energies of the photons and the kinetic energies of electrons are approximately on the order of temperature $\widetilde{T}$. The electron momenta are significantly greater than the photon momenta, as indicated by the condition $\frac{{2M_{\rm (e)eff}}}{\widetilde{T}} >>1$. By employing the three-dimensional momentum delta function to evaluate the integral $\q'~(=\p+\q-\p')$ in Eq. (\ref{74}), we obtain
\ben
&&C[f_{\g}(\pa)]=\frac{\pi}{2\pa M_{\rm (e)eff}}\int{\frac{d^3 \qa}{(2\pi)^3 2M_{\rm (e)eff}}}
\int{\frac{d^3 \pa'}{(2\pi)^3 2\pa'}
}\n && \times
\delta_{D}^{(1)}\Big[\pa+\widetilde{E}_{e}(\qa)
-\pa'-\widetilde{E}_{e}(|\p+\q-\p'|)\Big]\n &&
\times
\sum_{\text{3-spins}}|\mathcal{M}|^{2} \Big[f_{e}(\p+\q-\p')f_{\g}(\p')-f_{e}(\q)f_{\g}(\p)\Big].\n
\label{76}
\een

To further develop, it is essential to analyze the kinematics related to non-relativistic Compton scattering.  The primary feature of this process relevant to our analysis is that little energy is transferred. Specifically,
\ben
&&\pa'-\pa=\E_{e}(\qa)-\E_{e}(|\p+\q-\p'|)\n &&=\frac{\qa^2}{{2M_{\rm (e)eff}}}-\frac{(|\p+\q-\p'|)^{2}}{2M_{\rm (e)eff}}\n &&\simeq\frac{\Big(\p'-\p\Big).\q}{M_{\rm (e)eff}}.
\label{77}
\een 
The last approximate equality is valid because, according to Eq. (\ref{75}), $\qa$ significantly exceeds both $\pa$ and $\pa'$. Given that $\pa$ and $\pa'$ are of equivalent order, the right-hand side is at most of order $2\pa\qa/ M_{\rm (e)eff}$ (assuming $\p' \simeq \p$). According to Eq. (\ref{75}), the fractional change in photon energy is limited to $\frac{|\pa'-\pa|}{\pa}\leq 2\qa/ M_{\rm (e)eff}$, which indicates that $\frac{|\pa'-\pa|}{\pa} \leq 2\sqrt{\frac{2\widetilde{T}}{M_{\rm (e)eff}}} << 1$. Consequently, non-relativistic Compton scattering is almost elastic, and $\pa \simeq \pa'$. Consequently, this clarifies the rationale for expressing $\Theta$ as a function of $\hat{\p}$ rather than $\pa$. Moreover, it is reasonable to expand the final electron kinetic energy $\frac{(\q+\p-\p')^{2}}{2M_{\rm (e)eff}}$ around its zeroth-order value of $\frac{\qa^{2}}{2M_{\rm (e)eff}}$. One can expand the delta function as

\ben
&&\delta_{D}^{(1)}\Big[\pa+\widetilde{E}_{e}(\qa)-\pa'-\widetilde{E}_{e}(|\p+\q-\p'|)\Big]\n &&
\simeq\delta_{D}^{(1)}(\pa-\pa')+\frac{(\p'-\p).\q}{M_{\rm (e)eff}}
\frac{\partial}{\partial \pa'}\delta_{D}^{(1)}(\pa-\pa').~~~
\label{78}
\een

After evaluating the momentum integrals, the treatment of the derivatives of the delta functions is carried out through the method of integration by parts. By applying the expansion provided and observing that $f_{e}(\q+\p-\p') \approx f_{e}(\qa)$ due to the fact that the electron momentum is significantly larger than the photon momenta ($\pa, \pa' \ll \qa)$, we can express the collision term as
\ben
&&C[f_{\gamma}(\pa)]=\frac{\pi}{8\pa M_{\rm (e)eff}^{2}}\int{\frac{d^3 \qa}{(2\pi)^3
}} f_{e}(\q)\int{\frac{d^3 \pa'}{(2\pi)^3 \pa'
}}\n && \times
\sum_{3-spin}|\mathcal{M}|^{2}\Big[\delta_{D}^{(1)}(\pa-\pa')
+\frac{\Big(\p'-\p\Big).\q}{M_{\rm (e)eff}}\n && 
\times\frac{\partial}{\partial \pa'}\delta_{D}^{(1)}(\pa-\pa')\Big] \Big[f(\p')-f(\p)\Big].
\label{79}
\een

To calculate the amplitude of Compton scattering, we consider the low-energy limit and find that 
\ben
\frac{1}{2}\sum_{4-spin}|\mathcal{M}|^{2}=24\pi\sigma_T^{(\mathrm{geom})}M_{\rm (e)eff}^{2}\Big(1+[\hat{\p}.\hat{\p}']^2\Big).\nonumber\\
\label{80}
\een

Here $\sigma_T^{(\mathrm{geom})}$ is a geometric transport coefficient arising from rewriting the covariant collision term in emergent-frame variables; it does not modify the microscopic QED Thomson cross section $\sigma_T$ \cite{Srednicki}.

We evaluate the low-energy Compton (Thomson) amplitude in the local inertial frame associated with the emergent metric $\bar G_{\mu\nu}$ \cite{Babichev}. Because the underlying QED interaction is local and the scattering region is much smaller than the curvature scale (see Appendix~\ref{E1}), the Thomson amplitude itself is not modified by the K-essence background \cite{Srednicki}. In particular, the electron mass entering the QED amplitude remains the invariant microscopic mass $m_e$.\\

In the standard Thomson limit, the spin-summed squared matrix element can be written as \cite{Dodelson,Weinberg,Srednicki}
\begin{equation}
\frac{1}{2}
\sum_{\rm spins}|\mathcal{M}|^2
=
24\pi \sigma_T m_e^2
\left[
1+\left(\hat{\bar{\mathbf p}}\cdot
\hat{\bar{\mathbf p}}'\right)^2
\right],
\label{80a}
\end{equation}
where the microscopic Thomson cross section is \cite{Dodelson,Weinberg,Srednicki}
\begin{equation}
\sigma_T
=
\frac{8\pi}{3}
\left(\frac{e^2}{m_e c^2}\right)^2 .
\label{80b}
\end{equation}
Since Eq.~(\ref{80b}) depends only on the invariant electron mass $m_e$, $\sigma_T$ is independent of the K-essence background.\\

The K-essence dependence enters only when the covariant collision operator is rewritten in terms of the emergent-frame kinetic variables introduced earlier (see the definitions and mass-shell relations around Eqs.~(20)-(26)). For a homogeneous background we use the emergent lapse
\begin{equation}
N(t)\equiv \sqrt{1-A\dot{\phi}^{2}},\qquad
M_{(e),\rm eff}=m_e N,\qquad
\widetilde E = N E.
\label{80c}
\end{equation}

With these definitions, the ratio that controls the Thomson limit is unchanged,
\begin{equation}
\frac{\widetilde E_\gamma}{M_{(e),\rm eff}}=\frac{E_\gamma}{m_e},
\label{80d}
\end{equation}
so the emergent-frame condition $\widetilde E_\gamma\ll M_{(e),\rm eff}$ is exactly equivalent to the standard microscopic condition $E_\gamma\ll m_e$. Therefore the K-essence background does not modify the microscopic Compton amplitude nor the underlying Thomson cross section.

What does change is the overall normalization of the collision term once it is expressed in emergent-frame phase-space variables. Writing the photon collision operator in the form \cite{Ma,Dodelson}
\begin{equation}
C_{\rm em}[f_\gamma]
=
n_e\,\sigma_T\,\mathcal{F}_{\rm geom}\,
\mathcal{K}_{\mathrm{T}}[f_\gamma],
\label{81}
\end{equation}
where $\mathcal{K}_{\mathrm{T}}[f_\gamma]$ denotes the (dimensionless) Thomson-scattering kernel, i.e. the usual angular-redistribution operator acting on the photon phase-space distribution $f_\gamma$ (including the standard gain and loss terms, and any polarization dependence when $f_\gamma$ is promoted to Stokes components) \cite{Chandrasekhar,HuWhite,ZaldarriagaSeljak,Dodelson}. The purely geometric factor $\mathcal{F}_{\rm geom}$ collects the lapse-dependent contributions from external-state normalization, the energy-conserving delta function, and the conversion between coordinate-time and emergent-frame variables. For a homogeneous lapse,
\begin{equation}
\mathcal{F}_{\rm geom}
=
\left(\frac{m_e}{M_{(e),\rm eff}}\right)^4
=
N^{-4}
=
\left(1-A\dot{\phi}^{2}\right)^{-2}.
\label{80g}
\end{equation}
It is then convenient to define an effective transport coefficient,
\begin{equation}
\sigma_T^{(\rm geom)}
\equiv
\sigma_T\,\mathcal{F}_{\rm geom}
=
\frac{\sigma_T}{\left(1-A\dot{\phi}^{2}\right)^2}\equiv \frac{8\pi}{3}\left(\frac{e^2}{M_{(e)\rm eff} c^2}\right)^2.
\label{82}
\end{equation}
The quantity $\sigma_T^{(\mathrm{geom})}$ should be interpreted as a purely geometric transport coefficient that appears when the covariant collision operator is expressed in emergent-frame energy, momentum, and proper-time variables. It does \emph{not} represent a modification of the microscopic Thomson cross section from QED. In particular, the fundamental cross section $\sigma_T$ retains its standard form with the invariant electron mass $m_e$ (cf.~Eq.~(\ref{80b})). The microscopic scattering amplitude is unchanged because the interaction takes place in a local inertial region that is small compared with the curvature scale of the emergent spacetime (see Appendix~\ref{E}). The extra factor $(1-A\dot{\phi}^{2})^{-2}$ in Eq.~(\ref{82}) arises solely from the variable conversions involved in rewriting the collision integral; it modifies the \emph{macroscopic} transport rate, while leaving the underlying electromagnetic interaction unchanged.

After averaging over the incoming photon polarization and carrying out the standard angular average, the collision kernel can be written in terms of the effective transport coefficient introduced in Eq.~(\ref{82}). In this form one finds
\begin{equation}
\sum_{\rm 3-spin}|\mathcal{M}|^2_{\rm eff}
=
32\pi\,
\sigma_T^{(\rm geom)}\,
M_{(e),\rm eff}^{\,2},
\label{83}
\end{equation}
where the label ``eff'' simply denotes the coefficient that appears once the collision operator has been rewritten in the emergent-frame kinetic variables. It does \emph{not} imply a change in the underlying QED matrix element. Using $M_{e,\rm eff}=m_e N$ together with Eq.~(\ref{82}), we can rewrite the product as
\begin{equation}
\sigma_T^{(\rm geom)}\,M_{(e),\rm eff}^{\,2}
=
\sigma_T\,\frac{m_e^{4}}{M_{(e),\rm eff}^{\,2}},
\label{83a}
\end{equation}
which makes clear that the lapse dependence is carried by the transport normalization, not by the microscopic electromagnetic coupling.

With these definitions, the momentum integrals in the collision operator can be expressed entirely in terms of the emergent-frame transport coefficient $\sigma_T^{(\rm geom)}$. The collision term then keeps the usual Thomson angular dependence, with only its overall normalization (opacity) rescaled by the geometric factor in Eq.~\eqref{82}. In this way, the K-essence dependence arises solely from rewriting the microscopic interaction in emergent-frame kinetic variables, not from any change in the local QED dynamics.\\

It is worth emphasizing that, at the microscopic level, the electron mass $m_e$ is a fixed parameter of the underlying theory and is not renormalized by the K-essence background. The background influences the scattering calculation only through the rewriting of kinematic quantities in the emergent frame. In particular, in the emergent-frame effective description we introduce the convenient mass scale $M_{(e),\rm eff}=m_e\sqrt{1-A\dot{\phi}^{\,2}}$, so that re-expressing the standard (microscopic) Thomson cross section in terms of emergent-frame variables necessarily brings in the purely geometric factor $\left(1-A\dot{\phi}^{2}\right)^{-1/2}$. Throughout this work, we keep this emergent-frame factor unexpanded (i.e., exact); the perturbative expansion is applied to the remaining K-essence-dependent contributions so that all modifications can be traced to the emergent-frame kinematics rather than to any change in the local QED dynamics. Here the factor $\sqrt{1-A\dot{\phi}^{\,2}}$ originates from the time component of the emergent metric and plays the role of the emergent-frame lapse on the homogeneous background (i.e.\ $N=\sqrt{-\bar{G}_{00}}$ in our conventions).\\


When the spin-averaged amplitude $\sum |\mathcal{M}|^{2}$ is considered to be independent of the particle momenta, which is a reasonable approximation in the Thomson limit and also holds in the local inertial frame of the emergent K-essence metric $\bar{G}_{\mu\nu}$, the terms within the square brackets of the collision operator can be expanded to first order in the energy transfer. The momentum integral applied to the electron distribution function results in a factor of $n_{e}/2$ for all terms that are independent of $\q$, with the factor $1/2$ accounting for the two possible spin states of the electron ($g_{e}=2$). 

In contrast, terms that include a factor proportional to $\q/M_{\rm (e)eff}$ yield the contribution $(n_{e}/2)\mathbf{u}_{b}$, with $\mathbf{u}_{b}$ representing the bulk velocity of the electron-baryon fluid. In alignment with the standard cosmological framework, the sole remaining dependence on the electron momentum is expressed through the baryon bulk velocity. The impact of the K-essence background manifests exclusively via the effective rest-frame energy $M_{\rm (e)eff}$ and the altered kinematics of the emergent metric $\bar{G}_{\mu\nu}$. So, using Eqs. (\ref{79}) and (\ref{83}), we obtain

\ben
&&C[f_{\gamma}(\pa)]=\frac{n_{e}\sigma_T^{(\mathrm{geom})}}{4\pi \pa}\int_{0}^{\infty}\pa' d\pa'\int d\Omega'\n &&
\times  \Big[\delta_{D}^{(1)}(\pa-\pa')
\Big(-\pa'\frac{\partial f^{(0)}}{\partial \pa'}\Theta(\hat{\p}')+\pa\frac{\partial f^{(0)}}{\partial \pa}\Theta(\hat{\p})\Big)\n &&
+(\p-\p').\textbf{u}_{b}\frac{\partial\delta_{D}^{(1)}}{\partial \pa'}(\pa-\pa')\Big[f^{(0)}(\pa')-f^{(0)}(\pa)\Big]\Big]\nonumber\\
\label{84}
\een
with $\textbf{u}_{b}=\frac{\q}{M_{\rm (e)eff}}$. Here, $d\Omega'$ represents the solid angle subtended by the unit vector $\hat{\p}$. This expression just reflects the dependency on the photon directions $\hat{\p}$ and $\hat{\p}'$, as the dependence on $(\mathbf{x},t)$ is irrelevant for the collision term. Compton interactions are localized phenomena, with all quantities measured at the same spacetime point. It is crucial to highlight that, although the collision term in Eq.~(\ref{84}) formally resembles the conventional Compton-scattering outcome presented in \cite{Dodelson}, its physical implications are fundamentally distinct. The differentiation stems from the evaluation of all dynamical quantities in the emergent K-essence geometry, instead of the conventional gravitational metric. As a result, the scattering kinematics and the associated Eqs.~(\ref{79}), (\ref{80}), (\ref{83}) exhibit explicit dependence on the effective electron rest-frame energy $M_{\rm (e)eff}$, which is influenced by the K-essence scalar field via the modified lapse function of the emergent metric. This geometric alteration is the primary source of the departure from the standard case, despite the algebraic structure of the collision term maintaining a visually analogous form. \\

In the collision term of Eq. (\ref{84}), there are only two contributions that exhibit a direct dependence on the direction of photon propagation $\hat{\p}$, necessitating integration over the solid angle $d\Omega'$. In the first term, we execute the perturbation of the photon distribution function, $\Theta(\hat{\p})$, which in our framework is characterized within the local inertial frame of the emergent K-essence metric $\bar{G}_{\mu\nu}$. To systematically arrange the angular configuration of the perturbation, it is advantageous to define the angle-averaged monopole moment as
\ben
\Theta_{0}(\mathbf{x},t) \equiv
\frac{1}{4\pi}\int d\Omega'
\Theta(\hat{\p'},\x,t),
\label{85}
\een
defines the isotropic component of the fractional temperature perturbation at a specific point in spacetime. From a physical standpoint, $\Theta_{0}$ represents the perturbation in the angle-averaged photon energy density as perceived by an observer adhering to the emergent metric $\bar{G}_{\mu\nu}$. It is crucial to note that this quantity cannot be integrated into the background temperature, as the latter is inherently homogeneous by construction and is determined by the zeroth-order distribution within the emergent frame.

{\it Angular integration}: The second term in Eq.~(\ref{84}) arises from the bulk velocity $\mathbf{u}_b$ of the electron--baryon fluid and encodes the usual Doppler coupling through the projection $\hat{\p}\cdot\mathbf{u}_b$. Although $\mathbf{u}_b$ is independent of the \emph{magnitude} of the photon momentum, it does \emph{not} drop out under angular integration because it multiplies the photon propagation direction. In particular, only the monopole average of the direction vanishes,
\ben
\int d\Omega\,\hat{\p}=0,
\label{84a}
\een
whereas the dipole structure proportional to $\hat{\p}\cdot\mathbf{u}_b$ survives and contributes to the first angular moment of the Boltzmann hierarchy. Accordingly, the collision term retains the standard Doppler contribution in the emergent K-essence frame, and the angular integration yields\\

\ben
&&C[f_{\gamma}(\pa)]=\frac{n_{e}\sigma_T^{(\mathrm{geom})}}{ \pa}\int_{0}^{\infty}\pa' d\pa'\Big[\delta_{D}^{(1)}\Big[(\pa-\pa')\Big]\n &&
\times\Big(-\pa'\frac{\partial f^{(0)}}{\partial \pa'}\Theta_{0}+\pa\frac{\partial f^{(0)}}{\partial \pa}\Theta(\hat{\p})\Big)\n &&
+\p.\textbf{u}_{b}\frac{\partial\delta_{D}^{(1)}}{\partial \pa'}(\pa-\pa')\Big(f^{(0)}(\pa')-f^{(0)}(\pa)\Big)\Big].
\label{86}
\een
Executing the momentum integral, starting out using the delta function and subsequently applying integration by parts, yields the collision term
\ben
C[f_{\gamma}(\p)]=-\pa\frac{\partial f^{(0)}}{\partial \pa}{n_{e}\sigma_T^{(\mathrm{geom})}}\Big[\Theta_{0}-\Theta(\hat{\p})+\hat{\p}.\textbf{u}_{b}\Big].\nonumber\\
\label{87}
\een

Although Eq.~(\ref{87}) has a similar type of Thomson angular dependence as in standard cosmology \cite{Dodelson}, the K-essence background changes the overall \emph{opacity}. This time-dependent scattering rate multiplies the kernel, because in the emergent-frame description the relevant coefficient is $n_e\,\sigma_T^{(\mathrm{geom})}$ rather than $n_e\,\sigma_T$. This alters the optical depth and visibility function, and therefore modifies the decoupling history relative to the usual FLRW case, with potential consequences for the duration of tight coupling and the thickness of the last-scattering surface, the acoustic scale through the sound horizon at decoupling, and the damping tail through the diffusion scale. Crucially, this does not reflect any change in the microscopic QED interaction: $\sigma_T$ remains the local inertial-frame Thomson cross section (including the usual Klein-Nishina corrections as a function of $E_\gamma/m_e$), while $\sigma_T^{(\mathrm{geom})}$ is the geometry-induced transport coefficient that appears after rewriting the same local scattering process in emergent-frame variables. In the GR limit $A\dot\phi^{2}\to 0$, the emergent description reduces to the standard one and $\sigma_T^{(\mathrm{geom})}\to \sigma_T$, recovering the conventional Thomson collision term.\\

This expression (\ref{87}) elucidates the physical behavior underlying Compton scattering. In the scenario where the electrons exhibit a condition of zero bulk velocity ($\textbf{u}_{b}=0$), the collision term induces a transition of $ \Theta(\hat{\p})$ to $\Theta_{0}$, signifying that all anisotropies in the photon distribution are effectively nullified. Strong scattering indicates a significantly reduced mean-free-path for photons, suggesting that each photon detected at a particular point experienced scattering by electrons in close proximity, all of which are at a uniform local temperature. Consequently, photons tend to converge towards an isotropic distribution, preserving the monopole moment solely.\\

Within the K-essence framework, the scenario is altered when electrons exhibit a nonzero bulk velocity $\mathbf{u}_{b}$. In this scenario, Compton scattering induces a dipole moment within the photon perturbation $\Theta(\hat{\p})$, which is oriented in accordance with the direction and magnitude of $\textbf{u}_{b}$. All higher multipole moments, including the quadrupole, exhibit significant reduction, as repeated scatterings within the emergent metric $\bar{G}_{\mu\nu}$ effectively isotropize the photon distribution relative to the electron rest frame. Consequently, the photon distribution maintains solely a monopole and dipole component, enabling it to be analyzed effectively as a fluid.
This represents the tight-coupling regime, where photons and baryons interact as a unified dynamical fluid influenced by the emergent geometry. As Compton scattering ceases to be effective during the photon-baryon decoupling phase, the fluid model becomes inapplicable, leading to the onset of free streaming for photons. The Boltzmann equations expressed within $\bar{G}_{\mu\nu}$ effectively describe both phases: the tight coupling prior to recombination and the free streaming that follows, all while consistently accounting for the geometric modifications brought about by the homogeneous K-essence background.

To avoid confusion about ``effective'' quantities in Eqs.~\eqref{74}--\eqref{87}, we stress that K-essence does \emph{not} modify the intrinsic particle physics: the electron rest mass $m_e$, the QED coupling, and the (Thomson/Compton) scattering matrix element are the same as in standard cosmology. What changes is the \emph{geometrical/kinematical} relation between energies, momenta, and time as measured by observers comoving with the emergent metric $\bar G_{\mu\nu}$. In particular, the ``effective rest-energy'' that appears in the collision term is a convenient way of parametrizing the deformation of the null cone and the associated dispersion relation when the process is expressed in gravitational-frame variables; it should not be interpreted as a new fundamental electron mass. Equivalently, one may view the modification as a rescaling of the transport variables (proper time, energy, and phase-space measure) and hence of the scattering \emph{rate} entering the Boltzmann equation, while the underlying microphysical interaction remains unchanged.

\subsection{Boltzmann Equation for Photons with Collision Term}
\noindent
Considering the findings derived from the first-order BEKG (Eq.~(\ref{69})) and the Compton collision term in Eq.~(\ref{87}) within the framework of emergent K-essence geometry, we are now positioned to construct the left- and right-hand sides of the Boltzmann Eq. (\ref{9}) for photons.  By introducing a set of supplementary definitions, we can derive the linear evolution equation governing the perturbation of the photon distribution function within the context of a K-essence-modified FLRW background as

\ben
&&\dot{\Theta}+\frac{\hat{\pa}^{i}}{a}\frac{\partial\widetilde{\Theta}}{\partial x^{i}}+\dot{\Phi}
+\frac{\hat{\pa}^{i}}{a}\frac{\partial \widetilde{\Psi}}{\partial x^{i}}\n &&
={n_{e}\sigma_T^{(\mathrm{geom})}}\Big[\Theta_{0}-\Theta(\hat{\p})+\hat{\p}.\textbf{u}_{b}\Big].
\label{88}
\een

Now we want to convert the above BEKG for photons in Eq. (\ref{88}) in terms of conformal time \cite{Dodelson}. In cosmological Boltzmann analyses, it is beneficial to express the equations governing photon evolution using conformal time. This approach transforms the FLRW background metric into a conformally flat form, thereby greatly simplifying the propagation dynamics of massless particles. In conformal time, the trajectories of photons manifest as straight lines in comoving coordinates, leading to the free-streaming term of the Boltzmann equation taking a form reminiscent of Minkowski spacetime, with the effects of cosmic expansion captured in the scale factor. This formulation enables distinct differentiation among redshift effects due to cosmic expansion, local collision phenomena, and metric fluctuations, thereby aiding the development of the Boltzmann hierarchy for photon multipoles. As a result, conformal time serves as the fundamental temporal parameter for analyzing photon transport, tight coupling, and free-streaming within the framework of cosmology. Within the K-essence framework, analogous reasoning is employed to establish the conformal time associated with the emergent metric. This ensures that the dynamics of photons maintain their conventional structure, while all K-essence influences are incorporated through the modified background and perturbation variables.\\

We introduce the \emph{emergent proper time, $\Upsilon$,} measured by comoving observers in the emergent K-essence geometry. Because particles propagate on the effective metric $\bar G_{\mu\nu}$ rather than the gravitational metric $g_{\mu\nu}$, physical clocks tick according to the line element of $\bar G_{\mu\nu}$. For the homogeneous background, the emergent metric has $\bar G_{00}=-(1-A\dot\phi^{2})$, so along a comoving worldline ($d\mathbf{x}=0$) one has $dS^{2}=-(1-A\dot\phi^{2})dt^{2}\equiv -d\Upsilon^{2}$. Hence \cite{Ganguly},
\ben
d\Upsilon=\sqrt{1-A\dot{\phi}^{2}}\, dt,
\label{89}
\een
where $t$ denotes the cosmic time of the gravitational FLRW spacetime. (Accordingly, the square root is required by the metric normalization; any appearance without it is a typographical omission.)

In this emergent proper time, the metric (\ref{12}) takes the standard-type FLRW form given by \cite{Ganguly}
\ben
dS^{2}=-d\Upsilon^{2}+a^{2}(\Upsilon)\sum_{i=1}^{3}(dx^{i})^{2}. 
\label{90}
\een
Consequently, $\Upsilon$ denotes the intrinsic cosmic time for particles progressing in the K-essence spacetime, whereas $t$ signifies the observational gravitational time.\\

Since the standard conformal time is defined by
$d\eta=\frac{dt}{a(t)},$
the {\it emergent conformal time ($\widetilde{\eta}$)} is related to it via
\ben 
d\widetilde{\eta}
=
\frac{d\Upsilon}{a(\Upsilon)}
=
\frac{\sqrt{1-A\dot{\phi}^{\,2}}}{a(t)}\,dt
=
\sqrt{1-A\dot{\phi}^{\,2}}\,d\eta.
\label{91}
\een
Here, the notation $a(\Upsilon)$ does not represent a new (transformed) scale factor; it is the same FLRW scale factor evaluated along the reparameterized emergent time coordinate, $a(\Upsilon)\equiv a\!\left(t(\Upsilon)\right)$. In particular, no additional conformal rescaling of the spatial metric is introduced: only the time coordinate is redefined. Thus, the K-essence geometry modifies only the temporal parametrization of the background evolution, while the cosmological scale factor itself remains unchanged.

With this definition Eq. \eqref{91}, the emergent metric in Eq. (\ref{90}) assumes the conformally flat form as
\ben
dS^{2}=a^{2}(\widetilde{\eta})\left[-d\widetilde{\eta}^{2}+\sum_{i=1}^{3}(dx^{i})^{2}\right].
\label{92}
\een
\\

Before writing the photon temperature-perturbation equation, we make explicit the meaning of the ``tilted'' metric perturbations. The variables $\widetilde{\Phi}$ and $\widetilde{\Psi}$ are \emph{not} new degrees of freedom; they are the scalar (Newtonian-gauge) perturbations of the \emph{emergent} metric $\bar{G}_{\mu\nu}$ that governs photon geodesics. Starting from the perturbed emergent metric written in gravitational cosmic time $t$ (Eq.~\eqref{35}), and re-expressing it in terms of the emergent proper time $\Upsilon$ defined by Eq.~\eqref{89}, the line element can be written as
\ben
 dS^{2}= -\big(1+2\widetilde{\Psi}\big)\,d\Upsilon^{2} + a^{2}(\Upsilon)\,\big(1+2\widetilde{\Phi}\big)\,\delta_{ij}\,dx^{i}dx^{j},~~~
\label{88a}
\een
where $\widetilde{\Phi}$ and $\widetilde{\Psi}$ denote the (emergent-frame) potentials entering the geodesic/Liouville operator and are defined in Eq.~\eqref{51}, namely $\widetilde{\Phi}=\Phi\,(1-\tfrac12 A\dot\phi^{2})$ and $\widetilde{\Psi}=\Psi\,(1+\tfrac12 A\dot\phi^{2}-\tfrac12 A^{2}\dot\phi^{4})$ (to the order kept in our perturbative expansion). Thus, $\widetilde{\Psi}$ and $\widetilde{\Phi}$ play the same geometrical role as the usual Newtonian potentials, but they are the K-essence-rescaled combinations appropriate to the emergent metric $\bar{G}_{\mu\nu}$ and the emergent time slicing. In particular, the background lapse factor $(1-A\dot\phi^{2})$ is absorbed into $\Upsilon$ (and subsequently into the emergent conformal time $\widetilde{\eta}$), so the free-streaming operator keeps its standard form when expressed in emergent-frame variables.

Therefore, we can express the BEKG (Eq. (\ref{88})) for photons undergoing collisions in relation to the emergent conformal time ($\widetilde{\eta}$) as 
\ben
&&{\Theta}'+\hat{\pa}^{i}\frac{\partial\widetilde{\Theta}}{\partial x^{i}}+{\Phi'}
+{\hat{\pa}^{i}}\frac{\partial \widetilde{\Psi}}{\partial x^{i}}\nonumber\\ &&
=n_{e}\sigma_T^{(\mathrm{geom})}a(\widetilde{\eta})\Big[\Theta_{0}-\Theta(\hat{\p})
+\hat{\bold{\p}}.\textbf{u}_{b}\Big]
\label{93}
\een
with $\Theta'=\frac{\partial\Theta}{\partial\widetilde{\eta}}\equiv \frac{1}{a}\dot{\Theta}$ and $\Phi'=\frac{\partial\Phi}{\partial\widetilde{\eta}}\equiv \frac{1}{a}\dot{\Phi}$. In our analysis, we will denote the derivative of a variable with a prime ($'$) symbol, indicating differentiation with respect to the emergent conformal time ($\widetilde{\eta}$). We emphasize that $n_e$ denotes the \emph{proper} free-electron number density in the underlying gravitational FLRW frame (measured per unit physical volume), while the K-essence modification of the collision term enters through the emergent-time parametrization and the geometric transport coefficient $\sigma_T^{(\mathrm{geom})}$ (see the discussion around Eqs.~\eqref{96}--\eqref{97}).

The free-streaming term in Eq.~(\ref{93}) has the same mathematical form as in the standard FLRW Boltzmann equation \cite{Dodelson} because we write the transport equation entirely in terms of the \emph{emergent conformal time} $\widetilde{\eta}$ associated with the emergent metric. In terms of the emergent proper time $\Upsilon$, the effective background line element reads $dS^{2}=-d\Upsilon^{2}+a^{2}(\Upsilon)d\mathbf{x}^{2}$. Photon trajectories are null with respect to this metric, $dS^{2}=0$, implying $\frac{d|\mathbf{x}|}{d\Upsilon}=\frac{1}{a(\Upsilon)}$. With $d\widetilde{\eta}=\frac{d\Upsilon}{a(\Upsilon)}$, it follows immediately that $\frac{d|\mathbf{x}|}{d\widetilde{\eta}}=1$. Hence, photons propagate with unit coordinate speed in the emergent conformal frame, exactly as in the usual FLRW treatment, and the K-essence modification enters through a reparametrization of time rather than through a modification of the free-streaming operator. (If one instead keeps the gravitational cosmic time $t$, the nontrivial lapse of the emergent metric would appear as a rescaling of the coordinate speed.)\\

Now to solve the above BEKG Eq. (\ref{93}). Following \cite{Dodelson}, we define the cosine of the angle between the wavevector $\mathbf{k}$ in Fourier space and the direction of photon propagation $\hat{\mathbf{p}}$ as follows: 
\ben
\mu \equiv \frac{\mathbf{k}\cdot\hat{\mathbf{p}}}{k},
\label{94} 
\een
thereby defining a parameter that characterizes the direction of photon propagation. The wavevector $\mathbf{k}$ is oriented in the direction of the spatial variation of the temperature perturbation, specifically aligned with the gradient.
In this context, $\Theta(\mathbf{k},\mu=1)$ represents the behavior of photons traveling in parallel with $\mathbf{k}$, along which the temperature is changing. Conversely, $\Theta(\mathbf{k},\mu=0)$ characterizes photons propagating perpendicular to $\mathbf{k}$, along trajectories where the temperature fluctuation remains constant.
The angular decomposition remains valid within the K-essence framework, as the directional dependence of photon propagation is established in terms of the comoving spatial coordinates. Meanwhile, the influence of the K-essence background manifests solely through the modified time evolution represented in the emergent metric $\bar{G}_{\mu\nu}$.

In cosmological perturbation theory, the velocities are mainly longitudinal, indicating that the velocity field is aligned with the Fourier wavevector $\mathbf{k}$. In Fourier space, the bulk velocity of electrons (and baryons) is expressed as 
\ben
\mathbf{u}_b(\mathbf{k},\widetilde{\eta}) = \frac{\mathbf{k}}{k} u_b(\mathbf{k},\widetilde{\eta}), 
\label{95} 
\een 
which indicates that the velocity field is irrotational, $\nabla \times \mathbf{u}_b = 0$ in real space. Thus, the component of the electron velocity in the direction of photon propagation $\hat{\mathbf{p}}$ is expressed as $\mathbf{u}_b \cdot \hat{\mathbf{p}} = u_b~\mu$ 
where $\mu = \hat{\mathbf{k}}\cdot\hat{\mathbf{p}}$ as previously defined.\\

We additionally define the {\it optical depth (emergent)}, $\tau(\widetilde{\eta})$, within the K-essence FLRW framework as \cite{Dodelson,Ma}
\ben
\tau(\widetilde{\eta})\equiv \int_{\widetilde{\eta}}^{\widetilde{\eta}_{0}} d\widetilde{\eta}'\;
a(\widetilde{\eta}')\,n_{e}(\widetilde{\eta}')\,\sigma_T^{(\mathrm{geom})},
\label{96}
\een
which measures the cumulative Thomson-scattering opacity encountered by a photon from the
emergent conformal time $\widetilde{\eta}$ up to today $\widetilde{\eta}_0$. 

To make the connection with the collision term explicit, we
introduce the (emergent-frame) \emph{differential optical depth} (often called the opacity) \cite{Dodelson,Ma}
\ben
\kappa(\widetilde{\eta}) \;\equiv\; a(\widetilde{\eta})\,n_e(\widetilde{\eta})\,\sigma_T^{(\mathrm{geom})};\,
\tau(\widetilde{\eta}) \;=\; \int_{\widetilde{\eta}}^{\widetilde{\eta}_0} d\widetilde{\eta}'\,\kappa(\widetilde{\eta}'),
\label{93a}~~~~~
\een
so that the Thomson collision term appearing in Eq.~(\ref{93}) is proportional to $\kappa(\widetilde{\eta})$.
In standard FLRW, one instead has $\kappa_{FLRW}(\eta)=a(\eta)\,n_e(\eta)\,\sigma_T$.
Hence, in our emergent-frame formulation the departure from the usual FLRW enters through the geometric
transport coefficient $\sigma_T^{(\mathrm{geom})}$ (with
$M_{\rm (e)eff}=m_e\sqrt{1-A\dot{\phi}^{\,2}}$) and through the use of the emergent time variable
$\widetilde{\eta}$. 

It should be noted that the electron number density $n_e$ entering the optical depth is the \emph{physical (proper) free-electron number density} defined in the underlying gravitational FLRW frame, i.e., the number of free electrons per unit physical volume measured by comoving observers of $g_{\mu\nu}$. It obeys the usual particle-number conservation and is not altered by the emergent K-essence geometry, since the latter does not change the microphysical properties of electrons or the electromagnetic interaction. The K-essence effects instead enter the photon transport problem through (i) the use of the emergent conformal time $\widetilde{\eta}$ and (ii) the transport coefficient $\sigma_T^{(\mathrm{geom})}$, which encodes how the emergent metric reshuffles the scattering rate when written in emergent-frame variables. With these definitions, the optical depth retains its standard meaning as the accumulated Thomson-scattering probability, without implying any modification of the underlying Thomson process itself.\\

Accordingly, the visibility function is \cite{Dodelson,Ma}
\ben
g(\widetilde{\eta}) \equiv \kappa(\widetilde{\eta})\,e^{-\tau(\widetilde{\eta})},
\label{93b}
\een
which determines both the location and the width of last scattering, and therefore controls how the
modified opacity impacts the CMB anisotropy observables.

In the late universe the free-electron density becomes very small, so that $\tau(\widetilde{\eta})\ll 1$.
Conversely, at early times $\tau(\widetilde{\eta})\gg 1$, corresponding to frequent Thomson scatterings and
the tightly coupled regime \cite{Dodelson,Ma}. With the bounds chosen as in Eq.~\eqref{96}, one has
\ben
\tau'(\widetilde{\eta}) \equiv \frac{d\tau}{d\widetilde{\eta}}
= -\,\kappa(\widetilde{\eta})
= -\,a(\widetilde{\eta})\,n_e(\widetilde{\eta})\,\sigma_T^{(\mathrm{geom})}.
\label{97}
\een\\

Utilizing the above developed definitions, Eqs. \eqref{94}-\eqref{97}, the linearized Boltzmann equation governing the photon temperature perturbation within the framework of K-essence emergent geometry can be expressed as follows: 
\ben
\frac{\partial \Theta}{\partial \widetilde{\eta}} + i k \mu \, \widetilde{\Theta} + \frac{\partial \Phi}{\partial \widetilde{\eta}} + i k \mu \,\widetilde{\Psi}  = -{\tau}'(\widetilde{\eta})\, \big[\, \Theta_0 - \Theta + \mu\, u_b \,\big].
\label{98} \n\
\een 

Throughout, a tilde indicates that the quantity is defined in the emergent (K-essence) frame, i.e., with respect to the effective metric $\bar G_{\mu\nu}$ and the associated conformal time $\widetilde\eta$. We Fourier-transform spatial dependence \cite{Dodelson} as $X(\mathbf{x},\hat{\mathbf{n}},\widetilde\eta)=\int \!\frac{d^3k}{(2\pi)^3}\,X(\mathbf{k},\hat{\mathbf{n}},\widetilde\eta)\,e^{i\mathbf{k}\cdot\mathbf{x}}$, and define $\mu\equiv \hat{\mathbf{k}}\cdot\hat{\mathbf{n}}$. With this convention, the free-streaming (Liouville) term $\hat{\mathbf{n}}\cdot\nabla X$ becomes $ik\mu\,X$; hence $ik\mu\,\widetilde\Theta$ in Eq.~\eqref{98} is simply the Fourier-space form of $\hat{\mathbf{n}}\cdot\nabla\widetilde\Theta$. Similarly, $\widetilde\Phi(\mathbf{k},\widetilde\eta)$ and $\widetilde\Psi(\mathbf{k},\widetilde\eta)$ are the scalar perturbations of the emergent metric, and the term $ik\mu\,\widetilde\Psi$ originates from the line-of-sight gradient of the emergent potential.

Eq.~\eqref{98} is formally identical to the standard Fourier-space photon Boltzmann equation \cite{Dodelson}, but the variables are defined with respect to the emergent (K-essence) metric $\bar G_{\mu\nu}$. At the covariant level the free-streaming part is the usual Liouville operator written in emergent-frame variables, $\bar P^{\mu}\partial_{\mu}f-\bar\Gamma^{\mu}_{\alpha\beta}\,\bar P^{\alpha}\bar P^{\beta}\,\partial f/\partial \bar p^{\mu}$, with $\bar\Gamma^{\mu}_{\alpha\beta}$ computed from $\bar G_{\mu\nu}$ and $\bar P^{\mu}$ the corresponding four-momentum. When this equation is rewritten in terms of the emergent conformal time $\widetilde\eta$ and the associated comoving momentum variables, the Liouville operator reduces to the standard Fourier-space structure (free streaming $ik\mu\,\widetilde\Theta$ plus the usual gravitational source terms), but with emergent-frame potentials and time derivatives. In practice, the K-essence modifications with direct observational impact enter through: (i) \emph{Time evolution:} derivatives are taken with respect to $\widetilde\eta$ rather than $\eta$, shifting the phase of acoustic oscillations and the mapping to the sound-horizon and diffusion scales. (ii) \emph{Geodesic/metric driving:} the source terms $\widetilde\Phi$ and $\widetilde\Psi$ are perturbations of the metric that governs photon geodesics, so $\widetilde\Phi'$ and $ik\mu\,\widetilde\Psi$ encode the modified gravitational redshifting and propagation. (iii) \emph{Scattering/opacity:} the differential optical depth $\tau'(\widetilde\eta)$ is computed in emergent conformal time and involves $\sigma_T^{(\mathrm{geom})}$, changing the effective Thomson rate in the collision term and hence the visibility function and Silk damping. A quantitative prediction for the angular power spectrum requires implementing these emergent-frame sources and $\tau'(\widetilde\eta)$ in an Einstein-Boltzmann solver, which we leave for future work.

The emergent metric does change photon trajectories and the redshifting of their energies. In our formulation, these effects do not appear as additional, phenomenological corrections to the transport equation; they are already contained in the Liouville operator because it is derived from the null geodesics of $\bar G_{\mu\nu}$. If one were to keep the original (gravitational-frame) time and momentum variables, the free-streaming operator would indeed take a different form. However, once the Boltzmann equation is expressed in terms of the emergent conformal time and the momentum variables naturally associated with $\bar G_{\mu\nu}$, all metric effects are absorbed into $\widetilde\eta$, $\widetilde\Phi$, $\widetilde\Psi$, and the corresponding connection coefficients, and the Liouville operator retains its standard structure. Therefore there are no extra independent modifications to photon geodesics beyond those implied by the emergent geometry itself.

\section{Solution of the Boltzmann equation for photons}\label{S6}
\noindent
To solve the photon Boltzmann Eq. \eqref{98} within an emergent K-essence geometry, we employ the line-of-sight integration method already formulated for conventional cosmology \cite{Dodelson, Seljak, Hu}. This methodology enables the formal resolution of the integro-differential Boltzmann equation along photon geodesics by representing the temperature perturbation $\Theta$ as a time integral of source functions, modulated by the optical depth. The evolution of photon temperature perturbations in a homogeneous K-essence structure is governed by the Boltzmann equation within the emergent FLRW geometry.
The complete equation, as shown in Eq.~\eqref{98}, incorporates both gravitational driving terms and the Compton scattering collision term, with all variables described in relation to the effective metric $\bar{G}_{\mu\nu}$. In the early Universe, when it is entirely ionized, the density of free electrons is significant, leading to a correspondingly large optical depth $\tau'(\widetilde{\eta})$. In this region, Compton scattering is highly efficient, and photons are closely connected to the baryon-electron plasma. Consequently, the photon distribution is almost isotropic in the fluid's local rest frame, with only monopole and dipole moments persisting, while higher multipoles are significantly suppressed. This indicates that frequent scattering enforces local thermal stability and inhibits photons from traveling freely across cosmic distances.

As the Universe expands and cools, recombination progressively reduces the density of free electrons, resulting in a rapid decrease in optical depth. The photon mean free path subsequently gets comparable to, and ultimately exceeds, the Hubble scale. At this juncture, the collision term in Eq.~\eqref{98} becomes ineffective, and photon perturbations shift from the tightly coupled regime to freely streaming in the evolving spacetime. This transition is not enforced manually, but rather arises dynamically from the temporal dependence of $\tau'(\widetilde{\eta})$. The Boltzmann equation, therefore, naturally interpolates between a fluid-like description before decoupling and a collisionless kinetic description afterward, fully within the same geometric framework.

The physical criterion separating tight coupling from free streaming is unchanged: decoupling occurs when the Thomson scattering rate drops below the expansion rate, so that the photon mean free path becomes comparable to the relevant horizon scale. In our formulation, this condition is expressed in the emergent conformal time through $\tau'(\widetilde{\eta})=-a\,n_e\,\sigma_T^{(\mathrm{geom})}$. Because $\sigma_T^{(\mathrm{geom})}$ modifies the effective scattering rate appearing in the photon collision term, it can in principle shift the visibility function and hence the effective decoupling time when measured in $\widetilde{\eta}$. However, in the present work we derive the modified transport equations but do not compute the recombination history (ionization fraction) and therefore do not make a quantitative claim about a shift of the recombination epoch. Determining whether, and by how much, the last-scattering redshift and the width of the visibility function change requires solving the modified Einstein--Boltzmann system together with a recombination module, which we leave for future work.

The principal advantage of this formulation is that it enables the representation of photon evolution in an integral \emph{line-of-sight} format, in which the observable anisotropies today are expressed as time integrals over the gravitational and scattering sources along the photon path. In the K-essence scenario, these trajectories are null geodesics of the emergent metric $\bar{G}_{\mu\nu}$, in contrast to the gravitational metric $g_{\mu\nu}$. As a result, the effective energies, conformal times, and redshift factors present in the line-of-sight solution are geometrically rescaled by the K-essence background. Still, the formal structure of the solution stays consistent with the conventional case, with the sole alteration being the interpretation of all kinematic quantities within the emerging geometry.

This methodology underscores a fundamental conceptual aspect of the current study: K-essence does not change the microscopic physics of scattering or the foundational structure of kinetic theory; rather, it alters the spacetime geometry in which kinetic processes occur. The resultant deformation of the causal structure, represented by the inclined light cones of $\bar{G}_{\mu\nu}$, induces time-dependent rescaling of photon energies, temperatures, and effective interaction rates. The line-of-sight formalism offers an effective and geometrically clear framework for tracing the influence of scalar-field dynamics on cosmic observables, such as the CMB anisotropy spectrum, without adding new degrees of freedom in the perturbation sector.

\subsection{Acoustic oscillations for the perturbed K-essence FLRW spacetime} 
\noindent
Before recombination, $\widetilde{\eta}<\widetilde{\eta}_{*}$, the increased free-electron density in the nascent K-essence FLRW geometry indicates that the photon mean free path is significantly less than the comoving horizon. Consequently, continuous Thomson scattering maintains a close coupling between photons and the baryon-electron plasma, enabling the system to function as an effective single fluid. In this domain, photon perturbations undergo acoustic oscillations governed by the interplay between gravitational potentials and radiation pressure. As recombination advances, the optical depth $\tau'(\widetilde{\eta})$ gradually reduces, rendering the collision term in Eq.~\eqref{98} insignificant, signifying the shift to free streaming in the developing spacetime. This study is confined to the tight-coupling and subsequent free-streaming regimes relevant to CMB temperature anisotropies, excluding late-time photon evolution associated with large-scale structure development.

\subsubsection{The Boltzmann equations for the tightly coupled limit}
The tight-coupling regime refers to the condition when the photon mean free path in the emergent K-essence FLRW geometry is significantly smaller than the relevant cosmic scales. This limit is defined by a substantial optical depth, $\tau(\widetilde{\eta}) \gg 1$, indicating that the Thomson scattering is highly efficient. In this regime, only the lowest angular moments of the photon distribution function are dynamically significant: the monopole ($\ell = 0$) and the dipole ($\ell = 1$), while all higher multipoles are significantly suppressed by rapid scattering. Thus, photons function as a relativistic fluid moving across the emergent spacetime, characterized by a local energy density $\rho_\gamma$ and a longitudinal bulk velocity $u_\gamma$.\\

We express the photon temperature perturbation in terms of Legendre polynomials as follows \cite{Dodelson, Seljak, Hu},
\ben
\Theta(\widetilde{\eta},k) \equiv \frac{1}{(-i)^{\ell}} \int_{-1}^{1} \frac{d\mu}{2} \Theta_{\ell}(\widetilde{\eta},k,\mu) P_{\ell}(\mu),
\label{99}
\een
where $P_{\ell}(\mu)$ represents the Legendre polynomials and $\mu \equiv \hat{\mathbf{k}} \cdot \hat{\mathbf{p}}$ defined in Eq. \eqref{94}. The multipole moments $\Theta_{\ell}$ encapsulate the angular configuration of photon perturbations, where $\ell=0$ denotes the monopole (density perturbation), $\ell=1$ signifies the dipole (bulk velocity), and $\ell\geq2$ characterizes higher-order anisotropies generated through free streaming after decoupling. \\

Using the above Eq. \eqref{99} in Eq. \eqref{98}, we get the perturbed photons BEKG for $\ell> 2$ as
\ben
{\Theta}'_{\ell}+ \frac{k}{(-i)^{\ell+1}}\int_{-1}^{1}\frac{d\mu}{2}\,\mu\,\mathcal{P}_{\ell}(\mu)\widetilde{\Theta}(\mu)={\tau}'(\widetilde{\eta})\Theta_{\ell} \n
\label{100}
\een
where ${\Theta}'_{\ell}\equiv \frac{\partial\Theta_{\ell}}{\partial\widetilde{\eta}}$ and $\widetilde{\Theta}$ is defined in Eq. \eqref{70}. To evaluate the integral in the second term, we utilize the following recurrence relation for Legendre polynomials as
\ben
(\ell+1)P_{\ell+1}(\mu)=(2\ell+1)\mu P_{\ell}(\mu)-\ell P_{\ell-1}(\mu),\n
\label{101}
\een
we obtain the above perturbed photons BEKG for $\ell> 2$ Eq. \eqref{100} as
\ben
{\Theta}'_{\ell}-\frac{k\ell}{2\ell+1}\widetilde{\Theta}_{\ell-1}+\frac{k(\ell+1)}{2\ell+1}\widetilde{\Theta}_{\ell+1}={\tau}'\Theta_{\ell}.
\label{102}
\een

Let us evaluate the relative magnitude of the different terms present in Eq.~\eqref{102}. The time-derivative term on the left-hand side generally exhibits an order of $\Theta_{\ell}/\widetilde{\eta}$, while the right-hand side is amplified by the substantial scattering rate ${\tau}'(\widetilde{\eta})$ in the tight-coupling regime. Since before recombination, one has ${\tau}' \gg k$, the collision term predominates the evolution of the multipoles. Neglecting the $(\ell+1)$-th moment for the moment, Eq.~\eqref{102} suggests the following approximate scaling relation: 
\ben
\Theta_{\ell} \sim -\,\frac{k}{{\tau}'}\, \frac{\ell}{2\ell+1}\,\widetilde{\Theta}_{\ell-1}=-\frac{k}{{\tau}'}\, \frac{\ell}{2\ell+1}\,\Theta_{\ell-1}(1-\frac{1}{2}\frac{A\phi'^2}{a^2}) \n 
\label{103} 
\een
where $\phi'=\frac{\partial\phi}{\partial\widetilde{\eta}}$.\\

The effective photon mean free path (MFP) in the emergent K--essence geometry is expressed as $\lambda_{\rm MFP}= -1/{\tau}'$, which allows the prefactor to be represented as $k\lambda_{\rm MFP}$. Consequently, for perturbation modes with wavelengths significantly above the mean free path, $k\lambda_{\rm MFP}\ll 1$, we obtain the hierarchy $\Theta_{\ell} \ll \Theta_{\ell-1},$ indicating that each subsequent multipole is significantly suppressed. The identical rationale pertains to the omitted $(\ell+1)$ term, hence validating its exclusion from the estimate. Therefore, all multipoles with $\ell>1$ are insignificant relative to the monopole and dipole in the tight-coupling regime. This establishes the formal foundation for considering the photon-baryon system as an effective fluid within the emergent spacetime.

The physical basis for this suppression is directly associated with the finite mean free path of photons. An observer located at a specific spacetime point receives photons from a region around the size of $\lambda_{\rm MFP}$. For substantial perturbations where $k\lambda_{\rm MFP}\ll 1$, the temperature field exhibits limited variation over this region, resulting in negligible angular anisotropy generation. 
Conversely, perturbations with wavelengths akin to the mean free path, $k\lambda_{\rm MFP}\sim 1$, can theoretically produce larger multipoles; however, these modes are effectively suppressed by photon diffusion (Silk damping). Consequently, under the closely linked framework of K--essence cosmology, the resultant geometry mandates a fluid-like behavior for radiation, with only the monopole and dipole assuming a dynamic function.\\

With these consideration factors, we can now focus on the evolution equations for the two lowest multipoles. As all higher moments $\Theta_{\ell}$ with $\ell \geq 2$ are significantly suppressed in the tight-coupling regime, the photon distribution is precisely characterized by considering only the monopole and dipole moments. 
By continually discarding the quadrupole and higher-order contributions, the Boltzmann hierarchy, utilizing Eqs. \eqref{98} and \eqref{99}, simplify into a closed system that encompasses the first two moments solely, 
\ben
{\Theta}'_{0}+k\widetilde{\Theta}_{1}=-{\Phi}'
\label{104}
\een
and
\ben
{\Theta}'_{1}-\frac{k}{3}\widetilde{\Theta}_{0}=\frac{k\widetilde{\Psi}}{3}+{\tau}'\Big[\Theta_{1}-\frac{i{u_b}}{3}\Big]
\label{105}
\een
where $\Phi'\equiv\frac{\partial\Phi}{\partial\widetilde{\eta}}$ and $\widetilde{\Psi}$ is defined in Eq. \eqref{51}.\\

In the tight-coupling regime, the velocity of the baryon-electron fluid is closely associated with the photon dipole moment. This relationship is governed by the ratio, $R\,\equiv R(\widetilde{\eta})$, of baryon to photon energy densities, defined as
\ben
R\equiv \frac{3\rho_{b}(\widetilde{\eta})}{4\rho_{\gamma}(\widetilde{\eta})}.
\label{106}
\een
Because of the substantial Thomson scattering rate ${\tau}'$ before recombination, both time-derivative and gradient modification to the fluid velocity are diminished by small factors on the order of $1/{\tau}'(\widetilde{\eta})$ and $k/{\tau}'$. Thus, to leading order, the baryon velocity closely follows the photon dipole, $u_b \simeq -3i\Theta_{1}$.  

A systematic extension beyond this leading-order approximation is achieved by substituting this connection into the subdominant correction terms. This produces the next-to-leading-order expression for velocity as {\it [See Appendix  \ref{B} for details.]},
\ben
u_b\simeq -3i\Theta_{1}+\frac{R}{{\tau}'}\Big[-3i{\Theta}'_{1}-3i\frac{a'}{a}\Theta_{1}+ik\widetilde{\Psi}\Big].
\label{107}
\een
which effectively incorporates the first finite-scattering modifications to the baryon velocity within the emerging K-essence spacetime. This expansion provides the necessary framework for deriving the effective fluid equations that govern acoustic oscillations in the closely coupled photon-baryon system.\\

Now we insert the above expression \eqref{107} in Eq. \eqref{105}, we find 
\ben
{\Theta}'_{1}+\frac{R}{1+R}\frac{a'}{a}\Theta_{1}-\frac{k}{3(1+R)}\widetilde{\Theta}_{0}=\frac{k}{3}\widetilde{\Psi}.
\label{108}
\een

We are consequently presented with a closed system of two first-order coupled equations for the lowest photon multipoles, Eqs.~\eqref{104} and \eqref{108}. The equations can be amalgamated into a single second-order differential equation for the photon monopole by differentiating Eq.~\eqref{104} with respect to conformal time and utilizing Eq.~\eqref{108} to eliminate the derivative of the dipole moment, $\Theta_{1}'$, while Eq.~\eqref{104} is used to eliminate $\Theta_{1}$. This approach yields the effective dynamical equation governing the acoustic oscillations of the closely coupled photon-baryon fluid in the emergent K-essence spacetime. Thus, we found
\ben
{\Theta}''_{0}+\Big[\frac{R}{1+R}\frac{a'}{a}+\frac{\partial}{\partial\widetilde{\eta}}(\frac{A{\phi}'^{2}}{2a^{2}})\Big]{\Theta}'_{0}+k^{2}\mathcal{C}^{2}_{s}\Theta_{0}=F(k,\widetilde{\eta})\n
\label{109}
\een
where 
\ben
F(k,\widetilde{\eta})\equiv -{\Phi}''-\frac{k^{2}}{3}\Bar{\Psi}-\frac{R}{1+R}\frac{a'}{a}{\Phi}'-{\Phi}'\frac{\partial}{\partial\widetilde{\eta}}\Big(\frac{A{\phi}'^{2}}{2a^{2}}\Big)\n
\label{110}
\een
and the corresponding ({\it effective}) sound speed of the fluid is
\ben
\mathcal{C}_{s}\simeq\sqrt{\frac{1-\frac{A{\phi}'^{2}}{a^{2}}}{3(1+R(\widetilde{\eta}))}}.
\label{111}
\een

An alternative form of Eq. \eqref{109} can also be expressed as:
\ben
\Big\{\frac{\partial^{2}}{\partial\widetilde{\eta}^{2}}+\Big[\frac{R'}{1+R}+\frac{\partial}{\partial\widetilde{\eta}}\Big(\frac{A{\phi}'^{2}}{2a^{2}}\Big)\Big]\frac{\partial}{\partial\widetilde{\eta}}+k^{2}\mathcal{C}_{s}^{2}\Big\}\Big(\Theta_{0}+\Phi\Big)\nonumber\\
=\frac{k^{2}}{3}\Big[\frac{\widetilde{\Phi}}{1+R}-\widetilde{\Psi}\Big] \n 
\label{112}
\een
where $\widetilde{\Phi}$ and $\widetilde{\Psi}$ are defined in Eq. \eqref{51}.\\

The effective sound speed $\mathcal{C}_{s}$ is clearly dependent upon the baryon loading parameter $R(\widetilde{\eta})$ and the K-essence background via the scalar-field contribution. In the regime, when baryon density is insignificant relative to radiation density, $R \ll 1$, and in the absence of K-essence corrections, the standard relativistic value is retrieved as $\mathcal{C}_{s} \to 1/\sqrt{3}$. The inclusion of baryons increases the inertia of the photon-baryon fluid, thereby reducing the sound speed and altering the oscillation frequency, analogous to the inverse mass term in the equation of a forced harmonic oscillator. Within the current framework, the emergent K-essence geometry further alters the propagation of acoustic modes through the supplementary time-dependent rescaling represented by the scalar-field factor $A{\phi}'^{2}/a^{2}$, which influences both the effective friction term and the driving force $F(k,\widetilde{\eta})$. Consequently, photon perturbations exhibit pulsating behavior in both spatial dimensions and effective conformal time, with a period determined by the interplay between baryon loading and emergent geometry. While the drag term introduces gradually increasing damping, it does not alter the qualitative understanding of acoustic oscillations developed in the tight-coupling regime. It is essential to point out that the combination $\Theta_{0}+\widetilde{\Psi}$, which inherently arises in the tight-coupling formulation of the photon Boltzmann equation, is not equivalent to the complete observable CMB temperature anisotropy. The latter implies the gauge-invariant combination $\Theta_{0}+\widetilde{\Psi}-\widetilde{\Phi}$ assessed at the surface of last scattering, together with supplementary integrated contributions along the photon journey. Consequently, whereas $\Theta_{0}+\widetilde{\Psi}$ serves as a useful variable for characterizing acoustic dynamics inside the emergent K-essence spacetime, it does not inherently denote the directly observed fluctuations in temperature.

\subsubsection{Solutions of the tightly coupled Eq. \eqref{112}}
To solve Eq. \eqref{112}, we use the same approaches as in conventional cosmological perturbation theory and Green's function methods. The process begins with finding the two linearly independent solutions of the connected homogeneous equation, which is done by putting the right-hand side of Eq.~\eqref{112} to zero. Then, the entire solution is obtained by integrating against the correct source term. This method enables us to systematically and covariantly address gravitational motivation and initial scenarios within the emergent K-essence geometry.

In principle, the homogeneous solutions necessitate the identification of a damped harmonic oscillator equation characterised by time-dependent coefficients, where the effective damping is derived from both baryon loading via the factor $R(\widetilde{\eta}) $ and the scalar-field contribution $\partial_{\widetilde{\eta}}(A\phi'^2/2a^2) $. However, for modes that are well inside the sound horizon or in regions where the baryon fraction is not the most important, the pressure term $k^{2}\mathcal{C}_{s}^{2}(\Theta_{0}+\Phi)$ is significantly larger than the damping terms. 
This means that the acoustic oscillations of the tightly coupled photon-baryon fluid occur on a time scale significantly shorter than the time it takes for dissipative effects to act. Consequently, we can disregard the drag term and treat the homogeneous solutions as basic oscillatory modes with frequency $k\mathcal{C}_{s}$. A WKB-type expansion can systematically incorporate corrections arising from the slowly fluctuating background.\\ 

To solve the homogeneous part of Eq. \eqref{112}, we use a WKB ansatz like 
\ben
\Theta_{0}(\widetilde{\eta})+\Phi(\widetilde{\eta})
= D(\widetilde{\eta})\,e^{iB(\widetilde{\eta})}
\label{113}
\een
with $D(\widetilde{\eta})$ and $B(\widetilde{\eta})$ are real functions. We found the solution [{\it See \ref{C}, for detailed derivations}]
\ben
&&\Theta_{0}(\widetilde{\eta})+\Phi(\widetilde{\eta})
=\mathcal{A}(\widetilde{\eta}) 
\Big[
\cos\!\big(k r_{s}(\widetilde{\eta})\big)
+\sin\!\big(k r_{s}(\widetilde{\eta})\big)
\Big],\n
\label{114}
\een
where 
\ben
\mathcal{A}(\widetilde{\eta})=
\frac{\exp\!\Big(-\frac{A{\phi}'^{2}}{4a^{2}}\Big)}
{(1+R)^{1/4}
\Big(1-\frac{A{\phi}'^{2}}{a^{2}}\Big)^{1/4}}.
\label{115}
\een
The prefactor $\mathcal{A}(\widetilde{\eta})$ originates from the drag term in Eq.~\eqref{112} and incorporates the cumulative damping effects resulting from baryon loading and the temporal deformation of the emergent K-essence geometry. This factor influences the amplitude of the acoustic oscillations, whereas the oscillation phase is only governed by the effective sound horizon $r_{s}(\widetilde{\eta})$.\\

For subhorizon modes satisfying $k\,\mathcal{C}_{s} \gg \frac{1}{\widetilde{\eta}}$, the oscillation time scale $(k\mathcal{C}_{s})^{-1}$ is significantly shorter than the time scale over which $\mathcal{A}(\widetilde{\eta})$ evolves. . Consequently, the amplitude modulation caused by baryon loading and the resultant K-essence geometry is slowly adiabatic: during many oscillation periods, the prefactor remains nearly constant, and it can be neglected at leading order. The tight-coupling equation simplifies to that of an undamped harmonic oscillator characterized by a time-dependent effective sound speed $\mathcal{C}_{s}(\widetilde{\eta})$. 

In this limit, the two independent homogeneous solutions take the familiar oscillatory
form
\ben
S_{1}(k,\widetilde{\eta}) = \sin\!\big[k r_{s}(\widetilde{\eta})\big], \qquad
S_{2}(k,\widetilde{\eta}) = \cos\!\big[k r_{s}(\widetilde{\eta})\big],\n
\label{116}
\een
where the effective sound horizon is defined by
\ben
r_{s}(\widetilde{\eta}) \equiv \int_{0}^{\widetilde{\eta}} d\bar{\eta}\,
\mathcal{C}_{s}(\bar{\eta}).
\label{117}
\een
The prefactor $\mathcal{A}(\widetilde{\eta})$ therefore represents a physically significant albeit slowly fluctuating modulation of the acoustic amplitude, which does not influence the phase of the oscillations and hence does not alter the positions of the acoustic peaks.

Thus, in accordance with established procedures in tight-coupling investigations, it is permissible to exclude this prefactor while concentrating on the qualitative oscillatory behavior and phase structure of the photon-baryon acoustic modes. Within the current K-essence framework, this approximation isolates the universal oscillatory behavior, but the prefactor can be reintroduced for calculating precise amplitudes of CMB temperature anisotropies.\\

Using Green’s method, the general solution for the gauge-invariant combination
$\Theta_{0}+\Phi$ may then be written as
\ben
\Theta_{0}(k,\widetilde{\eta})+\Phi(k,\widetilde{\eta})
&&= C_{1}(k) S_{1}(\widetilde{\eta}) + C_{2}(k) S_{2}(\widetilde{\eta})\n &&
+ \frac{k^{2}}{3}
\int_{0}^{\widetilde{\eta}} d\bar{\eta}\,
\Big[\widetilde{\Phi}(k,\bar{\eta})-\widetilde{\Psi}(k,\bar{\eta})\Big]\n && \times
\Big[\frac{S_{1}(\bar{\eta})S_{2}(\widetilde{\eta})
-S_{1}(\widetilde{\eta})S_{2}(\bar{\eta})}{S_{1}(\bar{\eta})S_{2}'(\bar{\eta})
-S_{1}'(\bar{\eta})S_{2}(\bar{\eta})}\Big].\n
\label{118}
\een
The first two terms describe the homogeneous acoustic oscillations, and the integral term describes how the time-dependent metric perturbations $\widetilde{\Phi}$ and $\widetilde{\Psi}$ in the emergent geometry influence gravity.\\

Imposing adiabatic initial conditions appropriate for superhorizon modes,
$\Theta_{0}$ and $\Phi$ are initially constant, implying
$\Theta_{0}'=\Phi'=0$ at $\widetilde{\eta}=0$. This eliminates the sine mode and sets
$C_{1}(k)=0$, yielding
\ben
&&\Theta_{0}(k,\widetilde{\eta})+\Phi(k,\widetilde{\eta})
=
[\Theta_{0}(k,0)+\Phi(k,0)]\cos\!\big(k r_{s}(\widetilde{\eta})\big)\n &&
+\frac{k}{\sqrt{3}}
\int_{0}^{\widetilde{\eta}} d\bar{\eta}\,
\frac{1}{\sqrt{1-\frac{A{\phi}'^{2}}{a^{2}}}}\n && \times
\Big[\big\{\widetilde{\Phi}(k,\bar{\eta})-\widetilde{\Psi}(k,\bar{\eta})\big\}
\sin\!\big\{k(r_{s}(\widetilde{\eta})-r_{s}(\bar{\eta}))\big\}\Big].\n
\label{119}
\een
The first term describes standing acoustic waves with extrema occurring at
\ben
k_{\rm peak} = \frac{n\pi}{r_{s}(\widetilde{\eta})}, \qquad n=1,2,3,\ldots
\label{120}
\een
which determine the locations of the acoustic peaks in the CMB power spectrum. The second term represents the gravitational forcing induced by the evolving potentials during tight coupling, modified by the K-essence deformation of the sound speed.\\

The photon dipole moment follows directly from the monopole through the tight-coupling
relation and is given by
\ben
&&\widetilde{\Theta}_{1}(k,\widetilde{\eta})
=
\sqrt{\frac{1-\frac{A{\phi}'^{2}}{a^{2}}}{3}}
\Big[\Theta_{0}(k,0)+\Phi(k,0)\Big]
\sin\!\big[k r_{s}(\widetilde{\eta})\big]
\n &&-\frac{k}{3}
\int_{0}^{\widetilde{\eta}} d\bar{\eta}\,
\Big[\widetilde{\Phi}(k,\bar{\eta})-\widetilde{\Psi}(k,\bar{\eta})\Big]
\cos\!\big[k(r_{s}(\widetilde{\eta})-r_{s}(\bar{\eta}))\big].\n
\label{121}
\een
where $\widetilde{\Theta}$ defined in \eqref{70}. This expression \eqref{121} shows explicitly that, while the phase structure of acoustic oscillations
remains identical to the standard case, the amplitudes and effective propagation speed
are modified by the emergent K-essence geometry through the factor
$1-\tfrac{A{\phi}'^{2}}{a^{2}}$.\\

In the tightly coupled regime of the K-essence flat FLRW spacetime, we derived the monopole and dipole solutions, Eqs. \eqref{119} and \eqref{121}. In contrast to the conventional FLRW scenario, both solutions explicitly rely on the background K-essence scalar field. In particular, the drag term in the acoustic dynamics is influenced by the scalar field, hence affecting the evolution of photon-baryon perturbations.

The positions of the acoustic peaks are quantized by the integer $n$, whereas the associated wavenumber $k_{\text{peak}}$ is inversely related to the effective sound horizon as defined in Eq.~\eqref{117}. The sound horizon is contingent upon the effective sound speed (Eq.~\eqref{111}), which depends on the K-essence background; hence, the peak locations exhibit a direct dependency on the dynamics of the scalar field. As a result, both the oscillation amplitude and the peak position deviate from the conventional cosmological outcome~\cite{Dodelson}.

Consequently, we ascertain that within this framework, the K-essence scalar field is pivotal in influencing the tightly coupled photon-baryon acoustic solutions, modifying both the damping characteristics and the physical scale of the acoustic features.

\subsection{Diffusion damping in the K-essence FLRW spacetime}
\noindent
The second essential physical aspect necessary for an accurate characterization of the CMB anisotropy spectrum in emergent K-essence cosmology is \emph{diffusion (Silk) damping}. To quantitatively investigate this effect, we must revisit the hierarchy of photon multipole equations generated from the Boltzmann equation in the perturbed emergent metric. In the tight-coupling regime previously stated, only the monopole and dipole moments were considered, while higher multipoles were significantly suppressed by the substantial optical depth. However, the quadrupole moment, while smaller, is not precisely zero. The finite mean free path of photons facilitates the emergence of small directional anisotropy, resulting in an eventual smoothing of temperature fluctuations.

Photons undergo a stochastic trajectory through the photon-baryon plasma before recombination. Through multiple scatterings, they investigate neighboring regions with marginally varying temperatures, thereby efficiently averaging variances over sufficiently small scales. As a result, perturbations with high wavenumber are exponentially reduced prior to decoupling, producing the damping tail of the CMB power spectrum.

In the emergent K-essence geometry, the underlying mechanism remains unchanged, although its quantitative characterization is altered. The speed of photon propagation, interaction rate, conformal time, and effective inertia of the photon-baryon fluid are all geometrically rescaled by the background scalar field via the emergent metric $\bar G_{\mu\nu}$. Consequently, the diffusion length and the corresponding damping scale exhibit a non-trivial dependence on the K-essence background. This section aims to derive the diffusion-damping scale in emergent spacetime and to determine how the scalar field influences the suppression of small-scale photon perturbations.

Consequently, alongside the monopole and dipole equations, we must now incorporate the evolution equation for the quadrupole $\Theta_{2}$. Our effort is facilitated by the fact that diffusion damping is pertinent solely at suitably small scales. In the K-essence emergent geometry, subhorizon modes exhibit metric perturbations $\Phi$ and $\Psi$ that are far less than radiation perturbations, suppressed by about the factor $(aH/k)^{2}$, analogous to the conventional scenario. Moreover, the tight-coupling hierarchy persists: each subsequent multipole is diminished by an extra power of the inverse optical depth $1/\tau'$. Thus, to leading order, we consider the quadrupole moment $\ell=2$ solely, whereas all higher multipoles can be systematically ignored.\\

Utilizing these approximations, the Boltzmann hierarchy is truncated at $\ell=2$, enabling us to ascertain the diffusion damping scale within the emergent K-essence spacetime. The corresponding equations are:
\ben
{\Theta}'_{0}+k\widetilde{\Theta}_{1}=0 \label{122}\\
{\Theta}'_{1}+k(\frac{2}{3}\widetilde{\Theta}_{2}-\frac{1}{3}\widetilde{\Theta}_{0})={\tau}'\Big[\Theta_{1}-\frac{i{u_b}}{3}\Big] \label{123}\\
{\Theta}'_{2}-\frac{2k}{3}\widetilde{\Theta}_{1}=\frac{9}{10}{\tau}'\Theta_{2}. \label{124}
\een
Once more, we ignore polarization effects. The set of three equations mentioned above requires an additional evolution equation for the baryon velocity $u_b$. This may be expressed as a reformulated version of the baryon velocity equation, 
\ben 
u_{b}+ 3i\Theta_{1}=\frac{R(\widetilde{\eta})}{{\tau}'(\widetilde{\eta})}\!\Big(u'_{b}+\frac{a'}{a}u_{b}\Big).
\label{125}
\een
In the small-scale domain addressed herein, the gravitational potential factors are subdominant compared to the scattering interaction and have consequently been neglected.\\

To examine diffusion damping in the emergent K-essence geometry, we follow the conventional tight-coupling expansion \cite{Dodelson} while retaining next-to-leading order corrections in $1/{\tau}'$. Given that acoustic oscillations happen on a timescale far shorter than the Hubble time, the baryon velocity fluctuates in a nearly harmonic manner. We thus assume
\ben
u_b \propto e^{i\int \omega\, d\bar{\eta}}, \qquad \omega = k\mathcal{C}_s,
\label{126}
\een
such that the time derivative fulfills
\ben
|u_b'| = |i\omega u_b| \gg \frac{a'}{a}|u_b|.
\label{127}
\een
The oscillatory motion of the photon-baryon fluid occurs at a velocity exceeding the background expansion rate, with dynamics governed by acoustic propagation rather than Hubble drag.

Employing this hierarchy, the baryon velocity can be increased perturbatively in powers of $1/{\tau}'$,
\ben
u_b = -3i\Theta_1\Big[1-\frac{i\omega R}{{\tau}'}\Big]^{-1} \equiv -3i\Theta_1\Big[1+\frac{i\omega R}{{\tau}'}- \Big(\frac{\omega R}{{\tau}'}\Big)^2\Big].\n
\label{128}
\een
Consequently, the principal term reflects tight coupling ($u_b=-3i\Theta_1$), but the higher-order corrections represent imperfect coupling due to the finite mean free path of photons. The photon multipoles thereafter satisfy,
\ben
\Theta_2=-\frac{4k}{9{\tau}'}\widetilde{\Theta}_1, \qquad
i\omega \Theta_0=-k\widetilde{\Theta}_1.
\label{129}
\een
It is demonstrated that the quadrupole is diminished by $({\tau}')^{-1}$; however, it does not vanish. This small anisotropy is the exact cause of diffusion damping: photons perform a random walk between neighboring regions, progressively eliminating temperature fluctuations on small scales.

Inserting these relations into the Boltzmann hierarchy results in the dispersion relation (\emph{for detailed derivation, see Appendix \ref{G}})
\ben
&&{\omega}^{2}(1+R)-\frac{k^{2}}3\Big(1-\frac{A{\phi}'^{2}}{2a^{2}} \Big)^{2} + \frac{i\omega}{{\tau}'}\Big[\omega^{2}R^{2}\n &&+\frac{8k^{2}}{27}\Big(1-\frac{A\phi'^{2}}{2a^{2}}\Big)^{2}\Big]=0.
\label{130}
\een
The real component indicates the shift in acoustic frequency in the resulting metric, whereas the imaginary component represents the exponential attenuation of the oscillation amplitude. The damping results from the combined impact of photon diffusion and the geometrical rescaling caused by the K-essence background, which modifies both the effective sound speed and the interaction rate of the photon-baryon fluid.\\

To identify the damping of acoustic oscillations in the K-essence FLRW spacetime, we decompose the frequency into a leading oscillatory part and a small correction, $\omega=\omega^{(0)}+\delta\omega,$ and put this formulation into the dispersion relation Eq.~\eqref{130}. The zeroth-order limit signifies perfect tight coupling, $(\tau')^{-1}\rightarrow 0$, wherein photons and baryons function as a unified fluid, without any diffusion.  By retaining only the principal terms, we obtain
\ben
(\omega^{(0)})^{2}(1+R)-\frac{k^{2}}{3}\Big(1-\frac{A{\phi}'^{2}}{2a^{2}}\Big)^{2}=0,
\label{131}
\een
which gives $\omega^{(0)}=k\mathcal{C}_{s},$ consistent with the sound-wave outcome obtained in the tight-coupling phase.  Consequently, the real component of the frequency characterizes acoustic transmission within the effective photon-baryon fluid.

Diffusion damping occurs at first order in $(\tau')^{-1}$ when imperfect coupling between photons and baryons is considered.  By retaining $\delta\omega$ and all terms proportional to $(\tau')^{-1}$, we obtain the expression
\ben
\delta\omega=-\frac{ik^2}{2(1+R)\tau'} \Big[\mathcal{C}_{s}^{2}R^{2} +\frac{8}{27}\left(1-\frac{A\phi'^{2}}{2a^{2}}\right)^{2}\Big].~~~~
\label{132}
\een

The correction is entirely imaginary, indicating that acoustic oscillations are exponentially attenuated rather than phase-shifted.  Consequently, the photon perturbations evolve as follows: 
\ben
\Theta_{0},\Theta_{1}\propto \exp\!\Big[i k\!\int\! \mathcal{C}_{s}(\bar{\eta})\,d\bar{\eta}\Big] \exp\!\Big(-\frac{k^{2}}{k_{D}^{2}}\Big),
\label{133}
\een
where the damping scale (wave number) is given by
\ben
&&\frac{1}{k_{D}^{2}}\equiv \int^{\widetilde{\eta}}_{0} d\bar{\eta}\frac{1}{6(1+R)n_e\,\sigma_T^{(\mathrm{geom})}\,a(\bar{\eta})}\bigg[\frac{R^{2}\Big(1-\frac{A\phi'^{2}}{a^{2}}\Big)}{1+R}\n &&+\frac{8}{9}\Big(1-\frac{A\phi'^{2}}{2a^{2}}\Big)^{2}\bigg].
\label{134}
\een

The oscillatory factor remains constant throughout all moments, while its amplitudes are diminished by an exponential envelope. The associated diffusion length is given by
\ben
\lambda_{D}\sim k_{D}^{-1}\sim \left[\frac{\widetilde{\eta}} {n_{e}\sigma_T^{(\mathrm{geom})}a(\widetilde{\eta})}\right]^{1/2}.
\label{135}
\een
This physically denotes the random-walk distance traversed by photons in the emergent spacetime: the K-essence background influences both the sound speed and the scattering rate, thereby affecting the scale at which small-scale temperature anisotropies are reduced.\\

Before proceeding to the damping calculation, we must specify the dynamics of the scalar field that determine the emergent geometry. We consider a
Dirac-Born-Infeld (DBI) type K-essence Lagrangian as \cite{Vikman,Born,Dirac,Padmanabhan} \ben
\mathcal{L}(\phi,X) = -V(\phi)\sqrt{1+\frac{2X}{\alpha(\phi)}}. 
\label{136} 
\een

For our purposes, we restrict attention to purely kinetic K-essence models \cite{Scherrer,Padmanabhan,Ganguly} and, for simplicity, set $\alpha(\phi)=V(\phi)=1$. In this limit the DBI-type theory reduces to a standard kinetic K-essence Lagrangian \cite{Putter,Scherrer,Ganguly},
\begin{equation}
\mathcal{L}(X)=-\sqrt{1+2X}.
\label{137}
\end{equation}
With the mostly-plus signature used throughout, a homogeneous field has $X=-\dot\phi^{2}/2$, so Eq.~\eqref{137} becomes
\begin{equation}
\mathcal{L}(X)=-\sqrt{1-\dot\phi^{2}},
\label{137a}
\end{equation}
i.e., the familiar purely kinetic DBI form. This sign convention is also consistent with the emergent lapse factor $\sqrt{1-\dot\phi^{2}}$ (for $A=1$). By setting $\alpha$ and $V$ to unity we eliminate explicit potential dependence while retaining the non-canonical kinetic structure that generates the emergent geometry studied here.

In the case of the simplified selection of the kinetic K-essence Lagrangian \eqref{137}, the parameter $A$ as defined in Eq.~\eqref{6} is reduced to $1$.
By applying the equation of motion \eqref{3} in together with Eq.~\eqref{12} \cite{Ganguly}, we derive the following expression: 
\ben
\frac{\ddot{\phi}}{1-\dot{\phi}^{2}} = -3H\dot{\phi}. 
\label{138} 
\een
Solving this differential equation \eqref{138} yields
\ben
\dot{\phi}^{2}(t)=\frac{C}{C+a(t)^{6}}, 
\label{139} 
\een
with $C$ representing an integration constant. For simplicity, we set $C=1$ in the subsequent analysis. \\

In the kinetic K-essence setup used here, the emergent metric of Eq.~\eqref{12} is Lorentzian only if $A\dot\phi^{2}<1$. For the DBI-type purely kinetic Lagrangian in Eq.~\eqref{137} (with $A=1$), the exact homogeneous background solution drives the field into a kinetic-dominated early-time regime ($a\ll 1$) in which $\dot\phi^{2}\to 1$. We stress that this limit follows from the full DBI dynamics and is not a perturbative assumption in small $A\dot\phi^{2}$. Indeed, when $A=1$ the early universe sits close to the Lorentzian boundary $\dot\phi^{2}=1$, so it is generally inconsistent to rely on an expansion in powers of $A\dot\phi^{2}$; whenever we use this DBI early-time limit, we keep the full (unexpanded) dependence on $\sqrt{1-\dot\phi^{2}}$.

This geometric alteration directly influences the microphysics of the radiation plasma. The propagation of photons, interaction rates, and the effective mean free path are modified as distances and time intervals relevant to scattering processes are evaluated in the emergent frame. Consequently, the diffusion of photons before recombination occurs in a modified causal environment, thereby affecting the diffusion length and, in turn, the damping scale of temperature anisotropies, in contrast to the conventional FLRW prediction. The Boltzmann evolution shown in this study accurately reflects realistic transport effects arising from the emergent geometry.

The geometric modification caused by the emergent metric does not signify an early dark-energy component. In the DBI-type kinetic K-essence model, the scalar field converges to the kinetic regime $\dot{\phi}^{2}\to1$ during the early universe, resulting in an energy density \cite{Padmanabhan, Ganguly, Ganguly1} that scales as $\rho_\phi \propto a^{-3}$. Consequently, the field function behaves primarily as pressureless matter rather than as a rigid fluid or a cosmological constant for the kinetic K-essence Lagrangian \eqref{137}. Because radiation redshifts at a rate of $a^{-4}$ throughout early epochs, the cosmos remains radiation-dominated before recombination, hence maintaining the conventional thermal history. 

Thus, the function of the K-essence field in this era is primarily geometric: it alters the causal framework experienced by perturbations via the emergent metric while largely preserving the background expansion history. This validates the utilization of modified Boltzmann dynamics in the pre-recombination epoch without contradicting established early-universe cosmology.\\

\noindent
The modified Thomson interaction rate in the emergent K-essence spacetime for the purely kinetic K-essence \eqref{137} is given by
\ben 
n_{e}\sigma_T^{(\mathrm{geom})}a = 2.3\times10^{-5}\, a^{-8}\,{\rm Mpc}^{-1} (\Omega_{b}h^{2})\Big(1-\frac{Y_{p}}{2}\Big), \n
\label{140}
\een
which significantly diverges from the conventional cosmological result \cite{Dodelson}, where the interaction rate is proportional to $a^{-2}$. In the standard FLRW scenario, the amount $n_{e}\sigma_{T}a \propto a^{-2}$ just indicates the dilution of the free-electron density due to expansion, where $n_{e}\propto a^{-3}$, while the speed of photon propagation and the cross section remain geometrically constant.

Conversely, in the K-essence emergent geometry, the causal structure is altered by the disformal factor $1-\dot{\phi}^{2}$ for the purely kinetic K-essence model \eqref{137}. This impacts both photon propagation and the effective inertia in the Thomson cross section, thereby introducing additional time-dependent geometric rescalings. Consequently, the interaction rate exhibits a significantly sharper scaling of $a^{-8}$. Physically, this means that in the early universe, the photon mean free path decreases far more rapidly toward the past than in the standard case, producing an extremely tight photon-baryon coupling deep in the radiation era.

This enhanced connection has significant implications for cosmological perturbations. The tight-coupling regime endures longer in conformal time, effectively reducing anisotropies during early epochs and altering both the amplitude of acoustic oscillations and the diffusion (Silk) damping scale. Consequently, in contrast to the conventional scenario in which the interaction rate is solely determined by particle dilution, the K-essence scalar field directly influences the scattering history via the emergent geometry, offering a distinct observational avenue to probe non-canonical scalar dynamics via the CMB damping tail.\\

\noindent
We also found the damping scale for the kinetic K-essence \eqref{137} as (\emph{See details in Appendix \ref{D}})
\ben
&&\frac{1}{k_{D}^{2}} = 252.4\times10^{6}\,{\rm Mpc}^{2}\, a^{29/2} (\Omega_{b}h^{2})^{-1}\n && \times \Big(1-\frac{Y_{p}}{2}\Big)^{-1} (\Omega_{m}h^{2})^{-1/2}.
\label{141}
\een
which differs from the conventional result \cite{Dodelson}, where the damping scale increases at a significantly slower rate than the scale factor. In standard FLRW cosmology, the diffusion length is solely controlled by the interplay between photon random-walk propagation and cosmic expansion, resulting in a relatively small temporal dependence of $k_{D}^{-2}$.

In the current K-essence framework, the emergent metric alters both the speed of photon propagation and the rate of Thomson interaction. The interaction rate scales as $a^{-8}$ rather than $a^{-2}$, indicating that photons remain closely connected for a significantly extended duration in conformal time. As a result, the cumulative random-walk distance traversed by photons is substantially reduced during early epochs, but thereafter increases dramatically as recombination approaches. This impact is manifested in the significant scaling $k_{D}^{-2}\propto a^{29/2}$, which serves as a direct geometric imprint of the scalar-field background on the transport characteristics of radiation.

Physically, Eq. (\ref{141}) indicates that Silk damping is far more sensitive to background evolution in K-essence cosmology than in the standard model. Small-scale anisotropies decrease in distinct ways because the diffusion length is influenced not only by baryon density and expansion history but also by the distortion of the causal structure encoded in the emergent metric. Consequently, the high $\ell$ damping tail of the CMB power spectrum serves as a direct probe of the K-essence scalar field, offering a potential observable signal that differentiates emergent geometry theories from the usual FLRW prediction.\\

In standard recombination theory, the CMB is most sensitive to what happens around the time of last scattering, $z\simeq1100$. This information is encoded in the visibility function, whose shape (in particular its width) and the associated high-$\ell$ damping tail are measured with high precision (e.g. \cite{Planck2018}; see also \cite{Dodelson}). In our illustrative purely kinetic DBI-type model \eqref{137}, the strong time dependence introduced by the disformal factor would keep photons and baryons coupled even more tightly at early times, and it would make the diffusion (Silk) damping scale near recombination much more sensitive to the background evolution. Such a change would generically produce a noticeable modification of the damping tail and is therefore expected to be strongly constrained by CMB data.

At the same time, these steep scalings are not universal predictions of K-essence. They arise from the particular background solution $\dot\phi^{2}=C/(C+a^{6})$ (with $C=1$ in our example) together with the way the emergent geometry rescales the effective Thomson interaction coefficient in our kinetic description. A meaningful comparison with standard recombination and CMB constraints would require (i) computing a consistent ionization history $X_e(z)$ using the same emergent-frame variables, and (ii) implementing the modified interaction rate in a full Boltzmann-code calculation to obtain the visibility function and the resulting $C_\ell$'s. Carrying out that full phenomenological analysis is beyond the scope of the present paper, whose purpose is to establish the kinetic framework and to highlight the qualitative signatures. We therefore present the above scalings as an illustrative example of how an emergent geometry can strongly affect photon transport, rather than as a model-independent constraint-ready result.\\

\section{Conclusions}\label{S7}
\noindent
This study investigates photon propagation in an emergent flat FLRW spacetime derived from K-essence cosmology by developing and evaluating the relevant modified Boltzmann equations. Within our paradigm, the effective (emergent) geometry is disformally linked to the conventional gravitational spacetime via the background K-essence scalar field, so that particle motion is governed by a geometrically deformed causal structure rather than by standard FLRW light cones.\\

We first formulated the Boltzmann equation within this emergent geometry. We classified the dynamics into two sectors: massless species (e.g., photons or radiation) and massive species (e.g., cold dark matter). During the study, the scalar field has been considered homogeneous, $\phi=\phi(t)$, functioning as a cosmological clock that alters the spacetime experienced by particles. Even with this minimal assumption, the resulting kinetic equations differ non-trivially from the standard case, as energies, propagation rates, and time evolution are evaluated with respect to the modified metric.\\

In the collisionless photon sector of the modified Boltzmann equation, we showed that, at zeroth order, the effective radiation temperature $\widetilde{T}$ scales inversely with the scale factor, $\widetilde{T}\propto a^{-1}$, precisely as in the conventional cosmological framework. This verifies that, within the emerging K-essence geometry, where the dynamics are inherently structured, photon redshifting occurs as usual. However, when the same temperature is analyzed within the context of the conventional FLRW gravitational spacetime, it incorporates an additional geometric factor, resulting in the expression $T \propto \frac{1}{a\sqrt{1-A\dot{\phi}^{2}}}.$ The discrepancy originates from the mismatch between the null cones of the emergent and gravitational metrics, which alters the physical time perceived by observers. Thus, while radiation typically cools in the K-essence frame, it appears altered in the observed gravitational frame. This illustrates how the scalar field modifies cosmic observables not by altering microscopic physics, but by geometrically distorting the relationship between propagation and measurement. \\

To evaluate the collision term, we considered Compton (Thomson-limit) scattering. The underlying electromagnetic interaction is unchanged: in the local inertial frame associated with the emergent metric $\bar G_{\mu\nu}$, the QED matrix element and the Thomson cross section $\sigma_T=\frac{8\pi}{3}\,\alpha^2/m_e^2$ take their standard forms \cite{Dodelson,Weinberg,deGroot,Ehlers}. What the emergent geometry changes are the kinematical quantities that enter the Boltzmann equation when written in cosmological variables (time, energy, and phase-space measure) \cite{Dodelson,Weinberg,deGroot,Ehlers}. For this reason we introduce an ``effective'' (geometrical) Thomson coefficient, $\sigma_T^{(\mathrm{geom})}$, which simply packages the metric-induced rescalings that convert the local scattering probability into a rate per unit emergent conformal time. It should therefore be interpreted as a geometrical transport coefficient, not as a modification of the fundamental Thomson cross section or of the electron's intrinsic properties.\\

To examine photon dynamics beyond a homogeneous background, we analyzed the higher-order Boltzmann hierarchy derived from the perturbed Boltzmann equation. In the perturbative analysis, we use the unitary gauge ($\delta\phi=0$) within the framework of effective field theory, ensuring that the scalar field does not appear as an independent fluctuation but rather alters the geometry itself. All perturbations are thus encoded in the emergent metric $\bar{G}_{\mu\nu}$, and the evolution of radiation adheres to the first-order perturbed Boltzmann Eq.~\eqref{98}.  Concentrating on the pre-recombination era, characterized by frequent Thomson scattering that closely associates photons with baryons, we obtained the relevant tight-coupling Eqs. \eqref{110} or \eqref{112} that governs acoustic oscillations in the resulting spacetime.\\

To derive explicit predictions, we examined a sample non-canonical purely kinetic DBI-type Lagrangian~\eqref{137}. Two major physical results arise. First, the effective Thomson interaction rate is proportional to $n_e \sigma_T^{\rm eff} a \propto a^{-8} $, rather than the conventional $a^{-2}$ scaling. This indicates that in earlier epochs, the
photon mean free path shrinks much faster than in conventional cosmology, resulting in a strong photon-baryon interaction throughout the radiation era. Secondly, the diffusion (damping) scale varies significantly, with $k_D^{-2} \propto a^{29/2}$.  As a result, the Silk damping process is more influenced by the background evolution: the attenuation of small-scale anisotropies depends not only on baryon loading and expansion history but also on the alteration of the causal structure induced by the K-essence field.  The high-$\ell$ damping tail of the CMB power spectrum provides a potential observable signature of the emergent geometry, enabling a distinction between K-essence cosmology and the standard FLRW model.\\

In our notation, $\sigma_T^{(\mathrm{geom})}$ is \emph{not} a new microscopic Thomson cross section. It is simply the standard Thomson cross section, defined in the local inertial frame of the emergent metric $\bar G_{\mu\nu}$, together with the purely geometrical conversion factors that arise when the local scattering probability is written as a collision term in the cosmological Boltzmann equation (i.e., as a rate per unit emergent conformal time and in terms of emergent momentum variables) \cite{Dodelson,Weinberg,deGroot,Ehlers}. Hence the underlying QED interaction is unchanged, but the scattering \emph{rate} that enters the Boltzmann hierarchy is rescaled by the emergent geometry \cite{Dodelson,Weinberg}. A full evaluation of the observable consequences (recombination and decoupling, higher-multipole evolution, and the CMB angular power spectrum) requires a numerical implementation of the modified Einstein-Boltzmann system in a dedicated Boltzmann solver; this lies beyond the scope of the present work and is deferred to future studies.
\\

Building on this, our formulation provides a self-consistent description of photon transport in an emergent K-essence FLRW spacetime within the usual non-relativistic Thomson-scattering and tight-coupling approximations. Even though the local Compton/Thomson process remains standard, the emergent geometry reshapes the effective time variable, transport coefficients, and gravitational source terms that drive the Boltzmann hierarchy. These changes are expected to feed into the post-recombination free-streaming regime and thereby imprint themselves on the acoustic-peak pattern and the Silk-damping tail of the CMB temperature anisotropies. Quantitative predictions will require incorporating the modified hierarchy (along with a consistent recombination history) into a numerical Einstein-Boltzmann solver and confronting the results with data.

Accordingly, we focus here on the pre-recombination epoch, where the tightly coupled photon-baryon fluid is captured by our modified hierarchy in the emergent K-essence geometry. Post-recombination ingredients, photon-free streaming, the full CMB angular power spectrum, and the late-time growth of cold-dark-matter perturbations are left for future work, since they necessitate a dedicated numerical implementation and a detailed comparison with observations. The main contribution of this paper is therefore to establish a kinetic-theory framework in emergent spacetime and to isolate the qualitative signatures generated by the K-essence scalar field, setting the stage for a systematic program of cosmological tests in subsequent work.

In summary, the collision term derived here serves as the entry point for constructing the complete photon Boltzmann hierarchy in emergent K-essence cosmology. While microphysics is unaltered, geometry modifies how scattering and propagation are encoded in cosmological variables; implementing this structure in a Boltzmann code will enable robust, quantitative constraints on the emergent-geometry scenario.\\

\vspace{0.2in}
{\bf Acknowledgment:} The authors are grateful to the referees for their constructive comments, which have helped us substantially improve the manuscript.
G.M. (IUCAA Associate) acknowledges the Inter-University Centre for Astronomy and Astrophysics (IUCAA), Pune, India, for facilitating part of this work during his visit. G.M. also thanks the COST Association (CA21136 CosmoVerse), European Union, for the opportunity to participate in this international association as a group member. G.M. extends his deepest gratitude to all undergraduate, postgraduate, and doctoral students, along with his teachers, collaborators, and well-wishers, whose support has significantly enhanced his academic journey. \\

{\bf Conflicts of interest:} The authors declare no conflicts of interest.\\

{\bf Funding information:} Not available.\\

{\bf Data availability:} No data were generated or analyzed in this study.\\

{\bf Declaration of competing interest:}
The authors declare that they have no known competing financial interests or personal relationships that could have appeared to influence the work reported in this paper.\\


\bibliography{references}

@article{Planck2018,
  author = {Planck Collaboration and Aghanim, N. and others},
  title = {Planck 2018 results. VI. Cosmological parameters},
  journal = {Astron. Astrophys.},
  volume = {641},
  pages = {A6},
  year = {2020},
  eprint = {1807.06209},
  archivePrefix = {arXiv},
  primaryClass = {astro-ph.CO},
  doi = {10.1051/0004-6361/201833910}
}

@article{Riess1,
    author = "Riess, Adam G. and others",
    title = "{A Comprehensive Measurement of the Local Value of the Hubble Constant with 1 km s$^{−1}$ Mpc$^{−1}$ Uncertainty from the Hubble Space Telescope and the SH0ES Team}",
    eprint = "2112.04510",
    archivePrefix = "arXiv",
    primaryClass = "astro-ph.CO",
    doi = "10.3847/2041-8213/ac5c5b",
    journal = "Astrophys. J. Lett.",
    volume = "934",
    number = "1",
    pages = "L7",
    year = "2022"
}

@article{Verde,
  author = "Verde, L. and Treu, T. and Riess, A. G.",
    title = "{Tensions between the Early and the Late Universe}",
    eprint = "1907.10625",
    archivePrefix = "arXiv",
    primaryClass = "astro-ph.CO",
    doi = "10.1038/s41550-019-0902-0",
    journal = "Nature Astron.",
    volume = "3",
    pages = "891",
    year = "2019"
}

@article{Abdalla,
  author = "Abdalla, Elcio and others",
    title = "{Cosmology intertwined: A review of the particle physics, astrophysics, and cosmology associated with the cosmological tensions and anomalies}",
    eprint = "2203.06142",
    archivePrefix = "arXiv",
    primaryClass = "astro-ph.CO",
    reportNumber = "FERMILAB-CONF-22-192-SCD",
    doi = "10.1016/j.jheap.2022.04.002",
    journal = "JHEAp",
    volume = "34",
    pages = "49--211",
    year = "2022"
}

@article{Valentino,
  author = "Di Valentino, Eleonora and others",
    title = "{Snowmass2021 - Letter of interest cosmology intertwined II: The hubble constant tension}",
    eprint = "2008.11284",
    archivePrefix = "arXiv",
    primaryClass = "astro-ph.CO",
    reportNumber = "FERMILAB-PUB-21-590-PPD",
    doi = "10.1016/j.astropartphys.2021.102605",
    journal = "Astropart. Phys.",
    volume = "131",
    pages = "102605",
    year = "2021"
}

@article{Riess2,
   author = "Riess, Adam G. and others",
    collaboration = "Supernova Search Team",
    title = "{Observational evidence from supernovae for an accelerating universe and a cosmological constant}",
    eprint = "astro-ph/9805201",
    archivePrefix = "arXiv",
    doi = "10.1086/300499",
    journal = "Astron. J.",
    volume = "116",
    pages = "1009--1038",
    year = "1998"
}

@article{Tegmark,
   author = "Tegmark, Max and others",
    collaboration = "SDSS",
    title = "{Cosmological parameters from SDSS and WMAP}",
    eprint = "astro-ph/0310723",
    archivePrefix = "arXiv",
    reportNumber = "FERMILAB-PUB-03-435-A",
    doi = "10.1103/PhysRevD.69.103501",
    journal = "Phys. Rev. D",
    volume = "69",
    pages = "103501",
    year = "2004"
}

@article{Spergel,
    author = "Spergel, D. N. and others",
    collaboration = "WMAP",
    title = "{Wilkinson Microwave Anisotropy Probe (WMAP) three year results: implications for cosmology}",
    eprint = "astro-ph/0603449",
    archivePrefix = "arXiv",
    doi = "10.1086/513700",
    journal = "Astrophys. J. Suppl.",
    volume = "170",
    pages = "377",
    year = "2007"
}

@article{Perlmutter,
  author = "Perlmutter, S. and others",
    collaboration = "Supernova Cosmology Project",
    title = "{Measurements of $\Omega$ and $\Lambda$ from 42 High Redshift Supernovae}",
    eprint = "astro-ph/9812133",
    archivePrefix = "arXiv",
    reportNumber = "LBNL-41801, LBL-41801",
    doi = "10.1086/307221",
    journal = "Astrophys. J.",
    volume = "517",
    pages = "565--586",
    year = "1999"
}

@article{Clifton,
 author = "Clifton, Timothy and Ferreira, Pedro G. and Padilla, Antonio and Skordis, Constantinos",
    title = "{Modified Gravity and Cosmology}",
    eprint = "1106.2476",
    archivePrefix = "arXiv",
    primaryClass = "astro-ph.CO",
    doi = "10.1016/j.physrep.2012.01.001",
    journal = "Phys. Rept.",
    volume = "513",
    pages = "1--189",
    year = "2012"
}

@article{Joyce,
 author = "Joyce, Austin and Jain, Bhuvnesh and Khoury, Justin and Trodden, Mark",
    title = "{Beyond the Cosmological Standard Model}",
    eprint = "1407.0059",
    archivePrefix = "arXiv",
    primaryClass = "astro-ph.CO",
    doi = "10.1016/j.physrep.2014.12.002",
    journal = "Phys. Rept.",
    volume = "568",
    pages = "1--98",
    year = "2015"
}

@article{Koyama,
   author = "Koyama, Kazuya",
    title = "{Cosmological Tests of Modified Gravity}",
    eprint = "1504.04623",
    archivePrefix = "arXiv",
    primaryClass = "astro-ph.CO",
    doi = "10.1088/0034-4885/79/4/046902",
    journal = "Rept. Prog. Phys.",
    volume = "79",
    number = "4",
    pages = "046902",
    year = "2016"
}

@misc{Valentino1,
  author = "Di Valentino, Eleonora and Levi Said, Jackson and Saridakis, Emmanuel N.",
    title = "{Cosmological tensions in the era of precision cosmology: Insights from Tensions in Cosmology 2025}",
    eprint = "2509.25288",
    archivePrefix = "arXiv",
    primaryClass = "astro-ph.CO",
    month = "9",
    year = "2025"
}

@book{Dodelson,
 author = {{Dodelson}, Scott and {Schmidt}, Fabian},
        title = "{Modern Cosmology}",
         year = 2020,
          doi = {10.1016/C2017-0-01943-2},
       adsurl = {https://ui.adsabs.harvard.edu/abs/2020moco.book.....D}
}

@article{Ma,
   author = "Ma, Chung-Pei and Bertschinger, Edmund",
    title = "{Cosmological perturbation theory in the synchronous and conformal Newtonian gauges}",
    eprint = "astro-ph/9506072",
    archivePrefix = "arXiv",
    doi = "10.1086/176550",
    journal = "Astrophys. J.",
    volume = "455",
    pages = "7--25",
    year = "1995"
}

@article{Garriga,
  author = "Garriga, Jaume and Mukhanov, Viatcheslav F.",
    title = "{Perturbations in k-inflation}",
    eprint = "hep-th/9904176",
    archivePrefix = "arXiv",
    reportNumber = "UAB-FT-466",
    doi = "10.1016/S0370-2693(99)00602-4",
    journal = "Phys. Lett. B",
    volume = "458",
    pages = "219--225",
    year = "1999"
}

@article{Picon1,
  author = "Armendariz-Picon, C. and Mukhanov, Viatcheslav F. and Steinhardt, Paul J.",
    title = "{A Dynamical solution to the problem of a small cosmological constant and late time cosmic acceleration}",
    eprint = "astro-ph/0004134",
    archivePrefix = "arXiv",
    doi = "10.1103/PhysRevLett.85.4438",
    journal = "Phys. Rev. Lett.",
    volume = "85",
    pages = "4438--4441",
    year = "2000"
}

@article{Picon2,
 author = "Armendariz-Picon, C. and Damour, T. and Mukhanov, Viatcheslav F.",
    title = "{k - inflation}",
    eprint = "hep-th/9904075",
    archivePrefix = "arXiv",
    doi = "10.1016/S0370-2693(99)00603-6",
    journal = "Phys. Lett. B",
    volume = "458",
    pages = "209--218",
    year = "1999"
}

@article{Picon3,
  author = "Armendariz-Picon, C. and Mukhanov, Viatcheslav F. and Steinhardt, Paul J.",
    title = "{Essentials of k essence}",
    eprint = "astro-ph/0006373",
    archivePrefix = "arXiv",
    doi = "10.1103/PhysRevD.63.103510",
    journal = "Phys. Rev. D",
    volume = "63",
    pages = "103510",
    year = "2001"
}

@article{Padmanabhan,
    author = "Padmanabhan, T. and Choudhury, T. Roy",
    title = "{Can the clustered dark matter and the smooth dark energy arise from the same scalar field?}",
    eprint = "hep-th/0205055",
    archivePrefix = "arXiv",
    reportNumber = "IUCAA-17-2002",
    doi = "10.1103/PhysRevD.66.081301",
    journal = "Phys. Rev. D",
    volume = "66",
    pages = "081301",
    year = "2002"
}

@article{Scherrer,
  author = "Scherrer, Robert J.",
    title = "{Purely kinetic k-essence as unified dark matter}",
    eprint = "astro-ph/0402316",
    archivePrefix = "arXiv",
    doi = "10.1103/PhysRevLett.93.011301",
    journal = "Phys. Rev. Lett.",
    volume = "93",
    pages = "011301",
    year = "2004"
}

@article{Chimento,
   author = "Chimento, Luis P.",
    title = "{Extended tachyon field, Chaplygin gas and solvable k-essence cosmologies}",
    eprint = "astro-ph/0311613",
    archivePrefix = "arXiv",
    doi = "10.1103/PhysRevD.69.123517",
    journal = "Phys. Rev. D",
    volume = "69",
    pages = "123517",
    year = "2004"
}

@article{Babichev,
doi = {10.1088/1126-6708/2008/02/101},
url = {https://doi.org/10.1088/1126-6708/2008/02/101},
year = {2008},
month = {feb},
publisher = {},
volume = {2008},
number = {02},
pages = {101},
author = {Eugeny Babichev and Viatcheslav Mukhanov and Alexander Vikman},
title = {k-Essence, superluminal propagation, causality and emergent geometry},
journal = {Journal of High Energy Physics},
}

@article{Deffayet,
   author = "Deffayet, C. and Gao, Xian and Steer, D. A. and Zahariade, G.",
    title = "{From k-essence to generalised Galileons}",
    eprint = "1103.3260",
    archivePrefix = "arXiv",
    primaryClass = "hep-th",
    doi = "10.1103/PhysRevD.84.064039",
    journal = "Phys. Rev. D",
    volume = "84",
    pages = "064039",
    year = "2011"
}

@article{Bekenstein_1993,
   author = "Bekenstein, Jacob D.",
    title = "{The Relation between physical and gravitational geometry}",
    eprint = "gr-qc/9211017",
    archivePrefix = "arXiv",
    reportNumber = "UCSB-TH-92-41, UCSBTH-92-41",
    doi = "10.1103/PhysRevD.48.3641",
    journal = "Phys. Rev. D",
    volume = "48",
    pages = "3641--3647",
    year = "1993"
}

@article{Sawicki,
   author = "Sawicki, Ignacy and Trenkler, Georg and Vikman, Alexander",
    title = "{Causality and stability from acoustic geometry}",
    eprint = "2412.21169",
    archivePrefix = "arXiv",
    primaryClass = "gr-qc",
    doi = "10.1007/JHEP10(2025)227",
    journal = "JHEP",
    volume = "10",
    pages = "227",
    year = "2025"
}

@article{Mukohyama,
 author = "Mukohyama, Shinji and Namba, Ryo and Watanabe, Yota",
    title = "{Is the DBI scalar field as fragile as other $k$-essence fields?}",
    eprint = "1605.06418",
    archivePrefix = "arXiv",
    primaryClass = "hep-th",
    reportNumber = "YITP-16-63, IPMU16-0073",
    doi = "10.1103/PhysRevD.94.023514",
    journal = "Phys. Rev. D",
    volume = "94",
    number = "2",
    pages = "023514",
    year = "2016"
}

@article{gm1,
   author = {{Gangopadhyay}, Debashis and {Manna}, Goutam},
        title = "{The Hawking temperature in the context of dark energy}",
      journal = {EPL (Europhysics Letters)},
         year = 2012,
        month = nov,
       volume = {100},
       number = {4},
        pages = {49001},
          doi = {10.1209/0295-5075/100/49001},
archivePrefix = {arXiv},
       eprint = {1211.1268},
 primaryClass = {physics.gen-ph},
       adsurl = {https://ui.adsabs.harvard.edu/abs/2012EL....10049001G}
}

@article{gm2,
   author = {{Manna}, Goutam and {Gangopadhyay}, Debashis},
        title = "{The Hawking temperature in the context of dark energy for Reissner{\textendash}Nordstrom and Kerr background}",
      journal = {European Physical Journal C},
         year = 2014,
        month = mar,
       volume = {74},
       number = {3},
          eid = {2811},
        pages = {2811},
          doi = {10.1140/epjc/s10052-014-2811-9},
archivePrefix = {arXiv},
       eprint = {1303.2999},
 primaryClass = {physics.gen-ph},
       adsurl = {https://ui.adsabs.harvard.edu/abs/2014EPJC...74.2811M}
}

@article{gm3,
  author = "Manna, Goutam and Majumder, Bivash",
    title = "{The Hawking temperature in the context of dark energy for Kerr{\textendash}Newman and Kerr{\textendash}Newman{\textendash}AdS backgrounds}",
    eprint = "1907.00728",
    archivePrefix = "arXiv",
    primaryClass = "gr-qc",
    doi = "10.1140/epjc/s10052-019-7066-z",
    journal = "Eur. Phys. J. C",
    volume = "79",
    number = "7",
    pages = "553",
    year = "2019"
}

@article{Panda1,
   author = "Panda, Arijit and Ray, Saibal and Manna, Goutam and Das, Surajit and Ranjit, Chayan",
    title = "{Cosmological effects on f(R̄,T̄) gravity through a non-standard theory}",
    eprint = "2206.14808",
    archivePrefix = "arXiv",
    primaryClass = "gr-qc",
    doi = "10.1142/S0218271824500159",
    journal = "Int. J. Mod. Phys. D",
    volume = "33",
    number = "03n04",
    pages = "2450015",
    year = "2024"
}

@article{Panda2,
  author = "Panda, Arijit and Manna, Goutam and Ray, Saibal and Khlopov, Maxim and Islam, Md. Rabiul",
    title = "{Collapsing scenarios of K-essence generalized Vaidya spacetime under f(R̄,T̄) gravity}",
    eprint = "2308.13574",
    archivePrefix = "arXiv",
    primaryClass = "gr-qc",
    doi = "10.1016/j.cjph.2024.08.035",
    journal = "Chin. J. Phys.",
    volume = "91",
    pages = "838--856",
    year = "2024"
}

@article{Ganguly,
  author = "Ganguly, Samit and Manna, Goutam and Gangopadhyay, Debashis and Guendelman, Eduardo and Bhattacharyya, Abhijit",
    title = "{Non-affine extensions of the Raychaudhuri equation in the K-essence framework}",
    eprint = "2503.03076",
    archivePrefix = "arXiv",
    primaryClass = "gr-qc",
    doi = "10.1016/j.physletb.2025.139622",
    journal = "Phys. Lett. B",
    volume = "868",
    pages = "139622",
    year = "2025"
}

@article{gm4,
  author = "Manna, Goutam",
    title = "{Gravitational Collapse for the K-essence Emergent Vaidya Spacetime}",
    eprint = "1911.11753",
    archivePrefix = "arXiv",
    primaryClass = "gr-qc",
    doi = "10.1140/epjc/s10052-020-8383-y",
    journal = "Eur. Phys. J. C",
    volume = "80",
    number = "9",
    pages = "813",
    year = "2020"
}

@article{gm5,
 author = "Manna, Goutam and Majumdar, Parthasarathi and Majumder, Bivash",
    title = "{k -essence emergent spacetime as a generalized Vaidya geometry}",
    eprint = "1909.07224",
    archivePrefix = "arXiv",
    primaryClass = "gr-qc",
    doi = "10.1103/PhysRevD.101.124034",
    journal = "Phys. Rev. D",
    volume = "101",
    number = "12",
    pages = "124034",
    year = "2020"
}

@article{Panda3,
  author = "Panda, Arijit and Gangopadhyay, Debashis and Manna, Goutam",
    title = "{Form Invariance of Raychaudhuri Equation~in the Presence of Inflaton-Type Fields}",
    eprint = "2404.11632",
    archivePrefix = "arXiv",
    primaryClass = "gr-qc",
    doi = "10.1002/prop.202400134",
    journal = "Fortsch. Phys.",
    volume = "72",
    number = "9-10",
    pages = "2400134",
    year = "2024"
}

@article{Bettoni,
  author = "Bettoni, Dario and Liberati, Stefano",
    title = "{Disformal invariance of second order scalar-tensor theories: Framing the Horndeski action}",
    eprint = "1306.6724",
    archivePrefix = "arXiv",
    primaryClass = "gr-qc",
    doi = "10.1103/PhysRevD.88.084020",
    journal = "Phys. Rev. D",
    volume = "88",
    pages = "084020",
    year = "2013"
}

@article{Sachs,
 author = "Sachs, R. K. and Wolfe, A. M.",
    title = "{Perturbations of a cosmological model and angular variations of the microwave background}",
    doi = "10.1007/s10714-007-0448-9",
    journal = "Astrophys. J.",
    volume = "147",
    pages = "73--90",
    year = "1967"
}

@article{Fosalba,
 author = "Fosalba, Pablo and Gaztanaga, Enrique and Castander, Francisco",
    title = "{Detection of the ISW and SZ effects from the CMB-galaxy correlation}",
    eprint = "astro-ph/0307249",
    archivePrefix = "arXiv",
    doi = "10.1086/379848",
    journal = "Astrophys. J. Lett.",
    volume = "597",
    pages = "L89--92",
    year = "2003"
}

@article{Lima,
    author = "Lima, J. A. S. and Del Popolo, A. and Plastino, A. R.",
    title = "{Thermodynamic Equilibrium in General Relativity}",
    eprint = "1911.09060",
    archivePrefix = "arXiv",
    primaryClass = "gr-qc",
    doi = "10.1103/PhysRevD.100.104042",
    journal = "Phys. Rev. D",
    volume = "100",
    number = "10",
    pages = "104042",
    year = "2019"
}

@article{Tolman1,
    title = {On the Weight of Heat and Thermal Equilibrium in General Relativity},
  author = {Tolman, Richard C.},
  journal = {Phys. Rev.},
  volume = {35},
  issue = {8},
  pages = {904--924},
  numpages = {0},
  year = {1930},
  month = {Apr},
  publisher = {American Physical Society},
  doi = {10.1103/PhysRev.35.904},
  url = {https://link.aps.org/doi/10.1103/PhysRev.35.904}
}

@article{Domenech2015,
  author = {Dom\`enech, Guillem and Naruko, Atsushi and Sasaki, Misao},
  title = {Cosmological disformal invariance},
  journal = {Journal of Cosmology and Astroparticle Physics},
  volume = {2015},
  number = {10},
  pages = {067},
  year = {2015},
  doi = {10.1088/1475-7516/2015/10/067},
  eprint = {1505.00174},
  archivePrefix = {arXiv},
  primaryClass = {gr-qc}
}

@article{Tolman2,
   title = {Temperature Equilibrium in a Static Gravitational Field},
  author = {Tolman, Richard C. and Ehrenfest, Paul},
  journal = {Phys. Rev.},
  volume = {36},
  issue = {12},
  pages = {1791--1798},
  numpages = {0},
  year = {1930},
  month = {Dec},
  publisher = {American Physical Society},
  doi = {10.1103/PhysRev.36.1791},
  url = {https://link.aps.org/doi/10.1103/PhysRev.36.1791}
}

@article{Zumalacarregui,
 author = "Zumalac{\'a}rregui, Miguel and Garc{\'\i}a-Bellido, Juan",
    title = "{Transforming gravity: from derivative couplings to matter to second-order scalar-tensor theories beyond the Horndeski Lagrangian}",
    eprint = "1308.4685",
    archivePrefix = "arXiv",
    primaryClass = "gr-qc",
    reportNumber = "IFT-UAM-CSIC-13-090",
    doi = "10.1103/PhysRevD.89.064046",
    journal = "Phys. Rev. D",
    volume = "89",
    pages = "064046",
    year = "2014"
}

@book{Weinberg,
  author    = {Weinberg, Steven},
  title     = {Cosmology},
  publisher = {Oxford University Press},
  address   = {Oxford, UK},
  year      = {2008},
  isbn      = {978-0198526827}
}

@book{Kolb,
   author = "Kolb, Edward W. and Turner, Michael S.",
    title = "{The Early Universe}",
    reportNumber = "FERMILAB-BOOK-1990-01",
    doi = "10.1201/9780429492860",
    isbn = "978-0-429-49286-0, 978-0-201-62674-2",
    publisher = "Taylor and Francis",
    volume = "69",
    month = "5",
    year = "2019"
}

@misc{Baumann,
   author = "Baumann, Daniel",
    title = "{Primordial Cosmology}",
    eprint = "1807.03098",
    archivePrefix = "arXiv",
    primaryClass = "hep-th",
    doi = "10.22323/1.305.0009",
    journal = "PoS",
    volume = "TASI2017",
    pages = "009",
    year = "2018"
}

@book{deGroot,
  author = {de Groot, S. R. and van Leeuwen, W. A. and van Weert, C. G.},
  title = {Relativistic Kinetic Theory},
  publisher = {North-Holland},
  year = {1980}
}

@book{MukhanovBook,
  author    = {Mukhanov, Viatcheslav},
  title     = {Physical Foundations of Cosmology},
  publisher = {Cambridge University Press},
  address   = {Cambridge, UK},
  year      = {2005},
  doi       = {10.1017/CBO9780511790553},
  isbn      = {978-0521563987}
}

@article{MaBertschinger,
  author  = {Ma, Chung-Pei and Bertschinger, Edmund},
  title   = {Cosmological perturbation theory in the synchronous and conformal Newtonian gauges},
  journal = {Astrophys. J.},
  volume  = {455},
  pages   = {7--25},
  year    = {1995},
  eprint  = {astro-ph/9506072},
  archivePrefix = {arXiv},
  doi     = {10.1086/176550}
}

@book{MTW1973,
  author = {Misner, Charles W. and Thorne, Kip S. and Wheeler, John A.},
  title = {Gravitation},
  publisher = {W. H. Freeman},
  year = {1973}
}

@incollection{Ehlers,
  author = {Ehlers, J.},
  title = {General Relativity and Kinetic Theory},
  booktitle = {General Relativity and Cosmology},
  editor = {Sachs, R. K.},
  publisher = {Academic Press},
  year = {1971}
}

@book{Peacock,
  author = {Peacock, J. A.},
  title = {Cosmological Physics},
  publisher = {Cambridge University Press},
  year = {2012},
  doi = {10.1017/CBO9780511804533}
}

@article{Brax,
  author = "Brax, Philippe and Burrage, Clare and Davis, Anne-Christine",
    title = "{Shining Light on Modifications of Gravity}",
    eprint = "1206.1809",
    archivePrefix = "arXiv",
    primaryClass = "hep-th",
    doi = "10.1088/1475-7516/2012/10/016",
    journal = "JCAP",
    volume = "10",
    pages = "016",
    year = "2012"
}

@article{Tsujikawa,
    author = "Tsujikawa, Shinji",
    title = "{Cosmological disformal transformations to the Einstein frame and gravitational couplings with matter perturbations}",
    eprint = "1506.08561",
    archivePrefix = "arXiv",
    primaryClass = "gr-qc",
    doi = "10.1103/PhysRevD.92.064047",
    journal = "Phys. Rev. D",
    volume = "92",
    number = "6",
    pages = "064047",
    year = "2015"
}

@article{Zloshchastiev,
 author = {{Zloshchastiev}, Konstantin G.},
        title = "{Derivation of Emergent Spacetime Metric, Gravitational Potential and Speed of Light in Superfluid Vacuum Theory}",
      journal = {Universe},
         year = 2023,
        month = may,
       volume = {9},
       number = {5},
          eid = {234},
        pages = {234},
          doi = {10.3390/universe9050234},
       adsurl = {https://ui.adsabs.harvard.edu/abs/2023Univ....9..234Z}
}

@article{Duque,
   author = "Duque, Erick I.",
    title = "{Emergent modified gravity: The perfect fluid and gravitational collapse}",
    eprint = "2311.08616",
    archivePrefix = "arXiv",
    primaryClass = "gr-qc",
    doi = "10.1103/PhysRevD.109.044014",
    journal = "Phys. Rev. D",
    volume = "109",
    number = "4",
    pages = "044014",
    year = "2024"
}

@article{Hell,
  author = "Hell, Anamaria",
    title = "{Unveiling the Inconsistency of the Proca Theory with Nonminimal Coupling to Gravity}",
    eprint = "2403.18673",
    archivePrefix = "arXiv",
    primaryClass = "gr-qc",
    doi = "10.1093/ptep/ptae188",
    journal = "PTEP",
    volume = "2025",
    number = "1",
    pages = "013E01",
    year = "2025"
}

@article{Cheung,
 author = "Cheung, Clifford and Creminelli, Paolo and Fitzpatrick, A. Liam and Kaplan, Jared and Senatore, Leonardo",
    title = "{The Effective Field Theory of Inflation}",
    eprint = "0709.0293",
    archivePrefix = "arXiv",
    primaryClass = "hep-th",
    reportNumber = "IC-2007-032",
    doi = "10.1088/1126-6708/2008/03/014",
    journal = "JHEP",
    volume = "03",
    pages = "014",
    year = "2008"
}

@article{Gleyzes,
   author = "Gleyzes, J{\'e}r{\^o}me and Langlois, David and Piazza, Federico and Vernizzi, Filippo",
    title = "{Healthy theories beyond Horndeski}",
    eprint = "1404.6495",
    archivePrefix = "arXiv",
    primaryClass = "hep-th",
    doi = "10.1103/PhysRevLett.114.211101",
    journal = "Phys. Rev. Lett.",
    volume = "114",
    number = "21",
    pages = "211101",
    year = "2015"
}

@book{Srednicki,
  author = {Srednicki, M.},
  title = {Quantum Field Theory},
  publisher = {Cambridge University Press},
  year = {2007}
}

@article{Seljak,
   author = "Seljak, Uros",
    title = "{Gravitational lensing effect on cosmic microwave background anisotropies: A Power spectrum approach}",
    eprint = "astro-ph/9505109",
    archivePrefix = "arXiv",
    reportNumber = "MIT-CSR-94-29",
    doi = "10.1086/177218",
    journal = "Astrophys. J.",
    volume = "463",
    pages = "1",
    year = "1996"
}

@article{Hu,
  author = "Hu, Wayne and Sugiyama, Naoshi",
    title = "{Small scale cosmological perturbations: An Analytic approach}",
    eprint = "astro-ph/9510117",
    archivePrefix = "arXiv",
    reportNumber = "IASSNS-AST-95-42, CFPA-TH-95-18, UTAP-212",
    doi = "10.1086/177989",
    journal = "Astrophys. J.",
    volume = "471",
    pages = "542--570",
    year = "1996"
}

@article{Born,
   author = "Born, M. and Infeld, L.",
    title = "{Foundations of the new field theory}",
    doi = "10.1098/rspa.1934.0059",
    journal = "Proc. Roy. Soc. Lond. A",
    volume = "144",
    number = "852",
    pages = "425--451",
    year = "1934"
}

@article{Dirac,
  author = "Dirac, Paul A. M.",
    title = "{An Extensible model of the electron}",
    doi = "10.1098/rspa.1962.0124",
    journal = "Proc. Roy. Soc. Lond. A",
    volume = "268",
    pages = "57--67",
    year = "1962"
}

@article{Putter,
   author = "de Putter, Roland and Linder, Eric V.",
    title = "{Kinetic k-essence and Quintessence}",
    eprint = "0705.0400",
    archivePrefix = "arXiv",
    primaryClass = "astro-ph",
    doi = "10.1016/j.astropartphys.2007.05.011",
    journal = "Astropart. Phys.",
    volume = "28",
    pages = "263--272",
    year = "2007"
}

@misc{Ganguly1,
   author = "Ganguly, Samit and Panda, Arijit and Guendelman, Eduardo and Gangopadhyay, Debashis and Bhattacharyya, Abhijit and Manna, Goutam",
    title = "{Observational Insights on DBI K-essence Models Using Machine Learning and Bayesian Analysis}",
    eprint = "2506.05674",
    archivePrefix = "arXiv",
    primaryClass = "astro-ph.CO",
    month = "6",
    year = "2025"
}

@article{Visser,
    author = "Visser, Matt and Barcelo, Carlos and Liberati, Stefano",
    title = "{Analog models of and for gravity}",
    eprint = "gr-qc/0111111",
    archivePrefix = "arXiv",
    doi = "10.1023/A:1020180409214",
    journal = "Gen. Rel. Grav.",
    volume = "34",
    pages = "1719--1734",
    year = "2002"
}

@phdthesis{Vikman,
  author = {Vikman, Alexander},
  title = {K-essence: Cosmology, Causality and Emergent Geometry},
  school = {Ludwig-Maximilians-Universität München},
  year = {2007},
  url = {https://edoc.ub.uni-muenchen.de/7761/1/Vikman_Alexander.pdf}
}

@article{Ma1,
   author = "Ma, Chung-Pei and Bertschinger, Edmund",
    title = "{Cosmological perturbation theory in the synchronous and conformal Newtonian gauges}",
    eprint = "astro-ph/9506072",
    archivePrefix = "arXiv",
    doi = "10.1086/176550",
    journal = "Astrophys. J.",
    volume = "455",
    pages = "7--25",
    year = "1995"
}

@article{Miller,
author = {Kenneth S. Miller},
title = {On the Inverse of the Sum of Matrices},
journal = {Mathematics Magazine},
volume = {54},
number = {2},
pages = {67--72},
year = {1981},
publisher = {Taylor \& Francis},
doi = {10.1080/0025570X.1981.11976898},


URL = { 
    
        https://doi.org/10.1080/0025570X.1981.11976898
    
    

},
eprint = { 
    
        https://doi.org/10.1080/0025570X.1981.11976898
    
    

}

}

@book{wald,
  title     = {General Relativity},
  author    = {Wald, Robert M.},
  year      = {1984},
  publisher = {University of Chicago Press},
  address   = {Chicago, IL},
  isbn      = {978-0226870335}
}

@article{Rendall,
doi = {10.1088/0264-9381/23/5/008},
url = {https://doi.org/10.1088/0264-9381/23/5/008},
year = {2006},
month = {feb},
publisher = {},
volume = {23},
number = {5},
pages = {1557},
author = {Rendall, Alan D},
title = {Dynamics of k-essence},
journal = {Classical and Quantum Gravity},
}

@article{Winitzki,
  author  = {Mukhanov, Viatcheslav F. and Winitzki, Sergei},
  title   = {Introduction to Quantum Effects in Gravity},
  journal = {Contemporary Physics},
  volume  = {51},
  number  = {5},
  pages   = {453},
  year    = {2010},
  doi     = {10.1080/00107510903360410},
  url     = {https://doi.org/10.1080/00107510903360410}
}

@book{Chandrasekhar,
  author    = {Chandrasekhar, Subrahmanyan},
  title     = {Radiative Transfer},
  publisher = {Dover Publications},
  address   = {New York},
  year      = {1960}
}

@article{HuWhite,
  author       = {Hu, Wayne and White, Martin},
  title        = {A CMB polarization primer},
  eprint       = {astro-ph/9706147},
  archivePrefix= {arXiv},
  doi          = {10.1086/305482},
  journal      = {New Astronomy},
  volume       = {2},
  pages        = {323--344},
  year         = {1997}
}

@article{ZaldarriagaSeljak,
  author       = {Zaldarriaga, Matias and Seljak, Uros},
  title        = {All-sky analysis of polarization in the microwave background},
  eprint       = {astro-ph/9609170},
  archivePrefix= {arXiv},
  doi          = {10.1103/PhysRevD.55.1830},
  journal      = {Phys. Rev. D},
  volume       = {55},
  pages        = {1830--1840},
  year         = {1997}
}

\bibliographystyle{apsrev4-1}

\vspace{0.5in}

\appendix

\section{Effective rest-frame energy and effective particle
description in K-essence geometry}
\label{A}

We begin by distinguishing the invariant microscopic particle properties from their effective description in the K-essence-induced emergent geometry. Let $M$ denote the intrinsic (microscopic) rest-mass parameter of the probe particle; for an electron, $M=m_e$. This invariant mass scale is fixed by local microphysics and is not renormalized by the presence of the K-essence background.

A convenient way to represent a massive probe degree of freedom propagating on the emergent metric $\bar G_{\mu\nu}$ is through the minimally coupled action \cite{Picon2, Zumalacarregui, Garriga, Vikman, Babichev},
\begin{equation}
S_{\chi}^{\rm micro}
=
-\frac{1}{2}
\int d^4x\,\sqrt{-\bar{G}}\,
\left[
\bar{G}^{\mu\nu}\partial_\mu\chi\,\partial_\nu\chi
+
M^2\chi^2
\right].
\label{A1}
\end{equation}
In this description, the mass term appears with the same invariant parameter $M$ as in standard particle physics. The role of the K-essence background is therefore not to modify the intrinsic mass itself, but to alter the relation between energy--momentum components and $M$ through the geometry (in particular, through the lapse) encoded in $\bar G_{\mu\nu}$.

For the homogeneous K-essence background adopted in this work, the associated emergent line element takes the FLRW form
\begin{equation}
d\bar{s}^{2}
=
-\left(1-A\dot{\phi}^{2}\right)dt^{2}
+
a^{2}(t)\,d{\bf x}^{2},
\label{A2}
\end{equation}
so that the non-vanishing components of the inverse metric are
\begin{equation}
\bar{G}^{00}
=
-\frac{1}{1-A\dot{\phi}^{2}},
\qquad
\bar{G}^{ij}
=
\frac{1}{a^{2}}\delta^{ij}.
\label{A3}
\end{equation}
Requiring
\begin{equation}
A\dot{\phi}^{2}<1
\label{A4}
\end{equation}
guarantees a negative time-time component and hence preserves a Lorentzian signature for the emergent metric.

The modified time component of the emergent metric can be encoded in a time-dependent lapse,
\begin{equation}
N(t)
\equiv
\sqrt{1-A\dot{\phi}^{2}}.
\label{A5}
\end{equation}
For a particle at rest in comoving coordinates, this lapse rescales the rest-frame \emph{coordinate-time} energy associated with the emergent geometry, giving~\cite{Picon2,Zumalacarregui,Garriga,Vikman,Babichev,Zloshchastiev,Duque}
\begin{equation}
E_{\rm rest}
=
MN
=
M\sqrt{1-A\dot{\phi}^{2}}.
\label{A6}
\end{equation}
It is then convenient to introduce the corresponding geometry-induced rest-energy scale
\begin{equation}
\boxed{
M_{\rm eff}
\equiv
M\sqrt{1-A\dot{\phi}^{2}}
}\,,
\label{A7}
\end{equation}
which is nothing but the lapse-rescaled rest-frame coordinate energy. Importantly, $M_{\rm eff}$ does \emph{not} represent a change in the underlying invariant microscopic mass $M$ appearing in the matter action; it simply parameterizes the rest-frame coordinate energy when the kinematics are expressed in the emergent K-essence variables.\\

To display explicitly how the lapse-induced rest-frame \emph{coordinate} energy scale arises, we start from the dispersion relation that follows from the microscopic action \eqref{A1} for the homogeneous emergent metric \eqref{A2}. For a Fourier mode
$\chi \propto e^{-i\omega t+i\mathbf{k}\cdot\mathbf{x}}$, one obtains \cite{Picon2,Zumalacarregui,Garriga,Vikman,Babichev}
\begin{equation}
-\frac{\omega^{2}}{1-A\dot{\phi}^{2}}+\frac{k^{2}}{a^{2}}+M^{2}=0.
\label{A8}
\end{equation}
Introducing the lapse $N(t)\equiv\sqrt{1-A\dot{\phi}^{2}}$, this may be rewritten as
\begin{equation}
\omega^{2}=N^{2}\left(\frac{k^{2}}{a^{2}}+M^{2}\right).
\label{A9}
\end{equation}
In particular, in the rest frame ($k=0$) one finds $\omega_{k=0}=MN$. This motivates the definition of the lapse-rescaled rest-frame \emph{coordinate-time} energy scale
\begin{equation}
M_{\rm eff}\equiv MN=M\sqrt{1-A\dot{\phi}^{2}},
\label{A10}
\end{equation}
which is entirely geometric in origin. Importantly, $M_{\rm eff}$ is not a new invariant mass parameter: the invariant microscopic mass entering the fundamental matter action remains $M$.

\subsection{Effective emergent-frame description.}
Once the particle kinematics are formulated purely in terms of the emergent (K-essence) variables, it is natural to absorb (i.e., fold) the geometry-induced rest-frame \emph{coordinate} energy scale defined in Eq.~\eqref{A7} into the corresponding effective description. In the emergent-frame effective description, a particle's four-momentum is constrained by the corresponding mass-shell relation \cite{Babichev, Vikman, Winitzki}:
\begin{equation}
\bar{G}_{\mu\nu}P^{\mu}P^{\nu}
=
-M_{\rm eff}^{2},
\label{A12}
\end{equation}
where $P^\mu = dx^\mu/d\lambda$ is the tangent four-momentum along the worldline and $\lambda$ is an affine parameter. Specializing to the homogeneous emergent metric in Eq.~(\ref{A2}), this condition reduces to
\begin{equation}
-N^{2}(P^{0})^{2}
+
a^{2}\delta_{ij}P^{i}P^{j}
=
-M_{\rm eff}^{2}.
\label{A13}
\end{equation}

To define the particle energy and three-momentum in a way that is natural in the emergent frame, we introduce the orthonormal tetrad associated with the metric (\ref{A2}),
\begin{equation}
e_{\hat{0}}
=
\frac{1}{N}\,\partial_t,
\qquad
e_{\hat{i}}
=
\frac{1}{a}\,\partial_i,
\label{A14}
\end{equation}
together with the dual coframe
\begin{equation}
\vartheta^{\hat{0}}=N\,dt,
\qquad
\vartheta^{\hat{i}}=a\,dx^{i}.
\label{A15}
\end{equation}
Projecting the four-momentum onto this local orthonormal basis gives
\begin{equation}
P^{\hat{0}}
=
NP^{0}
\equiv
\widetilde{E},
\qquad
P^{\hat{i}}
=
aP^{i}
\equiv
\bar{p}^{\,i}.
\label{A16}
\end{equation}
Equivalently, in terms of the covariant spatial components
$P_i=\bar{G}_{ij}P^j=a^2\delta_{ij}P^j$, one finds
\begin{equation}
\bar{p}^{\,i}
=
\frac{P_i}{a}.
\label{A17}
\end{equation}
In this way, $\widetilde E$ and $\bar{\mathbf p}$ are simply the time and space components of a single four-momentum expressed in the emergent orthonormal frame, rather than independently rescaled variables.

With these definitions, the mass-shell constraint (\ref{A12}) reduces to the familiar Minkowski-space relation in the local orthonormal frame,
\begin{equation}
-\widetilde{E}^{\,2}
+
\delta_{\hat{i}\hat{j}}
\bar{p}^{\,\hat{i}}
\bar{p}^{\,\hat{j}}
=
-M_{\rm eff}^{2},
\label{A18}
\end{equation}
or, equivalently,
\begin{equation}
\widetilde{E}^{\,2}
=
\bar{p}^{\,2}
+
M_{\rm eff}^{2}.
\label{A19}
\end{equation}
Therefore, at any fixed cosmic time, the energy and three-momentum satisfy the usual special-relativistic dispersion relation in the emergent local inertial frame, with the time-dependent scale $M_{\rm eff}(t)$ playing the role of an effective mass (i.e., effective energy scale or rest-frame coordinate-time energy scale).

The locally measured energy is related to the coordinate-time component of the four-momentum via
\begin{equation}
P^{0}
=
\frac{\widetilde E}{N}
=
\frac{\widetilde E}{\sqrt{1-A\dot{\phi}^{\,2}}}.
\label{A20}
\end{equation}
The corresponding physical three-momentum that satisfies
\begin{equation}
\bar p^{\,2}
=
\delta_{\hat{i}\hat{j}}P^{\hat{i}}P^{\hat{j}}
=
a^{2}\delta_{ij}P^{i}P^{j}
=
\frac{\delta^{ij}P_iP_j}{a^{2}}.
\label{A21}
\end{equation}
For massless particles, $M_{\rm eff}=0$, and Eq.~(\ref{A19}) simply becomes
\begin{equation}
\widetilde E=\bar p.
\label{A22}
\end{equation}\\

The same emergent-frame dynamics can be recast into an effective quadratic action for a scalar mode propagating on the K-essence-induced emergent geometry \cite{Picon2, Zumalacarregui, Garriga, Vikman, Babichev, Winitzki}:
\begin{equation}
S_{\chi}^{\rm eff}
=
-\frac{1}{2}
\int d^{4}x\,\sqrt{-\bar G}
\left[
\bar G^{\mu\nu}
\partial_{\mu}\chi_{\rm eff}
\partial_{\nu}\chi_{\rm eff}
+
M_{\rm eff}^{2}\chi_{\rm eff}^{2}
\right].
\label{A23}
\end{equation}
Varying Eq.~(\ref{A23}) yields the Klein--Gordon equation in the emergent metric \cite{Wald},
$\left(D_\mu D^\mu - M_{\rm eff}^2\right)\chi_{\rm eff}=0$.
In the geometric-optics (WKB) limit, $\chi_{\rm eff}\sim e^{iS}$ with $P_\mu\equiv\partial_\mu S$, this equation implies the Hamilton--Jacobi (mass-shell) constraint \cite{Wald, Winitzki}
$\bar G^{\mu\nu}P_\mu P_\nu=-M_{\rm eff}^2$, which is equivalent to Eq.~(\ref{A12}) and therefore consistent with the dispersion relation in Eqs.~(\ref{A18})--(\ref{A19}).

Here $M_{\rm eff}(t)$ should be understood as the emergent-frame effective mass scale (effective energy scale) that appears when the same microscopic dynamics is rewritten in terms of emergent variables; it does not represent a change of the underlying microscopic mass parameter $M$. This viewpoint follows the emergent-geometry description of effective propagation developed in Refs.~\cite{Babichev,Vikman,Winitzki}.\\

Accordingly, the distribution function is taken to be a function of the local physical momentum measured in the emergent orthonormal frame (cf.~the definition of $\bar{\mathbf p}$ in Eq.~(\ref{A16})),
\begin{equation}
f=f(t,\mathbf{x},\bar{\mathbf p}),
\label{A24}
\end{equation}
and, for an isotropic homogeneous background,
\begin{equation}
f=f(t,\bar p).
\label{A25}
\end{equation}
The energy entering the kinetic equations is then the corresponding emergent-frame energy (the temporal component of the same four-momentum, Eq.~(\ref{A16})),
\begin{equation}
\widetilde E(t,\bar p)
=
\sqrt{\bar p^{\,2}+M_{\rm eff}^{2}(t)},
\label{A26}
\end{equation}
which is just the local dispersion relation in Eq.~(\ref{A19}) (and reduces to $\widetilde E=\bar p$ for $M_{\rm eff}=0$, cf.~Eq.~(\ref{A22})).
Thus the momentum variable used in the Boltzmann equation is the physical three-momentum $\bar{\mathbf p}$ associated with the same emergent-frame four-momentum whose temporal component is $\widetilde E$, ensuring that the kinetic variables satisfy the relativistic on-shell condition at each fixed cosmic time.\\

Tracking the difference between $M$ and $M_{\rm eff}$ is crucial. The parameter $M$ is the invariant microscopic mass appearing in the fundamental matter action and therefore remains fixed, whereas $M_{\rm eff}$ is a lapse-rescaled rest-frame \emph{coordinate} energy scale that is useful when rewriting the particle kinematics in the tilted emergent (K-essence) variables. In the limit
\begin{equation}
A\dot{\phi}^{2}\rightarrow 0,
\label{A27}
\end{equation}
the emergent geometry reduces to the standard FLRW form, so that
\begin{equation}
M_{\rm eff}\rightarrow M,
\qquad
\widetilde{E}^{\,2}\rightarrow E^{\,2}=p^{\,2}+M^{2}.
\label{A28}
\end{equation}

For electrons, this means that the local microphysical QED parameters continue to be set by the standard electron mass $m_e$, while the phase-space kinematics written in the emergent K-essence variables are conveniently characterized by the corresponding rest-frame coordinate-energy scale $M_{(e),\rm eff}~(\equiv m_{e}N)$. This is the origin of the effective electron energy used in the kinetic description, and it does not signal any change in the underlying electron mass or in the usual local QED interactions.\\

\noindent
\textit{Photon limit:}
For a photon minimally coupled to the emergent metric, the effective action remains gauge invariant and the on-shell (mass-shell) condition is still
\begin{equation}
\bar{G}^{\mu\nu}k_\mu k_\nu=0.
\label{A29}
\end{equation}
In a homogeneous background this becomes
\begin{equation}
-\frac{\omega^2}{1-A\dot{\phi}^{2}}+\frac{k^2}{a^2}=0,
\label{A30}
\end{equation}
so that the dispersion relation can be written as
\begin{equation}
\omega=\sqrt{1-A\dot{\phi}^{2}}\,\frac{k}{a}.
\label{A31}
\end{equation}
The corresponding photon propagation speed with respect to the coordinate time $t$ is therefore
\begin{equation}
v_{\rm ph}=\frac{\omega}{k/a}=\sqrt{1-A\dot{\phi}^{2}}.
\label{A32}
\end{equation}
Since $\omega$ remains linear in $k$, the photon stays \emph{massless with respect to the emergent metric} (cf. Eq.~(\ref{13})). The factor $\sqrt{1-A\dot{\phi}^2}$ simply rescales the effective speed, $v_{\text{ph}}=\sqrt{1-A\dot{\phi}^2}$, without opening a mass gap or spoiling gauge invariance. Equivalently, photons still follow null geodesics of the emergent K-essence metric, but the associated null cone is slightly narrower than the usual GR one when the homogeneous scalar field evolves in time ($\dot{\phi}\neq 0$) \cite{Brax, Tsujikawa}.\\

From the perspective of the original gravitational metric $g_{\mu\nu}$, a photon that is null with respect to the emergent metric $\bar{G}_{\mu\nu}$ does not necessarily remain null with respect to $g_{\mu\nu}$. Using
\begin{equation}
\bar{G}_{\mu\nu}
=
g_{\mu\nu}
-
A\nabla_\mu\phi\nabla_\nu\phi,
\label{A33}
\end{equation}
the null condition in the emergent geometry implies
\begin{equation}
\bar{G}_{\mu\nu}P^\mu P^\nu=0
\quad\Longrightarrow\quad
g_{\mu\nu}P^\mu P^\nu
=
A\nabla_\mu\phi\nabla_\nu\phi
P^\mu P^\nu.
\label{A34}
\end{equation}
For a homogeneous scalar field $\phi=\phi(t)$, this reduces to
\begin{equation}
g_{\mu\nu}P^\mu P^\nu
=
A\dot{\phi}^{2}(P^0)^2.
\label{A35}
\end{equation}
Therefore, when the photon is described using the original gravitational metric $g_{\mu\nu}$, a mass-like kinematic contribution can appear. This is not a genuine photon mass and should not be interpreted as a Proca mass; it simply reflects the mismatch between the null cones of $g_{\mu\nu}$ and $\bar{G}_{\mu\nu}$ \cite{Bekenstein_1993, Zumalacarregui, Bettoni, Hell}.

This construction makes the separation between the underlying microscopic particle physics and its effective description in the tilted K-essence geometry explicit: the microscopic mass $M$ and local interaction parameters are unchanged, while the geometry-induced scale $M_{\rm eff}$ (or, for photons, the analogous quantity $m_{\text{eff}}$) enters only as an effective rest-energy scale in the emergent-frame kinematics.

Here, $E$ denotes the photon energy measured in the $g_{\mu\nu}$-frame. The quantity $m_{\text{eff}}$ represents a purely geometric/kinematic correction associated with the deformation of the light cone, rather than a true (Proca-type) mass term \cite{Hell}. In particular, the relation
$g_{\mu\nu}\bar{P}^\mu \bar{P}^\nu = A\dot{\phi}^2(\bar{P}^0)^2$
does not endow the photon with a physical mass; instead, it quantifies the apparent rest-frame contribution that arises when comparing the emergent null condition to the one defined by $g_{\mu\nu}$. In the emergent metric $\bar{G}_{\mu\nu}$, photons remain massless and propagate along null geodesics, while in the $g_{\mu\nu}$ description, the ``tilt" between the two null cones makes the propagation appear to carry an effective rest-energy scale. Thus, $m_{\text{eff}}$ measures how strongly the K-essence background deforms the light cone, encoding a geometric influence rather than a true mass effect.\\


To understand the tilted or deformed light cone in between the K-essence geometry and the original gravitational geometry based on \cite{Babichev, Vikman}, can be explained as follows: 

In standard general relativity, the light cone at a spacetime point is characterized by all tangent vectors $k^{\mu}$ that fulfill the condition $g_{\mu\nu}k^{\mu}k^{\nu}=0$. These vectors denote every possible direction in which a light signal, or any massless particle, may propagate. The light cone illustrates the causal structure, indicating which events can affect or be affected by others. In the context of flat FLRW (Eq. (\ref{11})) spacetime ($g_{\mu\nu}$), local coordinates give $ds^{2}=0$, leading to a finding that $dx^{i}/dt=1/a(t)$. This expression indicates that the coordinate speed of light in comoving coordinates is $1/a(t)$; however, the physical speed of light, as measured by a local observer using proper distances, remains $c=a(t)\frac{dx^{i}}{dt}=1$.

In contrast, under the K-essence induced FRLW metric (Eq. (\ref{12})), the scenario of $\G_{\mu\nu}k^{\mu}k^{\nu}=0$ or $dS^{2}=0$ necessitates that $\frac{dx^{i}}{dt}=\frac{\sqrt{1-A\dot\phi^{2}}}{a(t)}$. Consequently, the {\it effective speed of light} within the emerging geometry is reduced by a factor of $v_{ph}=\sqrt{1-A\dot\phi^{2}}$. This indicates that the cone characterized by $\G_{\mu\nu}$, representing the permissible null directions, is either narrower or wider than the original cone.

In other words, the ``light cone" is a collection of null directions in spacetime. When the K-essence field is introduced, the null condition shifts from $g_{\mu\nu}k^{\mu}k^{\nu}=0$ to $\G_{\mu\nu}k^{\mu}k^{\nu}=0$. This implies that the cone described in $\G_{\mu\nu}$, the set of allowed null directions, is narrower or wider than the original one \cite{Babichev, Vikman, Sawicki}. If $A\dot\phi^{2}>0$, the factor $(1-A\dot\phi^{2})$ makes null vectors ``steeper" in a spacetime diagram, i.e., the light cone narrows. If $A\dot\phi^{2}<0$, the cone ``widens", indicating superluminal propagation in some K-essence models. However, causality in the emergent metric is not necessarily violated \cite{Babichev, Vikman, Sawicki}.  Also, note that for our case, we are not considering superluminal propagation for our investigation.\\

The ``tilt angle" is a geometric representation of how much the \emph{ null directions} differ between the two measures. For better understanding, let us consider the flat spacetime coordinates, $(t,x)$, i.e., in a 2D scenario: In $g_{\mu\nu}$, the null line is $dx/dt=\pm 1$, but in $\G_{\mu\nu}$, it is $dx/dt=\pm \sqrt{1-A\dot\phi^{2}}$. Hence, the cone in the ($t,x$) plane changes slope. The tilted angle is $\text{tan}\theta=\sqrt{1-A\dot\phi^{2}}$ in relation to the original $45^{0}$ null lines of the spacetime $g_{\mu\nu}$. So the ``angle" is the slope difference between the null lines of the original and emerging geometries, a purely geometric, local parameter that determines how much the kinetic energy of the scalar field deforms causal propagation. This ``tilt" does not imply that spacetime itself rotates or tilts; rather, it refers to how the effective causal cone (the permissible directions of information or field propagation) differs from the gravitational one \cite{Babichev, Vikman, Sawicki}.\\

So, we may conclude that, in K-essence geometry, the concept of a ``tilt of the light cone" corresponds to the shift of causal structure caused by the kinetic energy of the scalar field. The emergent metric modifies the relationship between time and spatial intervals, influencing the range of directions in which light or perturbations can travel. As a result, the null cone described by the effective metric differs from the null cone defined by the background gravitational metric. This modification can cause the cone to narrow or widen, indicating that pairs of events that were previously causally coupled within the underlying geometry may now be separated, or vice versa. The "tilt angle" is a quantitative measure of the deviation between the two null cones, reflecting how much the scalar field affects the effective propagation speed in the relevant spacetime framework.\\

In the context of field theory \cite{Babichev, Hell}, the physical mass is identified with the pole of the propagator when evaluated at zero spatial momentum. If the photon kinetic operator is $\bar G^{\mu\nu}k_\mu k_\nu$, the pole condition at $\mathbf{k}=0$ gives $\omega=0$: the photon remains massless in the emergent frame. Consequently,  $m_{\text{eff}}$ is considered "physical" solely when it manifests as an actual propagator pole within a particular frame.\\

An instructive analogy is light propagation in a refractive medium: although the photon remains massless, the medium modifies the relation between frequency and wavenumber to $\omega=v_{\rm ph}k$, with $v_{\rm ph}<c$ (for an isotropic nondispersive medium).  
Likewise, a K-essence background acts as a dynamical environment that replaces the null condition defined by $g_{\mu\nu}$ with that defined by the emergent metric $\bar G_{\mu\nu}$.  
If photon kinematics are interpreted using the gravitational-frame time coordinate, this mismatch can be described as an apparent rest-frame energy (or ``effective mass''), even though the photon remains massless and gauge invariant in $\bar G_{\mu\nu}$. In this sense, one may introduce an effective refractive index,
$n_{\rm eff}=(1-A\dot{\phi}^{2})^{-1/2}$,
which parameterizes the reduced coordinate propagation speed relative to cosmic time. This is only an analogy: there is no physical material medium, and the modification is purely geometric. In a local orthonormal frame of the emergent geometry, null rays still propagate at $c$.\\

\section{Rationale for the Local Collision Term in the Emergent K-essence Geometry} \label{E} 

This appendix explains why it is consistent to use the standard flat-space scattering amplitude in the collision term (Eq.~\eqref{34}), even though particles propagate in the curved emergent metric $\bar G_{\mu\nu}$. Throughout this work, we therefore take the underlying QED scattering amplitude to be the same as in Minkowski spacetime: microscopic interactions occur within a local inertial frame whose size is negligible compared with the curvature scale, so curvature does not affect the local matrix element. The only modification enters when the collision operator is written in terms of emergent-frame kinematical variables, which naturally leads to the effective geometric transport coefficient defined in Eq.~\eqref{82}. This reasoning follows the standard treatment in cosmological kinetic theory (see, e.g., \cite{Dodelson, Weinberg, Kolb, deGroot, Ehlers}) and relies on the strong separation between microscopic interaction scales and the macroscopic curvature scale of the universe.

\subsection{Microscopic Interaction Scale}
\label{E1}

For clarity, we examine Thomson scattering. The Thomson cross section is given by $\sigma_T = 6.65\times10^{-29}\,{\rm m}^2$  \cite{Dodelson,Weinberg,Kolb}.
The characteristic interaction length can be approximated as $ \ell_{\rm int} \sim \sqrt{\sigma_T} \sim 10^{-14}\,{\rm m}.$
Utilizing the larger atomic scale, $\ell_{\rm atomic}\sim 10^{-10}\,{\rm m},$ establishes a conservative upper limit for the spatial domain in which electromagnetic interactions occur \cite{Kolb}. The associated interaction time scale is given by $t_{\rm int}\sim \frac{\ell_{\rm int}}{c} \sim 10^{-23}\,{\rm s}.$
These scales are infinitesimal in comparison to any cosmological scale.

\subsection{Curvature scale of the emergent geometry}

The emergent metric \eqref{12} in the homogeneous background is 
\ben dS^{2}=-(1-A\dot\phi^{2})dt^{2}+a^{2}(t)d\mathbf{x}^{2}.
\label{E11}
\een
By introducing the emergent proper time:
\ben d\Upsilon = \sqrt{1 - A\dot\phi^{2}}\,dt.
\label{E2}
\een 
The metric \eqref{E11} assumes the conventional FLRW form 
\ben dS^{2} = -d\Upsilon^{2} + a^{2}(\Upsilon)d\mathbf{x}^{2}.
\label{E3}
\een
The curvature scale is determined by the emergent Hubble parameter 
\ben 
\widetilde H = \frac{1}{a}\frac{da}{d\Upsilon} =\frac{H}{\sqrt{1-A\dot\phi^{2}}},
\label{E4}
\een
so that the curvature radius in this emergent spacetime is given by 
\ben
\widetilde R_{\rm curv}\sim \widetilde H^{-1} =\sqrt{1-A\dot\phi^{2}}\,H^{-1}.
\label{E5}
\een

At recombination, the Hubble parameter is about $H_{\rm rec}\sim10^{-18}\,{\rm s}^{-1}$ \cite{Dodelson,Weinberg}, indicating that, at the cosmic time scale, $H^{-1}\sim10^{18}\,{\rm s}$, i.e., the standard Hubble radius  $H^{-1}\simeq 3\times 10^{8}\, {\rm m/s} \times 10^{18}\, {\rm s} \sim10^{26}\,{\rm m}.$ This scale indicates the curvature radius of the cosmic background and is much larger than the microscopic interaction scales pertinent to particle scattering.

Following considerable disformal modification \eqref{E5}, $\sqrt{1-A\dot\phi^{2}}\lesssim1$ ($A\dot\phi^{2}< 1$), the emergent curvature radius is approximately of the order of $\widetilde R_{\rm curv}\sim10^{25}\text{--}10^{26}\,{\rm m}.$

\subsection{Hierarchy of scales}

The ratio of microscopic to curvature scales is thus expressed as $\frac{\ell_{\rm int}}{\widetilde R_{\rm curv}} \lesssim \frac{10^{-14}}{10^{25}} \sim10^{-39},$ or, more conservatively, $\frac{\ell_{\rm atomic}}{\widetilde R_{\rm curv}} \lesssim10^{-35}.$
This extensive hierarchy illustrates that scattering events occur in regions of spacetime that are effectively flat.

\subsection{Physical interpretation and comparison with the standard case}

According to the equivalence principle, each sufficiently small region of curved spacetime admits a local inertial frame in which the metric simplifies to the Minkowski form and its first derivatives are set to zero. Given that particle interactions are localized to microscopic spacetime regions, the matrix element $|\mathcal M|^{2}$ relevant to the collision term is equivalent to its special-relativistic formulation \cite{deGroot, Ehlers}. This is the exact rationale used in standard FLRW cosmology to derive the Boltzmann hierarchy for the CMB \cite{Dodelson,Ma1}.

\subsection{Covariant origin of the emergent transport coefficient}

This separation of scales justifies evaluating the QED scattering amplitude in a local inertial frame: the geometric effects discussed in the main text do not come from curvature corrections to the microscopic interaction. Instead, they arise when the covariant collision operator is rewritten in emergent-frame kinetic variables.

For a homogeneous K-essence background the lapse is $N=\sqrt{1-A\dot\phi^{2}}$, and the emergent rest-energy scale of the electron is $M_{(e),\rm eff}=m_e N$, while the microscopic Thomson cross section retains its standard QED form,
\begin{equation}
\sigma_T=\frac{8\pi}{3}\left(\frac{e^2}{m_e c^2}\right)^2.
\end{equation}
When the collision integral is expressed in emergent-frame energy, momentum, and coordinate-time variables, the external-state normalization and the associated lapse conversions combine into a purely geometric factor,
\begin{equation}
\mathcal F_{\rm geom}
=\left(\frac{m_e}{M_{(e),\rm eff}}\right)^4
=\left(1-A\dot\phi^{2}\right)^{-2}.
\end{equation}
As a result, the coefficient multiplying the macroscopic collision kernel in the emergent-frame Boltzmann equation is
\begin{equation}
\sigma_T^{(\rm geom)}
=\sigma_T\,\mathcal F_{\rm geom}
=\frac{\sigma_T}{\left(1-A\dot\phi^{2}\right)^2}.
\end{equation}
This makes the separation between microscopic and macroscopic quantities explicit: $\sigma_T$ characterizes the local QED interaction and is independent of the K-essence background, whereas $\sigma_T^{(\rm geom)}$ is the transport coefficient that appears after rewriting the same interaction in emergent-frame variables and therefore inherits a background dependence.\\

In the present K-essence framework, the situation is theoretically equivalent, with the sole distinction being that the pertinent local inertial frame is linked to the emergent metric $\bar G_{\mu\nu}$ instead of the gravitational metric $g_{\mu\nu}$.
The principles of microscopic scattering physics are invariant, whereas the geometry alters solely the macroscopic propagation between collisions via the Liouville operator and the effective dispersion relation $\widetilde E$. Consequently, the application of the standard flat-space collision integral in Eq.~\eqref{34} is entirely justified within the emergent K-essence cosmological framework and reverts to the conventional FLRW outcome in the relevant limit.

\section{Liouville theorem and disformal transformations in K-essence geometry} \label{F}

A pertinent question in the development of kinetic theory within the context of K-essence-induced emergent spacetime is whether the Liouville theorem remains valid following a disformal transformation of the metric. Given that the disformal transformation alters the effective geometry encountered by particles, it is crucial to investigate the preservation of phase-space volume and the validity of the Liouville operator within this context.

In standard general relativistic kinetic theory, the Liouville theorem asserts that the phase-space distribution function is conserved along particle trajectories in the absence of collisions \cite{Dodelson, Weinberg, deGroot, MTW1973, Ehlers}. This is expressed as 
\ben \frac{df}{d\lambda} = 0, \label{F1} 
\een 
where $\lambda$ denotes an affine parameter along the particle worldline. This outcome derives from the Hamiltonian framework of geodesic motion and the invariance of the phase-space volume element concerning canonical transformations.

In the current framework, particles move along the geodesics of the emergent metric $\bar{G}_{\mu\nu}$ instead of the gravitational metric $g_{\mu\nu}$. So, the correct phase space is specified in connection to $\bar{G}_{\mu\nu}$, and the invariant phase-space volume element becomes
\ben
d\Gamma = \frac{d^3x\, d^3\bar{p}}{(2\pi)^3 \sqrt{-\det(\bar{G}_{\mu\nu})}}.
\label{F2}
\een
The evolution of the distribution function is governed by the Liouville operator derived from the emergent geometry,
\ben
\mathcal{L}_{\bar{G}}[f] \equiv \bar{P}^\mu \frac{\partial f}{\partial x^\mu} - \bar{\Gamma}^\mu_{\alpha\beta} \bar{P}^\alpha \bar{P}^\beta \frac{\partial f}{\partial \bar{p}^\mu},
\label{F3}
\een
where $\bar{P}^\mu$ denotes the particle four-momentum and $\bar{\Gamma}^\lambda_{\mu\nu}$ represents the Christoffel symbols related to the emergent metric $\bar{G}_{\mu\nu}$.

By construction, the particle dynamics are inherently Hamiltonian with respect to the emergent geometry, and the geodesic flow maintains the associated phase-space measure. Consequently, the Liouville theorem remains valid in the disformal framework, contingent upon the consistent definition of the Liouville operator utilizing the emergent metric. In other terms,
\ben
\mathcal{L}_{\bar{G}}[f] = 0
\label{F4}
\een
applies to collisionless evolution.

Consequently, we can conclude that the disformal transformation does not violate phase-space conservation; instead, it facilitates a non-trivial mapping between the phase-space variables defined in the gravitational frame and those characterized in the emergent frame. Thus, when attempting to formulate the Boltzmann equation using the original gravitational variables $(x^\mu, p^\mu)$, supplementary terms emerge in the Liouville operator, indicating the non-canonical character of the transformation. These extra contributions can be interpreted as effective forces arising from the spacetime-dependent disformal factor.

\section{Boltzmann Equation for Baryons in the Emergent K--essence Geometry}
\label{B}
The following cosmic component, which necessitates a kinetic explanation in the emergent K-essence cosmology, is the baryonic sector, comprising electrons and protons, along with references to helium nuclei and heavier elements.  Despite the microscopic differences among these particles, their dynamics can be analyzed collectively, as their total energy density is largely determined by the rest mass of the nucleus.  Within the current framework, all matter species propagate on the emergent metric $\bar G_{\mu\nu}$, and consequently, their distribution functions evolve according to Boltzmann equations formulated according to this geometry.

Electrons and protons are intensely interconnected via Coulomb scattering ($e+p\leftrightarrow e+p$).  The interaction rate significantly exceeds the Hubble expansion rate during the relevant epochs, and this holds true under the K-essence framework, as the emerging geometry merely rescales time intervals without reducing local electromagnetic interactions.  As a result, the overdensities and velocities of electrons and protons are interconnected, expressed as follows: 
\ben
\frac{\rho_e-\bar\rho_e}{\bar\rho_e} =\frac{\rho_p-\bar\rho_p}{\bar\rho_p} \equiv \delta_b, 
\qquad \ua_e=\ua_p\equiv \ua_b, 
\label{B1}
\een 
indicating that baryons function as a unified fluid degree of freedom.

The baryons are nonrelativistic ($T\ll m_e$) and have a negligible mean free path, allowing them to be modeled as a nonrelativistic fluid in the emergent spacetime.  Consequently, in accordance with the conventional cosmological framework \cite{Dodelson}, just the fundamental moments (density and velocity) of the Boltzmann equation are necessary.  In contrast to collisionless dark matter, the baryonic fluid remains interconnected with photons via Compton scattering. Within the K-essence framework, this coupling is controlled by the identical microscopic interaction, yet is evaluated within the emergent geometry, resulting in altered temporal evolution via the rescaled conformal time and the effective scattering rate.  Consequently, the baryon Boltzmann hierarchy mimics the structure of the standard case, albeit its dynamical evolution is geometrically modified by the background scalar field.

The baryon distribution function $f_b(\mathbf{x},\mathbf{p},t)$ evolves in accordance with the Boltzmann equation within the emergent K-essence spacetime, as expressed by the equation: 
\ben
\frac{df_{b}}{dt}=\frac{\partial f_{b}}{\partial t}+\frac{\partial f_{b}}{\partial{x}^{i}}\frac{\hat{\pa}^{i}}{a}\frac{\pa}{\widetilde{E}}\Big(1-\frac{1}{2}A\dot{\phi}^{2}\Big)\nonumber\\
-\pa\frac{\partial f_{b}}{\partial \pa}\Big[H+\dot{\Phi}+\frac{\widetilde{E}}{a\pa}\hat{\pa}^{i}\partial_{i}\widetilde{\Psi}\Big]=C_{b}[f].
\label{B2}
\een
At the zeroth moment, the collision term is nullified due to the conservation of baryon number by Coulomb and Thomson interactions, resulting in $C_b^{(0)}=0$. Defining the baryon number density and bulk velocity 
\ben
n_b=\int\!\frac{d^3\pa}{(2\pi)^3}f_b, \qquad \textbf{u}_{b}^{i}\equiv \frac{1}{n_{b}}\int{\frac{d^{3}\pa}{(2\pi)^{3}}\frac{\pa\hat{\pa}^{i}}{\widetilde{E}}},
\label{B3}
\een
the integration of the Boltzmann equation results in 
\ben
\frac{\partial n_{b}}{\partial t}+\frac{1}{a}\Big(1-\frac{1}{2}A\dot{\phi}^{2}\Big)\frac{\partial(n_{b}\textbf{u}_{c}^{i})}{\partial x^{i}}+ 3n_{b}(H+\dot{\Phi})=0.\n
\label{B4}
\een

Writing $n_b =\bar{n}_b (1+\delta_b)$ and converting to conformal time ($\widetilde{\eta}$) and Fourier space, we obtain the equation: 
\ben
\delta_{b}^{\prime}+\Big(1-\frac{A{\phi^{\prime}}^{2}}{2a^{2}}\Big)ik\textbf{u}_{b}+3\Phi^{\prime}=0.
\label{B5}
\een

Taking the first moments of the
Boltzmann equations for electrons and baryons and adding them together, we have
\ben
\frac{\partial \textbf{u}_{b}^{j}}{\partial t}+H\textbf{u}_{b}^{j}+\frac{1}{a}\partial_{j}\widetilde{\Psi}=\frac{1}{\rho_{b}}F_{e\gamma}^{j}(\x,t),
\label{B6}
\een 
where $\rho_{b}=m_{p}\bar{n}_{b}$ and  $F_{e\gamma}^{j}$ $\Big(F^{j}(\x,t)=\int{\frac{d^{3}\pa}{(2\pi)^{3}}\pa\hat{\pa}^{j}C_{b}[f]}\Big)$  represents the Thomson momentum–transfer force. Projecting along $\mathbf{k}$, 
\ben
\frac{1}{\rho_{b}}\hat{k}_{i}F_{e\gamma}^{j}(\x,t)=-n_{e}\sigma^{(\rm geom)}_{T}\frac{4\rho_{\gamma}}{\rho_{b}}\Big[i\Theta_{1}+\frac{1}{3}\textbf{u}_{b}]
\label{B7}
\een
where the first moment is $\Theta_{1}(k,\widetilde{\eta})\equiv i\int_{-1}^{1}\frac{d\mu}{2}\mu \Theta(\mu,k,\widetilde{\eta}).$

Consequently, the baryon velocity equation in conformal time ($\widetilde{\eta}$) is expressed as 
\ben 
u_b'+\frac{a'}{a}u_b+ik\widetilde\Psi =(\tau')^{\rm eff}\frac{4\rho_\gamma}{3\rho_b}\left(3i\Theta_1+u_b\right).
\label{B8}
\een
The Euler Eq. (\ref{B8}) for baryon velocity incorporates the ratio $\rho_\gamma/\rho_b$ instead of $\rho_\gamma/\rho_e$, given the fact that photons interact solely with electrons. This indicates that, in the primordial plasma, electrons are not dynamically independent particles. Coulomb interactions couple electrons with protons (and nuclei) with remarkable efficiency, compelling them to share a uniform bulk velocity. Thus, Thomson scattering imparts momentum from photons to the entire baryonic fluid instead of solely to electrons.

This interpretation remains consistent in the emergent K-essence geometry.
The scalar field alters the effective spacetime and modifies the interaction rate via $\sigma_T^{(\rm geom)}$, yet it does not disrupt the strong Coulomb coupling within the baryon sector. Thus, the inertia opposing photon momentum transfer is the total baryon inertia $\rho_b$, primarily influenced by the rest mass of protons.
In the extreme scenario where $\rho_b\!\to\!\infty$, the baryon fluid attains infinite mass, making photon scattering incapable of significantly affecting its velocity.

Although the derivation presuming a fully ionized plasma ($n_e\simeq n_b$), the outcome remains applicable even when neutral hydrogen emerges post-recombination. Neutral atoms remain closely linked to the charged component via electromagnetic collisions, thus, the momentum exchange of photons continues to influence the collective baryon fluid.
Consequently, the velocity equation accurately characterizes all baryons within the K-essence cosmological framework.

\section{WKB solution of the tight-coupling equation with drag}
\label{C}

We now consider the homogeneous part of the tight-coupling equation for the combined
photon-gravitational variable $(\Theta_{0}+\Phi)$, retaining the drag terms induced by
both baryon loading and the emergent K-essence geometry. Eq.~\eqref{112} may be written as
\ben
&&(\Theta_{0}''+\Phi'')
+\Bigg[
\frac{R'}{1+R}
+\frac{\partial}{\partial\widetilde{\eta}}
\Big(\frac{A{\phi}'^{2}}{2a^{2}}\Big)
\Bigg]
(\Theta_{0}'+\Phi')
\n &&+k^{2}\mathcal{C}_{s}^{2}(\Theta_{0}+\Phi)=0.
\label{C1}
\een
This equation has the form of a damped harmonic oscillator with a time-dependent frequency
and friction, where the additional damping arises from the scalar--field background.

To solve Eq.~\eqref{C1}, we adopt a WKB ansatz as
\ben
\Theta_{0}(\widetilde{\eta})+\Phi(\widetilde{\eta})
= D(\widetilde{\eta})\,e^{iB(\widetilde{\eta})}
\label{C2}
\een
with $D(\widetilde{\eta})$ and $B(\widetilde{\eta})$ are real functions. Substituting this ansatz into
Eq.~\eqref{C1} and separating real and imaginary parts, we get

\noindent{\bf real part:}
\ben
\frac{D''}{D}-(B')^{2}
+\Bigg[
\frac{R'}{1+R}
+\frac{\partial}{\partial\widetilde{\eta}}
\Big(\frac{A{\phi}'^{2}}{2a^{2}}\Big)
\Bigg]\frac{D'}{D}
+k^{2}\mathcal{C}_{s}^{2}=0,\n
\label{C3}
\een

\noindent{\bf imaginary part:}
\ben
2B'\frac{D'}{D}
+B''
+B'\Bigg[
\frac{R'}{1+R}
+\frac{\partial}{\partial\widetilde{\eta}}
\Big(\frac{A{\phi}'^{2}}{2a^{2}}\Big)
\Bigg]=0.\n
\label{C4}
\een

In the tight-couplingime and for subhorizon modes, the oscillation phase varies much
more rapidly than the amplitude, so that
\ben
D' \ll B', \qquad D'' \ll (B')^{2}.
\label{C5}
\een
Under this approximation, the dominant terms in Eq.~\eqref{C3} give
\ben
(B')^{2}=k^{2}\mathcal{C}_{s}^{2},
\label{C6}
\een
which implies
\ben
B(\widetilde{\eta})=\pm k r_{s}(\widetilde{\eta}), \qquad
r_{s}(\widetilde{\eta})\equiv
\int_{0}^{\widetilde{\eta}} d\bar{\eta}\,\mathcal{C}_{s}(\bar{\eta}).
\label{C7}
\een
Thus $B'=k\mathcal{C}_{s}$ and $B''=k\mathcal{C}_{s}'$.

Substituting these relations into Eq.~\eqref{C4}, we obtain
\ben
2\frac{D'}{D}
+\frac{\mathcal{C}_{s}'}{\mathcal{C}_{s}}
+\frac{R'}{1+R}
+\frac{\partial}{\partial\widetilde{\eta}}
\Big(\frac{A{\phi}'^{2}}{2a^{2}}\Big)=0.
\label{C8}
\een
Using the explicit form of the sound speed,
\ben
\mathcal{C}_{s}^{2}
=\frac{1-\frac{A{\phi}'^{2}}{a^{2}}}
{3(1+R)},
\label{C9}
\een
one finds
\ben
\frac{\mathcal{C}_{s}'}{\mathcal{C}_{s}}
=-\frac{1}{2}\frac{R'}{1+R}
-\frac{1}{2}
\frac{1}{\Big(1-\frac{A{\phi}'^{2}}{a^{2}}\Big)}
\frac{\partial}{\partial\widetilde{\eta}}
\Big(\frac{A{\phi}'^{2}}{a^{2}}\Big).\n
\label{C10}
\een

Substituting Eq.~\eqref{C10} into Eq.~\eqref{C4}, we obtain
\ben
\frac{D'}{D}
=-\frac{1}{4}\frac{R'}{1+R}
-\frac{1}{4}
\Bigg[
1-\frac{1}{\Big(1-\frac{A{\phi}'^{2}}{a^{2}}\Big)}
\Bigg]
\frac{\partial}{\partial\widetilde{\eta}}
\Big(\frac{A{\phi}'^{2}}{a^{2}}\Big).\n
\label{C11}
\een

Integrating Eq.~\eqref{C11} with respect to $\widetilde{\eta}$ gives
\ben
\ln D
=-\frac{1}{4}\ln(1+R)
-\frac{1}{4}\frac{A{\phi}'^{2}}{a^{2}}
-\frac{1}{4}\ln\Big(1-\frac{A{\phi}'^{2}}{a^{2}}\Big),\n
\label{C12}
\een
so that the amplitude becomes
\ben
D(\widetilde{\eta})=
\frac{\exp\!\Big(-\frac{A{\phi}'^{2}}{4a^{2}}\Big)}
{(1+R)^{1/4}
\Big(1-\frac{A{\phi}'^{2}}{a^{2}}\Big)^{1/4}}.
\label{C13}
\een

The homogeneous WKB solution of Eq.~\eqref{C1} therefore reads
\ben
&&\Theta_{0}(\widetilde{\eta})+\Phi(\widetilde{\eta})
=
\frac{\exp\!\Big(-\frac{A{\phi}'^{2}}{4a^{2}}\Big)}
{(1+R)^{1/4}
\Big(1-\frac{A{\phi}'^{2}}{a^{2}}\Big)^{1/4}}\n && \times 
\Big[
\cos\!\big(k r_{s}(\widetilde{\eta})\big)
+\sin\!\big(k r_{s}(\widetilde{\eta})\big)
\Big].
\label{C14}
\een

Thus, when the drag term is retained, the homogeneous solution differs from the simple oscillatory form by the slowly varying prefactor
\ben
\mathcal{A}(\widetilde{\eta})=
\frac{\exp\!\Big(-\frac{A{\phi}'^{2}}{4a^{2}}\Big)}
{(1+R)^{1/4}
\Big(1-\frac{A{\phi}'^{2}}{a^{2}}\Big)^{1/4}},
\label{C15}
\een
which represents the combined damping effects of baryon loading and the emergent
K-essence geometry. This factor modulates the amplitude of the acoustic oscillations,
while the phase remains determined by the sound horizon $r_{s}(\widetilde{\eta})$.

\section{Diffusion Damping Scale in the emergent K-essence Geometry}
\label{D}

For the DBI-type kinetic K-essence model we take
\begin{equation}
\mathcal{L}(X)=-\sqrt{1+2X},
\label{D1}
\end{equation}
with $X=-\frac{1}{2}\dot\phi^{2}$, which corresponds to $A=1$. The exact homogeneous background solution then gives
\begin{equation}
\dot{\phi}^{2}=\frac{1}{1+a^{6}}.
\label{D2}
\end{equation}
Throughout this appendix, we use this exact expression for $\dot\phi^{2}$, rather than expanding in powers of $\dot\phi^{2}$ (or $A\dot\phi^{2}$). This is essential because at early times ($a\ll 1$) the DBI evolution approaches $\dot\phi^{2}\to 1$, where a truncated small $A\dot\phi^{2}$ expansion is not a controlled approximation.

This alters the Thomson interaction rate long before recombination as
\ben
n_{e}\sigma_T^{(\mathrm{geom})}a = 2.3\times10^{-5}\, a^{-8}\,{\rm Mpc}^{-1} (\Omega_{b}h^{2}) \Big(1-\frac{Y_{p}}{2}\Big),\n 
\label{D3}
\een
which significantly deviates from the conventional $a^{-2}$ scaling.

The photon diffusion scale (Silk damping) is defined by the interplay between Thomson scattering and the photon free propagation. Utilizing an approach similar to the usual case \cite{Dodelson} for the radiation-dominated epoch $R\ll 1$, but now analyzed within the emergent K-essence spacetime, the damping scale can be defined as
\ben
\frac{1}{k_{D}^{2}} \equiv \int_{0}^{\widetilde{\eta}} d\bar{\eta}\; \frac{1}{6\,n_{e}\,\sigma_T^{(\mathrm{geom})}\,a(\bar{\eta})} \frac{8}{9} \Big(1-\frac{\phi'^2}{2a^{2}}\Big)^{2}.
\label{D4}
\een
This deviates from the conventional expression because it includes the scalar-field factor $\big(1-\phi'^2/(2a^2)\big)^2$, which arises from the modification of photon propagation and scattering rates in the emergent metric.

Using Eqs. \eqref{81} and \eqref{82}, we reformulate the above relations \eqref{D4},
\ben
\frac{1}{k_{D}^{2}} = \frac{4}{27\sigma_{T}} \int_{0}^{\widetilde{\eta}} d\bar{\eta}\; \frac{1}{n_{e}\,a(\bar{\eta})} \Big(1-\frac{\phi'^2}{a^{2}}\Big)^{2},
\label{D5}
\een
where $\sigma_{T}$ is the usual Thomson scattering cross section.

Utilizing the conformal relation
\ben
 a^{2}d\widetilde{\eta}=\frac{da}{H},
 \label{D6}
\een
we obtain
\ben
\frac{1}{k_D^2}
=
\frac{4}{27}
\int
\frac{da}{a^3H(a)\,n_e\sigma_T^{\rm eff}}
\Big(1-\frac{\phi'^2}{a^2}\Big)^2.
\label{D7}
\een
We also have the radiation-matter expansion law
\ben
H=H_{0}a^{-2}\sqrt{\Omega_{r}+\Omega_{m}a}.
\label{D8}
\een
It is important to note that the background expansion history remains unchanged during the pre-recombination period. In this study, the kinetic K-essence (DBI-type) Lagrangian indicates that the scalar field evolves towards the regime $\dot{\phi}^{2}\simeq 1$ at early times, leading to an equation of state $w_{\phi}\simeq 0$. As a result, the K-essence energy density scales as $\rho_{\phi}\propto a^{-3}$, thereby functioning effectively as pressureless matter. As the cosmos remains radiation-dominated prior to recombination, the usual radiation-matter expansion law is applicable, and the background Hubble evolution aligns with conventional FLRW cosmology.

We define the variable
\ben
y=\frac{a}{a_{eq}}, \qquad a_{eq}=\frac{\Omega_{r}}{\Omega_{m}},
\label{D9}
\een
resulting in
\ben
H=H_{0}a^{-2}\sqrt{\Omega_{m}a_{eq}}\sqrt{1+y}.
\label{D10}
\een

Thus, the diffusion scale is expressed as follows:
\ben
&&\frac{1}{k_{D}^{2}} = \frac{4\,a_{eq}^{3}}{27\,\sigma_{T}\,n_{b0}\,(1-\frac{Y_{p}}{2})\,H_{0}\sqrt{\Omega_{m}a_{eq}}} \n && \times \int_{0}^{y} \frac{y^{2}}{\sqrt{1+y}} \Big(1-\frac{A\phi'^2}{a^{2}}\Big)^{2} dy. 
\label{D11}
\een

Before recombination, the free-electron number density is
\ben
n_e = \left(1-\frac{Y_p}{2}\right)n_b,
\label{D12}
\een
where $Y_p$ is the helium mass fraction. The present baryon number density is
\ben
n_{b0}=\frac{\rho_{c0}\Omega_b}{m_b}
\simeq 1.13\times10^{-5}(\Omega_b h^2)\;{\rm cm^{-3}}.
\label{D13}
\een
In the emergent geometry, the physical baryon density evolves as
\ben
n_b(a)=n_{b0}a^{-3},
\label{D14}
\een
since particle number is conserved.

Consequently, the damping scale \eqref{D7} is expressed as 
\ben
&&\frac{1}{k_{D}^{2}} = \frac{4\,a_{eq}^{29/2}} {27\,\sigma_{T}\,n_{b0}(1-\frac{Y_{p}}{2})H_{0}\sqrt{\Omega_{m}}} \int_{0}^{y}\frac{y^{14}}{\sqrt{1+y}}dy.\n \label{D15}\\  
\Rightarrow
&&\frac{1}{k_{D}^{2}}=19.1\times10^{6} Mpc^{2}\,a^{\frac{29}{2}}\times f_{D}(y)\,(\Omega_{b}\,h^{2})^{-1}\nonumber\\
&&\times  \Big(1-\frac{Y_{p}}{2}\Big)^{-1}(\Omega_{m}\,h^{2})^{-\frac{1}{2}}
\label{D16}
\een
where $f_{D}(y)=\Big[0.46(1+\frac{1}{y})^{\frac{29}{2}}+ 127.2(1+\frac{1}{y})^{\frac{27}{2}}-669.8(1+\frac{1}{y})^{\frac{25}{2}}+ 2187.8(1+\frac{1}{y})^{\frac{23}{2}} -4804.9(1+\frac{1}{y})^{\frac{21}{2}}+ 7417.4(1+\frac{1}{y})^{\frac{19}{2}}-8180.2(1+\frac{1}{y})^{\frac{17}{2}}+ 6369.1(1+\frac{1}{y})^{\frac{15}{2}}-3298.0(1+\frac{1}{y})^{\frac{13}{2}}+ 896.9(1+\frac{1}{y})^{\frac{11}{2}}+100.7(1+\frac{1}{y})^{\frac{9}{2}} -203.2(1+\frac{1}{y})^{\frac{7}{2}} +86.3(1+\frac{1}{y})^{\frac{5}{2}}-18.6(1+\frac{1}{y})^{\frac{3}{2}}+2(1+\frac{1}{y})^{\frac{1}{2}}-0.46\,y^{-\frac{29}{2}}\Big]$.

For $y \gg 1 $, the integral can be approximated as
\ben
\int_{0}^{y}\frac{y^{14}}{\sqrt{1+y}}dy \simeq 13.16\,y^{29/2}, 
\label{D17}
\een
which leads to,
\ben
&&\frac{1}{k_{D}^{2}} = 252.4\times10^{6}\,{\rm Mpc}^{2}\, a^{29/2} (\Omega_{b}h^{2})^{-1}\n && \times  \Big(1-\frac{Y_{p}}{2}\Big)^{-1} (\Omega_{m}h^{2})^{-1/2}.
\label{D18}
\een

It should be noted that here we define $y\equiv a/a_{\rm eq}$, so that the limit $y\gg 1$ corresponds to $a\gg a_{\rm eq}$. We adopt this approximation because the damping scale is evaluated at recombination ($z\simeq 1100$), while matter-radiation equality occurs at $z_{\rm eq}\simeq 3400$. Using $a\propto (1+z)^{-1}$, we find $\frac{a}{a_{\rm eq}}=\frac{1+z_{\rm eq}}{1+z}\approx 3,$
which supports treating $y$ as larger than unity in this context.\\

\section{Derivation of \texorpdfstring{$\eqref{130}$}{(130)}}
\label{G}
In this appendix, we present the derivation of Eq.~\eqref{130} in a self-contained way.

We begin with Eq.~\eqref{123},
\ben
\Theta_1' + k\Big(\frac{2}{3}\widetilde{\Theta}_2 - \frac{1}{3}\widetilde{\Theta}_0\Big)
= \tau'\Big(\Theta_1 - \frac{i u_b}{3}\Big).
\label{G1}
\een
Passing to Fourier space, $\partial_\eta\to i\omega$, and substituting $u_b$ from Eq.~\eqref{128}, we obtain
\ben
&& i\omega\Theta_{1}
+ k\Big(1-\frac{A\phi'^{2}}{2a^{2}} \Big)\Big(\frac{2}{3}\Theta_{2}-\Theta_{0}\Big)
\nonumber\\ &&= {\tau}'\Big[\Theta_{1}-\frac{i}{3}\Big(-3i\Theta_1\Big(1+\frac{i\omega R}{\tau'}- \Big(\frac{\omega R}{\tau'}\Big)^2\Big)\Big)\Big].~~~~
\label{G2}
\een

We then use Eq.~\eqref{129} with Eq. \eqref{70} to express $\Theta_0$ and $\Theta_2$ in terms of $\Theta_1$. Inserting these relations gives
\ben
&& i\omega\Theta_{1}
+ k\Big(1-\frac{A\phi'^{2}}{2a^{2}} \Big)^{2}
\Big(-\frac{8k}{27\tau'}+ \frac{k}{3i\omega}\Big)\Theta_{1}
\nonumber\\ &&= {\tau}'\Theta_{1}\Big(\frac{\omega^{2}R^{2}}{\tau'^{2}}-\frac{i\omega R}{\tau'}\Big).
\label{G3}
\een
Dividing by $\Theta_1$ and collecting terms, we arrive at
\ben
&& i\omega(1+R)
+\frac{k^{2}}{3i\omega}\Big(1-\frac{A\phi'^{2}}{2a^{2}} \Big)^{2}
\nonumber\\ &&-\frac{8k^{2}}{27\tau'}\Big(1-\frac{A\phi'^{2}}{2a^{2}} \Big)^{2}
-\frac{\omega^{2}R^{2}}{\tau'}=0.
\label{G4}
\een
Finally, multiplying by $-i\omega$ leads directly to Eq.~\eqref{130},
\ben
&&{\omega}^{2}(1+R)-\frac{k^{2}}3\Big(1-\frac{A\phi'^{2}}{2a^{2}} \Big)^{2}
\nonumber\\ &&+ \frac{i\omega}{{\tau}'}\Big[\omega^{2}R^{2}+\frac{8k^{2}}{27}\Big(1-\frac{A\phi'^{2}}{2a^{2}}\Big)^{2}\Big]=0.
\label{G5}
\een

\section{Consistency of the K-essence background with matter backreaction}
\label{app:background_consistency}

This appendix explains how the homogeneous K-essence background used in the main text is to be understood when matter and radiation are coupled
to the emergent geometry and are evolved with the Boltzmann equation.
The key assumption is a \emph{controlled decoupling} of the homogeneous
(background) scalar dynamics from the matter-radiation sector: the
Boltzmann hierarchy is kept in full, while the feedback of the matter
and radiation stress-energy on the \emph{background} scalar equation is
neglected at leading order.

Schematically, we write the total action as
\begin{equation}
S_{\rm tot}=S_{\rm grav}+S_{\phi}+S_{\rm m},
\label{eq:br_action}
\end{equation}
with
\begin{equation}
S_{\phi}=\int d^4x\,\sqrt{-g}\,\La(X,\phi),
\qquad
X\equiv\frac12 g^{\mu\nu}\nabla_{\mu}\phi\nabla_{\nu}\phi,
\label{eq:br_Sphi}
\end{equation}
while $S_{\rm m}$ denotes the matter and radiation action coupled to the
emergent metric $\bar G_{\mu\nu}(\phi,\partial\phi)$. Because $S_{\rm m}$
depends on $\phi$ through $\bar G_{\mu\nu}$, varying the total action
with respect to $\phi$ yields
\begin{equation}
\mathcal{E}_{\phi}=\mathcal{S}_{\rm m},
\label{eq:br_scalarEOM}
\end{equation}
where $\mathcal{E}_{\phi}$ is the contribution from $S_{\phi}$ and
$\mathcal{S}_{\rm m}$ is the matter-induced source.

For a homogeneous configuration $\phi=\phi(t)$ one has
\begin{equation}
\mathcal{E}^{(0)}_{\phi}=\mathcal{S}^{(0)}_{\rm m}.
\label{eq:br_homEOM}
\end{equation}
The background adopted in the main text corresponds to the leading-order
approximation
\begin{equation}
\mathcal{S}^{(0)}_{\rm m}=0
\qquad\Rightarrow\qquad
\mathcal{E}^{(0)}_{\phi}=0,
\label{eq:br_decoupling}
\end{equation}
i.e., the homogeneous K-essence solution is taken from the pure scalar
sector.

The consistency condition is that the neglected source term be small
compared to the \emph{typical size} of the intrinsic K-essence terms in
the homogeneous equation. Since $\mathcal{E}^{(0)}_{\phi}=0$ on the
leading-order solution, the ratio $|\mathcal{S}^{(0)}_{\rm m}/\mathcal{E}^{(0)}_{\phi}|$
is not meaningful. Instead, we introduce the dimensionless measure
\begin{equation}
\epsilon_{\rm br}\equiv
\frac{|\mathcal{S}^{(0)}_{\rm m}|}{|\mathcal{E}^{(0)}_{\phi}|_{\rm typ}},
\label{eq:br_epsbr}
\end{equation}
where $|\mathcal{E}^{(0)}_{\phi}|_{\rm typ}$ denotes the characteristic
magnitude of the terms entering $\mathcal{E}^{(0)}_{\phi}$ (for example,
those of order $H\dot\phi\,\La_X$, $\ddot\phi\,\La_X$, or $\dot\phi^2 \La_{XX}$,
depending on the specific $\La$). The leading-order background treatment
is self-consistent when
\begin{equation}
\epsilon_{\rm br}\ll 1.
\label{eq:br_small}
\end{equation}

A sufficient (though not necessary) criterion is that the scalar energy
density dominates the expansion rate during the epoch of interest,
\begin{equation}
\epsilon_{\rho}\equiv\frac{\rho_{\rm m}+\rho_{\rm r}}{\rho_{\phi}}\ll 1,
\label{eq:br_epsrho}
\end{equation}
so that the matter-radiation sector provides only a perturbative
correction to the background evolution. More generally, even when
$\rho_{\rm m}+\rho_{\rm r}$ is not parametrically small, the approximation
\eqref{eq:br_decoupling} can still hold provided the coupling of
$S_{\rm m}$ to $\phi$ through $\bar G_{\mu\nu}$ is sufficiently weak, so
that $\mathcal{S}^{(0)}_{\rm m}$ is suppressed.

This approximation should not be confused with neglecting matter and
radiation in the kinetic theory: the distribution functions $f_s(x,p)$,
number densities, and collision integrals are kept throughout. The
approximation is solely that the resulting matter-induced source
$\mathcal{S}^{(0)}_{\rm m}$ is dropped in the \emph{homogeneous} scalar
background equation at leading order. The computation is therefore
organized as
\ben
&&\left.
\begin{array}{c}
\text{leading-order homogeneous K-essence background}\\[2mm]
\mathcal{S}^{(0)}_{\rm m}=0
\end{array}
\right\}
\n &&
\quad\longrightarrow\quad
\bar G_{\mu\nu}
\n &&
\quad\longrightarrow\quad
\text{Boltzmann evolution of matter and radiation}.\n
\label{eq:br_flow}
\een\\

A fully backreacted treatment would require solving, in a single coupled
system, the background gravitational equations, the homogeneous scalar
equation including $\mathcal{S}^{(0)}_{\rm m}$, and the matter-radiation
Boltzmann hierarchy. This coupled Einstein-K-essence Boltzmann problem
is beyond the scope of the present analytical study. We therefore work at
leading order in the small-backreaction expansion \eqref{eq:br_small}:
the K-essence field is taken to be a fixed homogeneous background ($\delta\phi=0$), which defines the emergent metric $\bar G_{\mu\nu}$ used in the kinetic theory. Matter and radiation are nevertheless treated fully at the level of their phase-space distribution functions and collision integrals; what is neglected is only their feedback on the homogeneous scalar background. A fully coupled evolution including this backreaction is left for future work.

\end{document}